\documentstyle[epsfig]{ioplppt}    
\begin{document}

\def\lapproxeq{\lower .7ex\hbox{$\;\stackrel{\textstyle <}{\sim}\;$}}
\def\gapproxeq{\lower .7ex\hbox{$\;\stackrel{\textstyle >}{\sim}\;$}}

{\bf\large DESY 96-049\hfill ISSN 0418-9833}\\

{\bf\large DTP/96/20}\\

{\bf\large March 1996}

\title{\begin{center} 
Working Group Report on the\\
STRUCTURE OF THE PROTON
\footnote{To be published in the Proc. of the Durham Workshop on 
``HERA Physics: Proton, Photon and Pomeron Structure'', J. Phys. G.}
\end{center}}

\author{B.\ Bade{\l}ek}
\address{Institute of Experimental Physics, Warsaw University, PL-00681, 
Warsaw, Poland,}
\address{and Physics Institute, University of Uppsala, S-751 21 Uppsala, Sweden} 
\vskip-0.5cm
\author{J.\ Bartels}
\address{II Institut f\"{u}r Theoretische Physik, Universit\"{a}t Hamburg, 
D-22761 Hamburg, Germany} 
\vskip-0.5cm
\author{N.\ Brook}
\address{Department of Physics, University of Glasgow, Glasgow, G12 8QQ, 
Scotland}
\vskip-0.5cm
\author{A.\ De Roeck}
\address{DESY, Notkestrasse 85, D-22603 Hamburg, Germany} 
\vskip-0.5cm
\author{T.\ Gehrmann}
\address{Department of Physics, University of Durham, Durham, DH1 3LE, England}
\vskip-0.5cm
\author{M.\ Lancaster}
\address{Department of Physics, University of Oxford, Oxford, OX1 3NP, England}
\vskip-0.5cm
\author{A.\ D.\ Martin}
\address{Department of Physics, University of Durham, Durham, DH1 3LE, England}
\vskip-0.5cm
\author{A.\ Vogt}
\address{DESY, Notkestrasse 85, D-22603 Hamburg, Germany} 
\address{On leave from Sektion Physik, Universit\"{a}t 
M\"{u}nchen, D-80333 Munich, Germany}

\begin{abstract}
We summarize the developments on the structure of the proton that
were studied at the Workshop on \lq\lq HERA Physics" that was
held in Durham in September 1995.  We survey the latest structure
function data; we overview the QCD interpretations of the
measurements of the structure functions and of final state
processes; we discuss charm production and the spin properties of
the proton.
\end{abstract}

\newpage
\setcounter{section}{0}
\setcounter{equation}{0}
\renewcommand{\theequation}{1.\arabic{equation}}
\setcounter{figure}{0}
\renewcommand{\thefigure}{1.\arabic{figure}}
\setcounter{table}{0}
\renewcommand{\thetable}{1.\arabic{table}}

\section{Introduction}
One of the highlights of the previous Durham Workshop on HERA physics in March 
1993 was the presentation \cite{ADR93}, for the first time, of the measurements 
of the proton structure function $F_2 (x, Q^2)$, which showed a dramatic rise 
with decreasing $x$.  They immediately ruled out a Regge-based description, 
which is so successful for hadronic and photoproduction 
$(Q^2 = 0)$ cross sections.  The Regge model predicts a much slower rise with 
decreasing $x$ than was observed in deep inelastic measurements of $F_2$.  The 
observation generated great activity at the 1993 Workshop, particularly as 
application of the BFKL equation \cite{BFKL} had anticipated a singular 
$x^{- \lambda}$ growth with decreasing $x$.  To quote from the 
Introduction to the Proceedings \cite{RCED}: 

\begin{quote}
\lq\lq The argument over the interpretation of the rise began immediately and 
continued 
all week --- was it evidence for the singular BFKL behaviour of the gluon; was 
it just evidence of the need for a different input parametrization; what did it 
imply for the Pomeron and diffractive scattering?  Although the protagonists 
tried very hard, it will take a lot more data and quite a few more Workshops to 
answer all these exciting questions!"
\end{quote}

Now two and a half years on, at the time of this Workshop (September 1995), the 
measurements by H1 and ZEUS have indeed improved remarkably.  Our knowledge 
of diffractive scattering has blossomed and it now has its own Working Group.  
Both the 
precision and the kinematic range of the HERA measurements of $F_2$ have 
greatly increased.  We now have measurements for $x < 10^{-4}$ and $Q^2 \sim 
1.5 \:$ GeV$^2$.  In fact one of the tasks undertaken at the 
Workshop was a study of the consistency between all the measurements of $F_2$ 
(see section 2).  The present surprise is that the strong rise of $F_2$ with 
decreasing 
$x$ appears to persist down to the lowest observed $Q^2$ ($Q^2 \sim 1.5 \:$ 
GeV$^2$) and yet the photoproduction measurements show that the slow rise 
of the conventional Regge description works well at $Q^2 = 0$.  The transition 
from the \lq\lq soft" to the \lq\lq hard" regime appears to be very rapid.  
It is appropriate 
to review our knowledge of the low $Q^2$ behaviour of $F_2$; section 3 
contains a summary of our discussions.
In fact sections 2 and 3 contain an overview of all deep inelastic data.
The high precision measurements of the fixed-target experiments
are now being extended, for $F_2(ep)$ at least, to much lower $x$ by the 
experiments at HERA. Specially compiled plots are presented to demonstrate
the complementary nature of these experiments, as well as to indicate their
relative errors. Also we summarize the information obtained about 
nuclear shadowing.

Partons satisfying the conventional GLAP(Altarelli-Parisi) \cite{GLAP} 
evolution seem to be well able to 
describe the new HERA 
and fixed-target deep inelastic data down to as low as $Q^2 \sim 1 \:$ 
GeV$^2$.  Could this be an indication of the precocious onset of 
perturbative QCD and the absence 
of higher twists?  Such a conclusion would be premature.  At larger $x$ ($x 
\gapproxeq 0.01$) there are many different types of high-precision fixed-target 
constraints on the individual parton distributions and, with the possible 
exception of the gluon, the partons are well determined.  On the other hand 
at small $x$ ($x \lapproxeq 10^{-3}$) we have, so far, only one type of 
structure function measurement, namely $F_2 (x, Q^2)$, and there 
are many different partonic descriptions at small $x$, particularly as we have 
to supply the (non-perturbative) partonic input at some scale $Q_0^2$ for the 
GLAP leading $\log Q^2$ limit, or at some $x_0$ for the BFKL leading $\log 
(1/x)$ 
limit, of the perturbative QCD evolution.  At small $x$ the dominant 
parton is the gluon and 
the description of the observables is driven by the behaviour of the gluon 
distribution.  According to perturbative QCD we expect the small $x$ behaviour 
of the (sea) quark and gluon distributions to be strongly correlated due to the 
$g \rightarrow q\overline{q}$ transition.  In section 4 we overview the 
different QCD contributions to $F_2$ at small $x$.  We see that $F_2$ is too 
inclusive a measurement to be 
easily able to distinguish the dominant underlying perturbative QCD 
component at small $x$.  Thus some 
of the questions of 1993 remain, but we now have a much clearer idea why the 
search for the answers is so difficult.

Unlike the conventional GLAP gluon distribution, the BFKL distribution $f 
(x, k_T^2)$ 
is unintegrated over the transverse momentum $k_T$ of the gluon.  The BFKL 
gluon 
has two main characteristics.  It has the familiar $x^{- \lambda}$ growth with 
decreasing $x$, but accompanied by a diffusion in $k_T$ --- typically 
a Gaussian distribution in $\log k_T^2$ about some initial transverse momentum 
which broadens as $\sqrt{\log (1/x)}$ as $x$ decreases.  The inclusive 
measurement $F_2$ integrates over this characteristic $k_T^2$ dependence and, 
worse, they sample contributions arising from diffusion to the 
(non-perturbative) 
low $k_T$ region (see section 4).  This suggests the value of studying deep 
inelastic final state processes which simultaneously expose both the 
characteristic 
$x$ and $k_T^2$ dependence of the BFKL gluon.  Section 5 discusses two such 
measurements, the energy flow in the central region and the production of 
forward jets.  Whereas the former one seems to be very sensitive to (unknown) 
hadronization effects and therefore, in its present form, cannot be considered 
to be a reliable test for BFKL dynamics, the second approach looks more 
promising:  
data of the 1993 HERA run are confronted with both the analytical BFKL 
prediction 
and fixed-order matrix element calculations, and the agreement with the BFKL 
calculation is quite encouraging.

We devote section 6 to charm production at HERA, processes which are directly 
driven by the 
gluon.  Both open charm, for instance the sizeable charm component $F_2^c$ of 
$F_2$, and $J/\psi$ production processes offer valuable information on the 
small $x$ behaviour of the gluon distribution.  We discuss the status of 
fixed-order calculations of $F_2^c$, present numerical predictions and try to 
answer in some detail to what extent the measurement of the charm component of 
$F_2$ can be used to pin down the gluon structure function.  Particular 
attention 
is also given to $J/\psi$ production which may be able to discriminate between 
different gluon distributions.

Last, but not least, in section 7 we summarize our discussions on the spin 
structure of the proton.  Many experiments, prompted by the original surprising 
result on the proton spin structure function by EMC, are now yielding much 
information.  After a brief introduction of the general background we give an 
overview of the present experimental situation and the status of the Bjorken 
and Ellis-Jaffe sum rules.  The small $x$ behaviour of $g_1$ seems to be of 
particular interest:  evaluation of the first moment $\Gamma_1$ requires an 
appreciable extrapolation of $g_1$, while recent QCD predictions indicate a 
rather strong rise of $g_1$ in the small $x$ region.  Finally section 8 
contains some brief conclusions.
\newpage

\setcounter{section}{1}
\setcounter{equation}{0}
\renewcommand{\theequation}{2.\arabic{equation}}
\setcounter{figure}{0}
\renewcommand{\thefigure}{2.\arabic{figure}}
\setcounter{table}{0}
\renewcommand{\thetable}{2.\arabic{table}}

\section{Overview of structure function data}
\subsection{Introduction}
Here we review the status of the data on structure functions and
derived quantities. Until a few years ago this knowledge came entirely
from fixed-target experiments. Now it is being complemented
and extended by the results from the HERA $ep$ collider, especially
in the region of low $x$ where the dynamics of a 
large number of confined partons has to be understood, as well as at
very high scales where the perturbative assumptions can be further
tested.
 
\begin{figure}[htb] 
\begin{center}
\epsfig{file=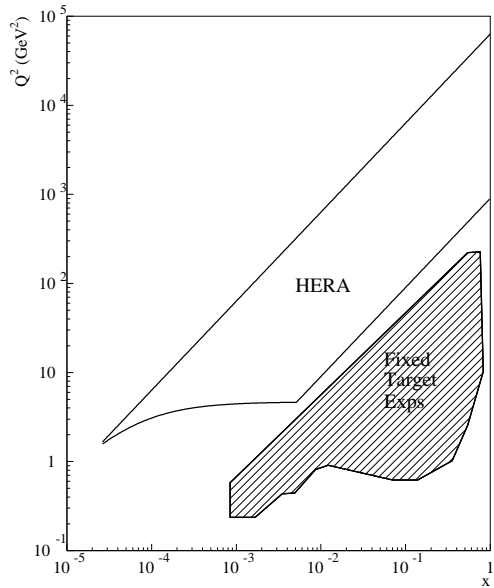,bbllx=50pt,bblly=100pt,bburx=550pt,bbury=700pt,%
height=8cm}
\end{center}
\caption{The kinematical region covered by the HERA and fixed-target
experiments.}
\label{sec2f1}
\end{figure}
  
At this Workshop the HERA experiments presented 
results from the 1994 data taking period~\cite{mark}.
During this period HERA collided 27.5 GeV positrons on 820 GeV protons,
as opposed to the 26.7 GeV electrons in 1992 and 1993.
The centre-of-mass energy of the $ep$ collision is 300 GeV.
The new $F_2$ results extend to larger $Q^2$ values due to
a tenfold increase in statistics, compared to the  data
collected in 1993  (and about a factor of hundred increase 
with respect to the 1992
data, on which the results shown at the previous Durham
workshop were based). Furthermore both the H1 and ZEUS experiments
have made a special effort to obtain measurements at lower
values of $Q^2$ so as to explore the region towards $Q^2\rightarrow 0$.
Values of $Q^2$ down to about 1 GeV$^2$ have been reached
as a result of detector upgrades,
operating HERA in a different collision mode called shifted
interaction vertex mode, and by using events with initial state 
QED radiation from the incoming lepton.
At the time of the Workshop all data were still
preliminary, but by now have become final to a large
extent~\cite{H1f294,zeuslowq2}. For the QCD
interpretation of the data, such as the extraction of the gluon
density, the 1993 data will be used.
 
In the HERA experiments, the scattered electron as well as the hadronic
final state are measured.
Apart from tantalizing questions on the dynamics of the hadronic
final state itself (see Section 5) this allows us to determine the kinematics
both from the scattered electron and from the hadronic final state. 
In practice, for the latter,
a mixture of the hadron and electron information is used, rather than
exclusively hadronic final state quantities.
 
The kinematic plane covered by HERA and the fixed-target measurements
 is shown in Fig.~\ref{sec2f1}.
Generally measurements at HERA can reach $Q^2$ ($x$) values two orders of
magnitude larger (smaller) than those reached by fixed-target experiments.
Fig.~\ref{sec2f1} shows that the two regions make contact and
thus the  continuity and normalization
of the data can be checked. New upgrades of the
HERA detectors will allow the exploration of even lower $Q^2$ in
the future.
 
The (unpolarised) fixed-target deep-inelastic
scattering programme will come to an end in 1996
except for the continuation of the CCFR neutrino experiments. Many experiments
(SLAC, BCDMS, EMC, NMC, E665, CDHSW, BEBC and CCFR) have contributed
to the heroic and successful effort to obtain a fundamental and precise
knowledge of properties of partons and of QCD.
Characteristics of a number  of them are listed in Table~\ref{table2.1}.
The fixed-target electron (muon) scattering experiments were almost always
inclusive, i.e. information on the  kinematic variables came only 
from measurements of
the incident and scattered leptons. In the charged-current neutrino
experiments the outgoing muon and the total energy of the produced
hadrons are measured,
and in neutral-current experiments only the latter is detected.
 
In deep-inelastic experiments the low $x$ region is correlated
with low values of $Q^2$, as shown in  Fig.~\ref{sec2f1}.
For fixed-targets the lowest values of
$x$ were reached by the NMC at CERN and E665 Collaboration at FNAL
applying special experimental techniques permitting
measurements of muon scattering angles as low as 1 mrad. These ``small $x$
triggers" and special off-line selection methods were also effective against
the background of muons scattered
elastically from target atomic electrons which produce a peak  at
$x=$0.000545. Systematic errors on $F_2$ in both experiments 
(in particular those on the
ratio of structure functions for different nuclei, $F_2^a/F_2^b$) were greatly
reduced as a result of exposing  several target materials at the time and/or
by a frequent exchange of targets in the beam.
 
\begin{table}
\caption{Fixed-target experiments contributing to the $F_2$
measurements. The $x$ and $Q^2$ ranges refer to the $F_2$ data;
structure function ratio measurements extend the low limits by 
approximately an order of magnitude.}
\begin{center}
\begin{tabular}{|c|c|c|l|l|}
\hline \hline
 Beam & Targets & Experiment &  $Q^2$(GeV$^2$) & $x$ \\ \hline
\hline
$e$ & p,d,A & SLAC & 0.6 --  ~~30 & 0.07~~~ -- 0.8 \\
$\mu$ & p,d,A & BCDMS & 7.5 -- 230 & 0.07~~~ -- 0.6 \\
$\mu$ & p,d,A & NMC & 0.5 --  ~~75 & 0.006~ -- 0.6 \\
$\mu$ & p,d,A & E665 & 0.2 --  ~~75 & 0.0008 -- 0.6 \\
$\nu$,$\bar \nu$& Fe & CCFR, CDHSW & 1.0 -- 500 & 0.015~ -- 0.6 \\
\hline \hline
\end{tabular}
\end{center}
\label{table2.1}
\end{table}
 
In the one-photon-exchange approximation, the differential
electroproduction cross section is 
related to the structure
function $F_2(x,Q^2)$ and the ratio $R(x,Q^2)$ of the cross sections
for the longitudinally and transversally polarised virtual photons by
\begin{equation}\fl
{d^2\sigma (x,Q^2)\over dQ^2dx}={4\pi \alpha^2\over Q^4x}
\left[1-y-
{Mxy\over 2E} + \left(1 - {2m^2\over Q^2}\right)
{y^2(1+4M^2x^2/Q^2)\over 2(1+R)}\right] F_2(x,Q^2)
\end{equation}
where $M$ and $m$ are the mass of the proton and the electron respectively,
and $E$ is the incident lepton energy.
For the HERA kinematics by neglecting $M$ and $m$ this expression reduces to:  
\begin{equation}
  \frac{d^2\sigma}{dx dQ^2} =\frac{4\pi\alpha^2}{Q^4x}
    \left[1-y+\frac{y^2}{2(1+R)}\right] F_2(x,Q^2).
\label{dsigma}
\end{equation}
The function $R(x,Q^2)$ has so far been measured only in fixed-target
experiments, but even here information is scarce. The usual
procedure to determine the $F_2(x,Q^2)$ is to assume
a value of $R(x,Q^2)$ (theoretical, experimental or a combination of these)
 and then to extract $F_2(x,Q^2)$ from the data,
using an iterative comparison of the experimental yield (corrected for
 acceptance,
inefficiency of the apparatus as well as for higher-order QED
processes) with the electroproduction cross section.
For the HERA measurements $R$
was calculated using the  QCD relation~\cite{Altmar} with the
GRV structure function parametrization.
At small $x$ and $Q^2$
the assumed $R$ values  can be as large as 1. Note that
a 20\% error on $R$ corresponds to about
a 2\% uncertainty on $F_2$ at $y=0.6$ for $R$ of about 0.6.
The effects due to $Z$ boson exchange 
for neutral current interactions in the presently covered high
$Q^2$ region for $F_2$ at HERA 
regime amount to a few percent only.
 New data on 
$R(x,Q^2)$ from fixed-target experiments will be discussed in
Section 2.5.
 
In this Section we present the status of the $F_2$ measurements
at HERA and in the fixed-target experiments, QCD fits to these data,
measurements of $R(x,Q^2)$ and
tests of the (flavour singlet and nonsinglet) sum rules.
 
\begin{figure}[htb]
\begin{center}
\epsfig{file=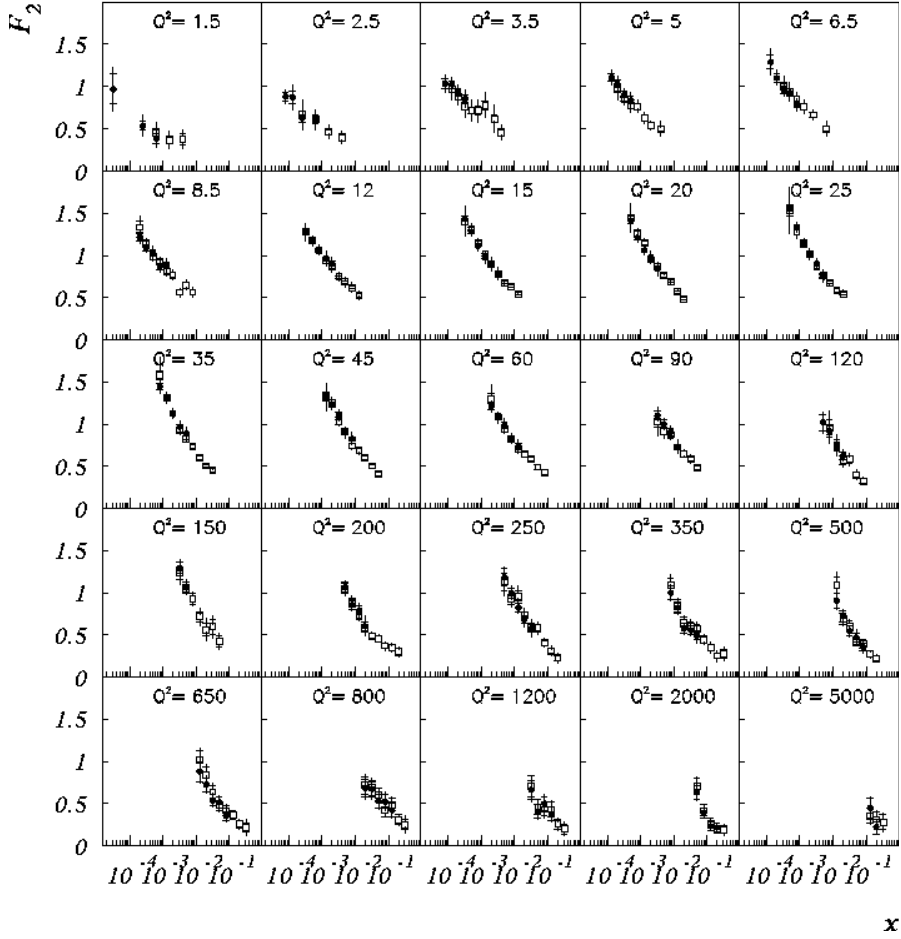,bbllx=35pt,bblly=155pt,bburx=550pt,bbury=690pt,%
width=13cm}
\end{center}
\caption{Measurement of the structure function with the electron (closed
circles) and the $\Sigma$ method (open squares)
by the H1 Collaboration. The inner error bar
is the statistical error. The full error bar represents the
statistical and systematic errors added in quadrature disregarding the
luminosity error (1.5\%).}
\label{sec2f4}
\end{figure}
 
\begin{figure}[htb]
\begin{center}
\epsfig{file=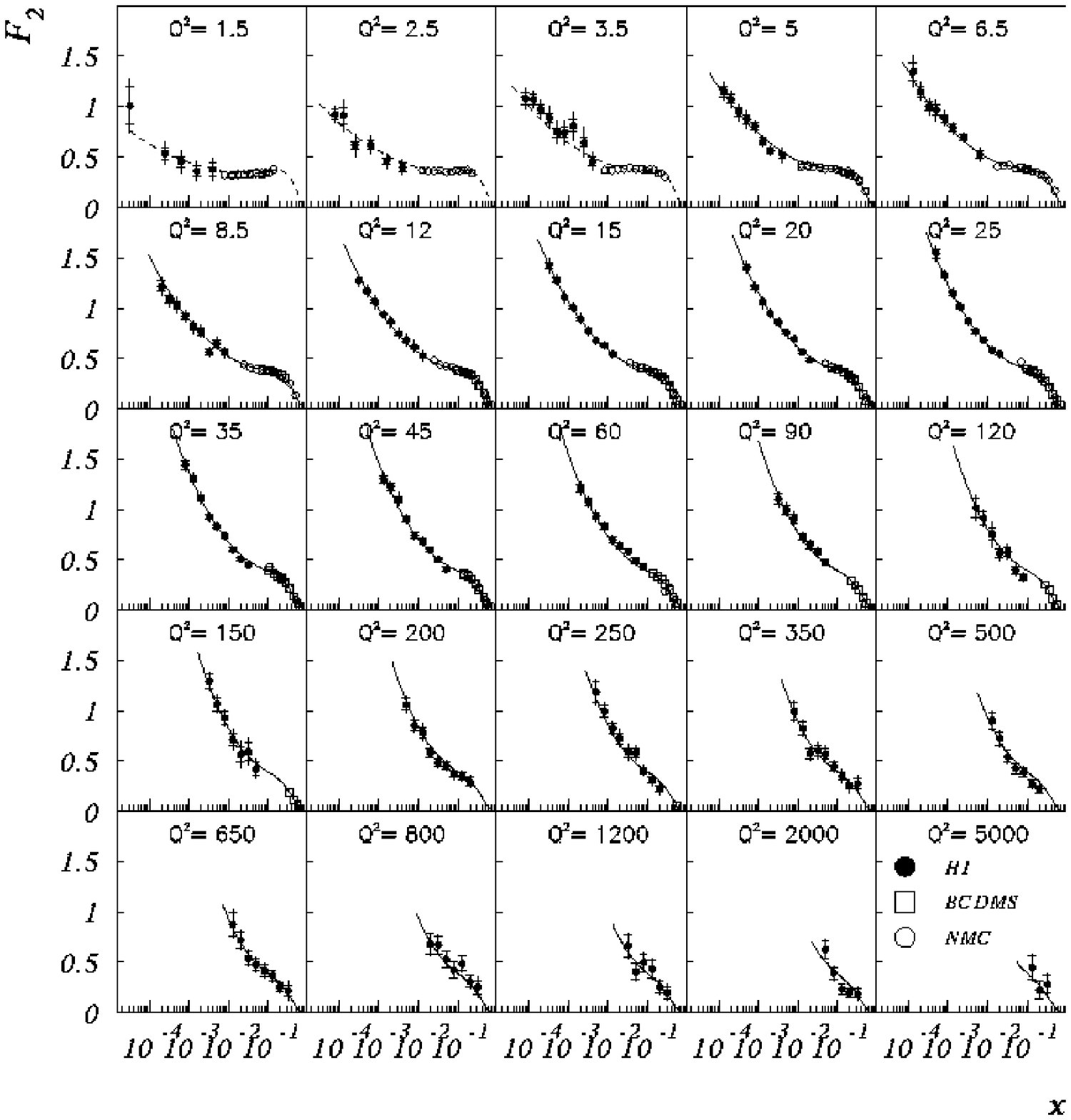,bbllx=15pt,bblly=170pt,bburx=550pt,bbury=670pt,%
width=13cm}
\end{center}
\caption{Measurement of the structure function $F_2(x,Q^2)$ as function
of $x$. The closed circles are H1 data, the open circles are data 
from NMC \protect \cite{f2nmc} and BCDMS \protect \cite{f2bcdms}.
The inner error bar is the statistical error. The full error bar represents the
statistical and systematic errors added in quadrature disregarding the
luminosity errors. The curves represent the NLO QCD fit, discussed in  
Section 2.4.}
\label{f2xall}
\end{figure}
 
\subsection{Measurement of $F_2(x,Q^2)$ at HERA}
 
Both the H1 and ZEUS experiments have released new data on
structure function measurements at small $x$.
It was noted above, that to determine
the kinematical variables $x$ and $Q^2$, we can use two out of 
four experimentally accessible quantities: the energy, $E_e^{\prime}$,
and angle, $\theta_e$, of the
scattered electron, and the energy, $E_h$, and average angle, $\theta_h$, 
of the hadron flow. The ultimate method is a global
fit of all observed quantities, which requires a level of understanding
of the detector response and of the error correlations that the
experiments have not yet achieved. In total four methods
are currently used in the analyses to reconstruct
the event kinematics. The electron method,
is the method used so far in all fixed-target experiments.
Here the basic formulae for $Q^2$ and $y$ are
\begin{equation}
  y_e   =1-\frac{E^{\prime}_e}{E_e} \sin^{2}\frac {\theta_e} {2}
   \hspace*{1cm}
   Q^2_e = 4E^{\prime}_eE_e\cos^2\frac{\theta_e}{2}
= \frac{E^{'2}_e \sin^2{\theta_e}}{ 1-y_e}. 
\label{kinematics1}
\end{equation}
The polar angle $\theta_e$ is defined with respect to the proton beam 
direction, referred to as ``forward" region. $E_e$ is the incident
electron energy. It remains at HERA
the most precise way to reconstruct $Q^2$ in the whole kinematic
range. However at low $y$ ($y < 0.1$) the measurement of $x$ becomes poor
and at large $y$ ($y > 0.8$) the radiative corrections which need to be
applied to the observed cross  section to extract
the Born cross section are very large.
The mixed method  used by the H1 Collaboration in 1992 takes $Q^2$ from the
electron  and $y$ from the hadronic variables ($y_{h}$) according to
\begin{equation}
y_{h} = \frac{\Sigma_h(E-P_z)_h}{2E_e} 
\label{yjb}
\end{equation}
where the sum includes all detected hadrons $h$, which have an energy $E$ and
longitudinal momentum component $P_z$.
The resolution of $y_{h}$ is better than the resolution of $y_e$ for low $y$
values but becomes inferior at large $y$ values.
For the double angle method (DA)~\cite{DOUBLEANGLE}
only the angles of the scattered electron and  the hadronic system
are used. The method is almost independent of energy scales in
the calorimeters but, at very low $y$,
the method is very sensitive to noise in the calorimeters.
The variables $y$ and $Q^2$ are reconstructed from
\begin{equation}\fl
y_{DA}=\frac{\sin\theta_e(1-\cos\theta_h)}
                {\sin\theta_h+\sin\theta_e-\sin(\theta_e+\theta_h)}
\hspace*{0.5cm}
Q^2_{DA}=4E^2_e\frac{\sin\theta_h(1+\cos\theta_e)}
                {\sin\theta_h+\sin\theta_e-\sin(\theta_e+\theta_h)}
\end{equation}
with
\begin{equation}
\tan\frac{\theta_h}{2}=\frac{\Sigma_h(E-P_z)_h}{P^2_{T,h}}.
\end{equation}
This method has been used by the ZEUS Collaboration.
A new method used by the H1 Collaboration~\cite{H1F2,bernardi},
called the $\Sigma$ method ($\Sigma$), determines $y$ and $Q^2$ from  
\begin{equation}
y_{\Sigma}=\frac{\Sigma_h(E-P_z)_h}
                {(E-P_z)_e+\Sigma_h(E-P_z)_h}
\hspace*{1cm}
Q^2_{\Sigma}=\frac{E^{\prime 2}_e \sin^2{\theta_e}}{1-y_{\Sigma}}
\end{equation}
where the sum runs over  all hadrons in the numerator and over all
hadrons plus the scattered electron in the denominator. In this
method the energy of the incident electron in the interaction is
reconstructed, which reduces drastically the sensitivity
to the main radiative process.
The resolution in $x$ at low $y$ is good enough to allow the
H1 Collaboration to reach $y=0.01$. The resolution at large $y$
is worse but less sensitive to radiative corrections
than when using only the measurement of the scattered electron. For
precision measurements of the structure function the different methods are
used to control the systematics of event smearing and radiative corrections.
In Fig.~\ref{sec2f4} the results using the electron and $\Sigma$ methods
are compared for the H1 data, showing a very good agreement.
 
\begin{figure}[htb] \centering \unitlength 1mm
 \epsfig{file=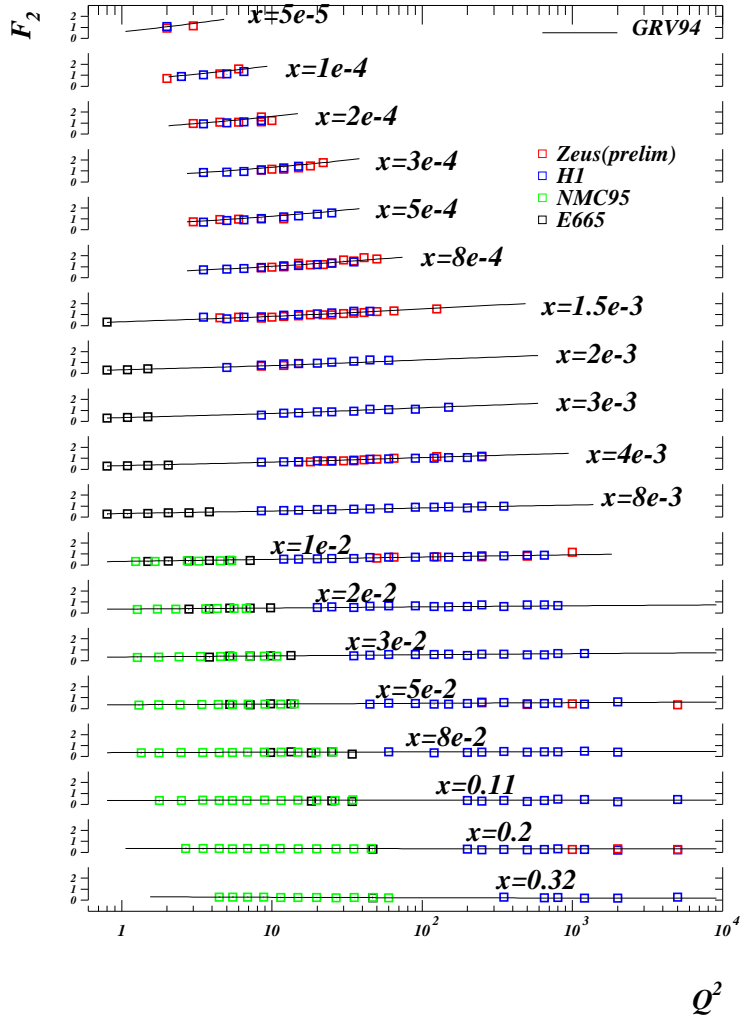,bbllx=0pt,bblly=50pt,bburx=575pt,%
bbury=850pt,height=14cm}
\caption{Measurement of the structure function $F_2(x,Q^2)$ as function
of $Q^2$ for the H1, ZEUS, NMC and E665 experiments. The curve is
the prediction of the GRV parton distributions.}
\label{sec2f5}
\end{figure}
 
\begin{figure}[htb]
\begin{center}
 \epsfig{file=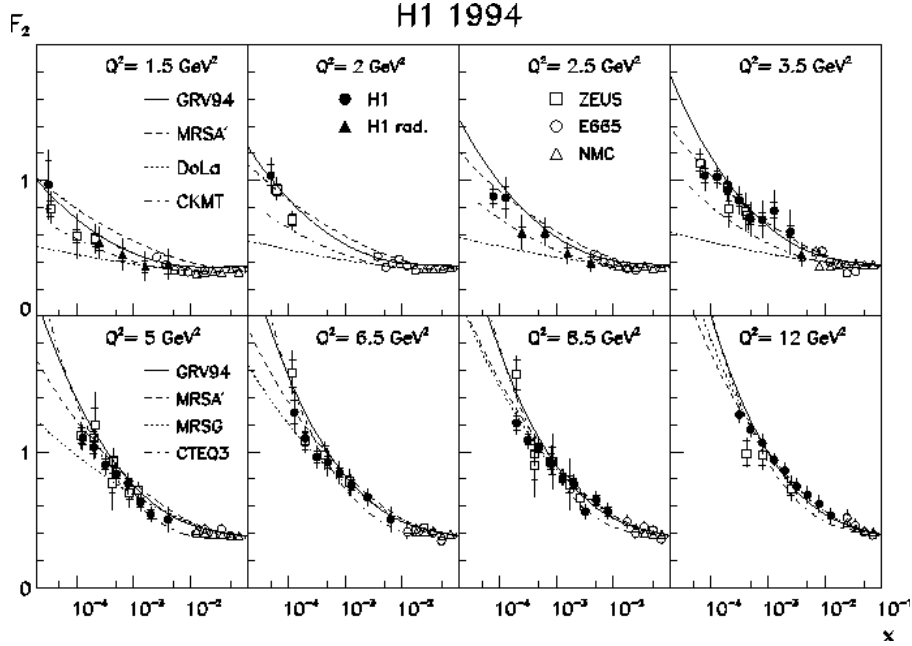,width=10cm,%
 bbllx=100pt,bblly=250pt,bburx=480pt,bbury=600pt}
\end{center}
\caption{Measurement of the proton structure function $F_2(x,Q^2)$
in the low $Q^2$ region by H1 (closed circles: non-radiative events;
closed triangles:  radiative events), together with results from the
ZEUS (open squares), E665 (open points) and NMC (open triangles) experiments.
Different parametrizations for $F_2$ are compared to the
data. The DOLA and CKMT curves are only shown for the upper row of $Q^2$
bins; CTEQ3M and MRSG are shown for the lower row; GRV
and MRSA$^{\prime}$ are shown for the full $Q^2$ range.
The $Q^2$ values of the ZEUS data shown for the bins $Q^2$ = 3.5, 5 and 6.5
GeV$^2$ are measurements at 3.0, 4.5 and 6 GeV$^2$ respectively.}
\label{sec2f7}
\end{figure}
 
\begin{figure}[htb] \centering \unitlength 1mm
 \epsfig{file=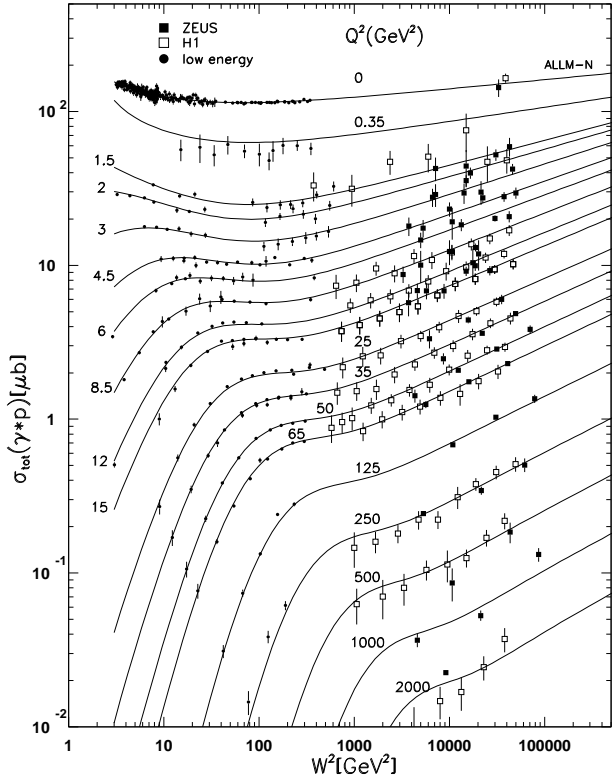,height=10cm}
\caption{Measurement of the proton structure function $F_2(W,Q^2)$
as function of $W^2$. The inner error bar is the statistical error. 
The full error represents the statistical and systematic errors added
in quadrature. Also data at $Q^2$ = 0 are shown. Note that this figure
contains the preliminary H1 data \protect{\cite{RT}}. The curves are an 
ALLM parametrization \protect{\cite{ALLM4}}.}
\label{sec2f8}
\end{figure}
 
\begin{figure}[htb]
\begin{minipage}[htb]{6cm}
\epsfxsize=5cm
\centering
 \epsfig{file=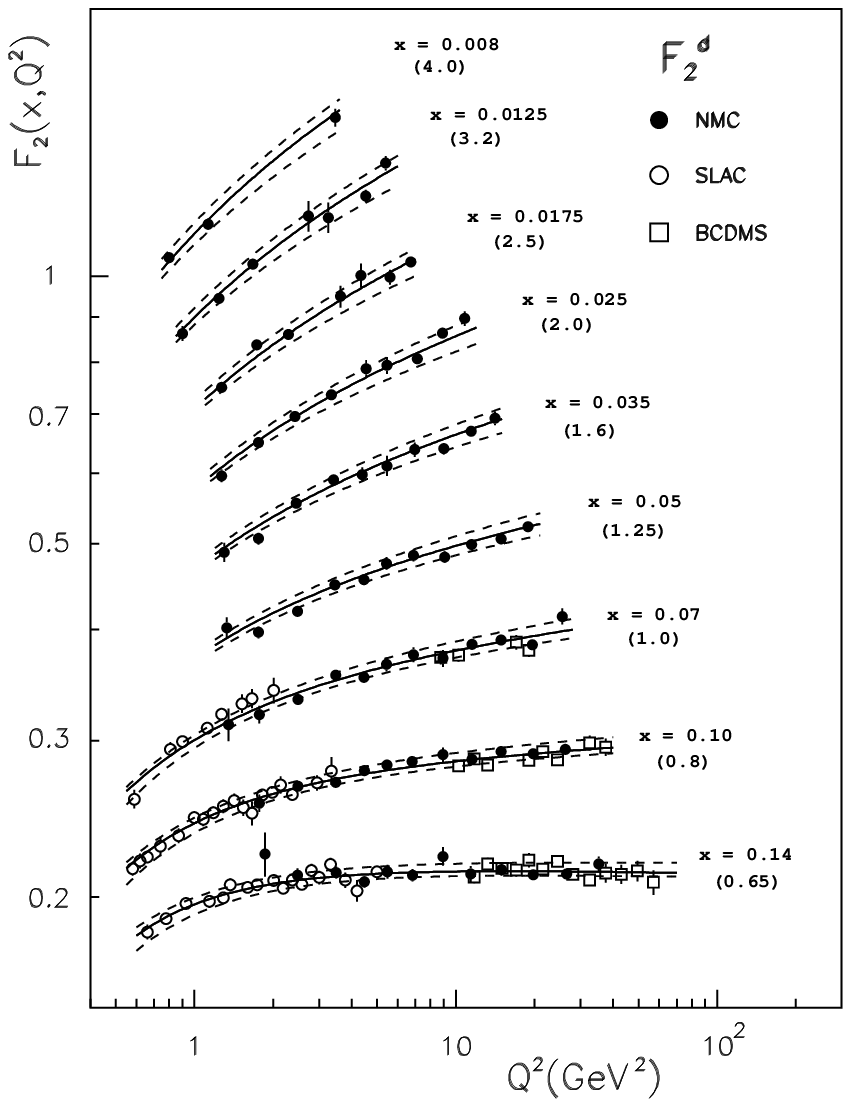,height=90mm,width=70mm}
\end{minipage}
\begin{minipage}[htb]{6cm}
\epsfxsize=5cm
\centering
 \epsfig{file=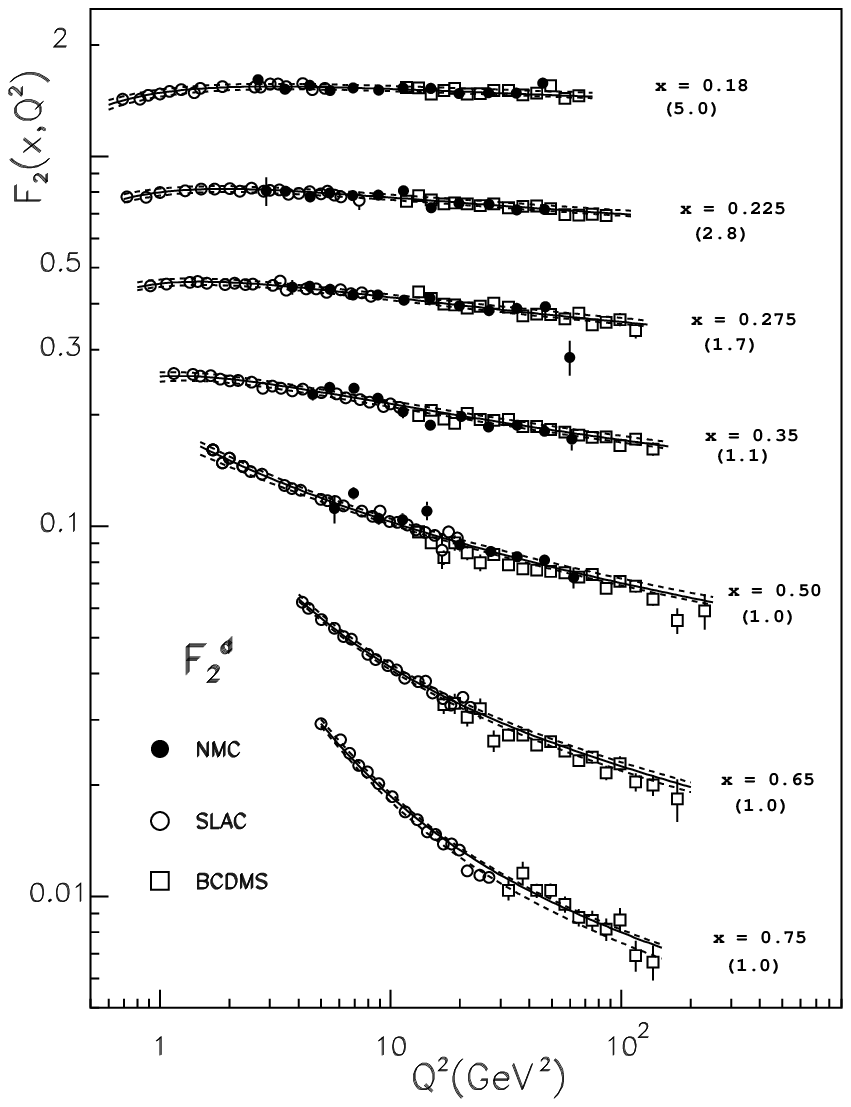,height=90mm,width=70mm}
\end{minipage}
\\
\caption{The data from NMC compared with the data from SLAC and BCDMS. 
The errors are statistical. The solid curves are the
results of a 15-parameter fit to all three data sets. The dashed
curves correspond to the extreme values of the parameters (from
\protect \cite{f2nmc}).}
\label{fig_f2nmc}
\end{figure}
 
Due to the inevitable beampipe hole for detectors at  a collider, the scattered
electron has to have a minimum angle to leave the beampipe and be
detected. From (\ref{kinematics1}) it follows that
this leads to a minimum requirement on $Q^2$, which is also visible
in Fig.~\ref{sec2f1}.  To study whether $F_2$ still
rises at lower values of $Q^2$, several ways to increase the acceptance
for low $Q^2$ were explored.

\begin{figure}[htb] 
 \epsfig{file=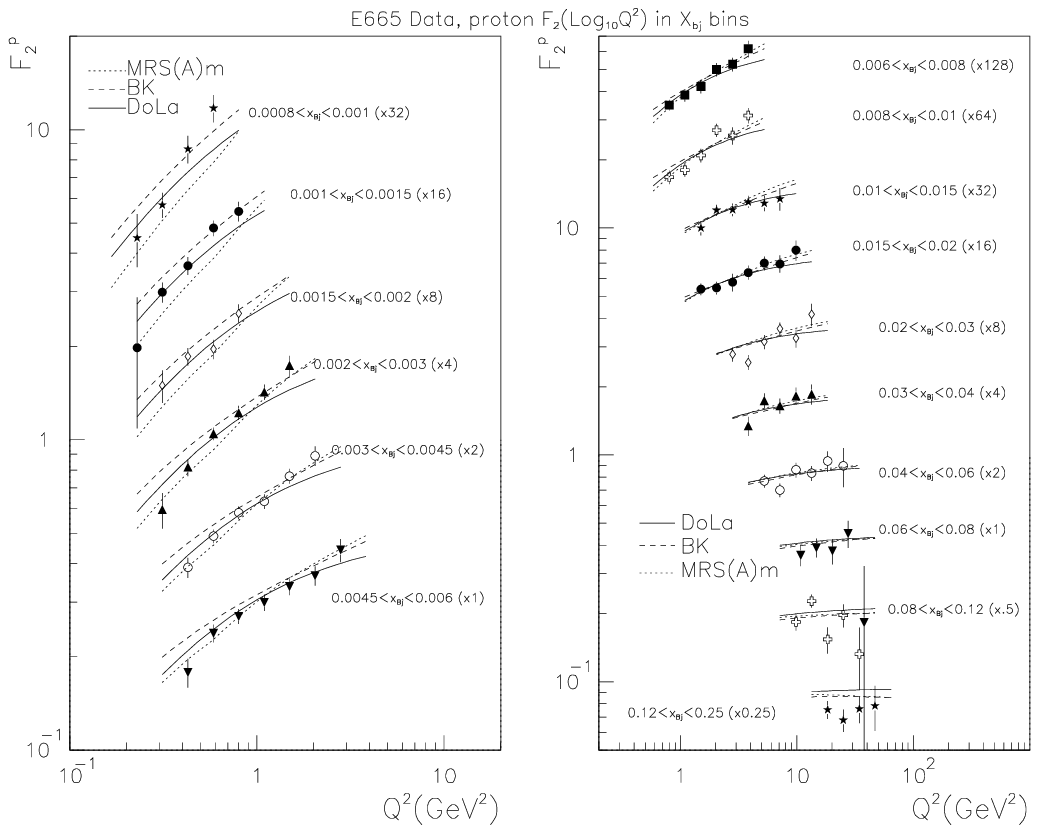,width=10cm,angle=-90}
\caption{Measurements of $F_2^p$ by the E665 Collaboration. The
errors are statistical and systematic added in quadrature, a normalisation
uncertainty (1.8$\%$) is not included. Curves show model calculations of
Martin, Stirling and Roberts, Bade\l ek and Kwieci\'nski, and Donnachie and
Landshoff (from \protect \cite{f2e665}).}
\label{fig_f2e665}
\end{figure}
 
\begin{figure}[htb]
\begin{center}
 \epsfig{file=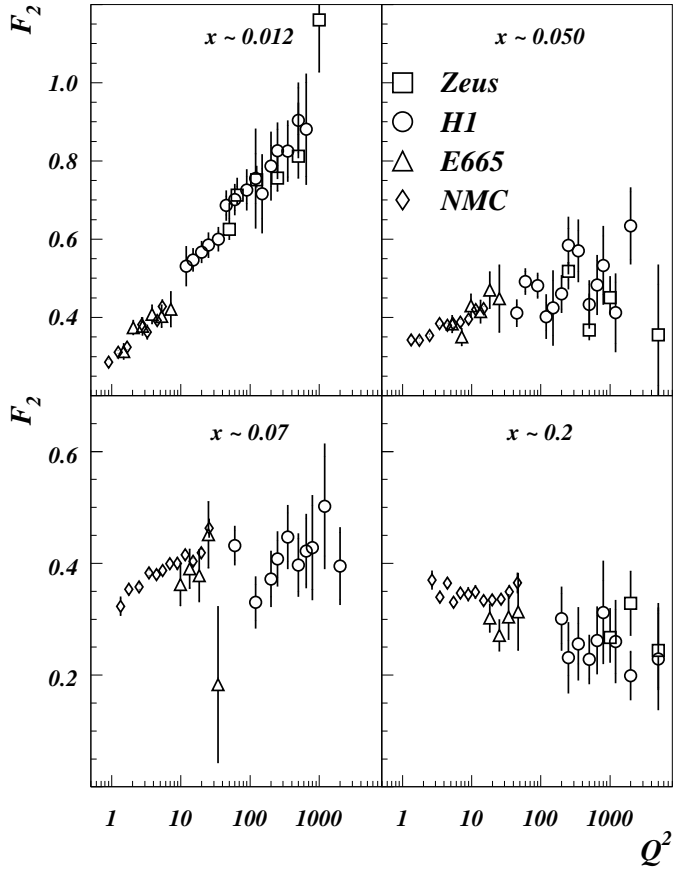,width=10cm,%
bbllx=0pt,bblly=50pt,bburx=550pt,bbury=650pt}
\end{center}
\caption{Detailed comparison of the structure function $F_2$
as function of $Q^2$ between H1, ZEUS,
E665 and NMC, for $x$ values around 0.2, 0.07, 0.05 and 0.012. The
error bars represent the full errors on the data points.}
\label{sec2f9}
\end{figure}

\begin{figure}[htb]
\begin{minipage}[htb]{5.7cm}
\epsfxsize=5cm
\centering
 \epsfig{file=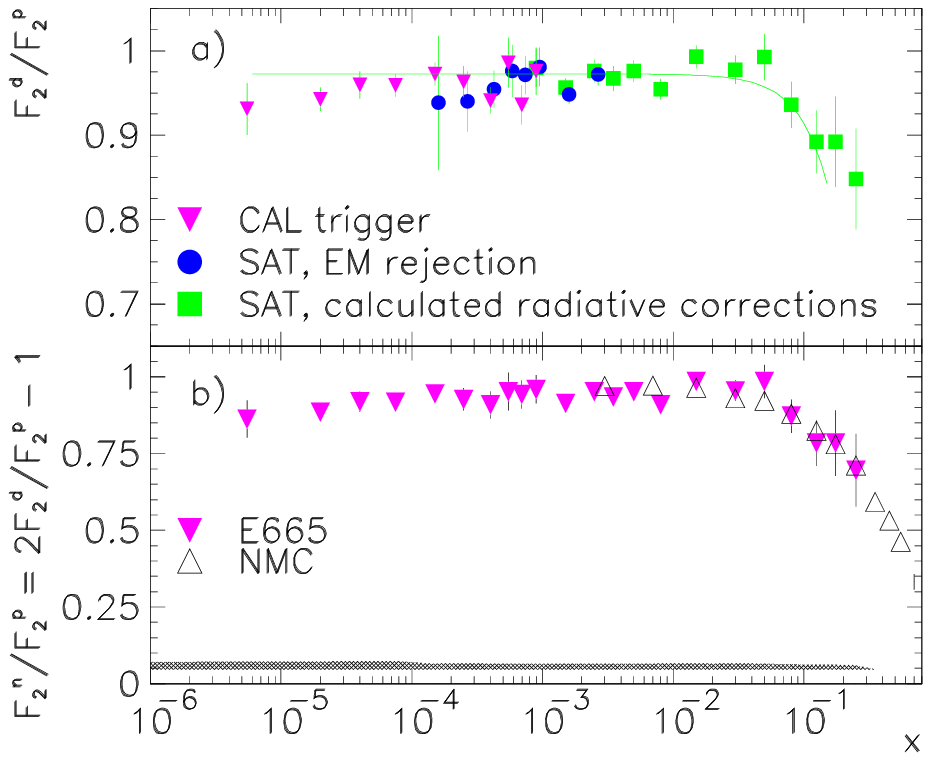,height=57mm,width=60mm}
\end{minipage}
\begin{minipage}[htb]{5.7cm}
\epsfxsize=5cm
\centering
 \epsfig{file=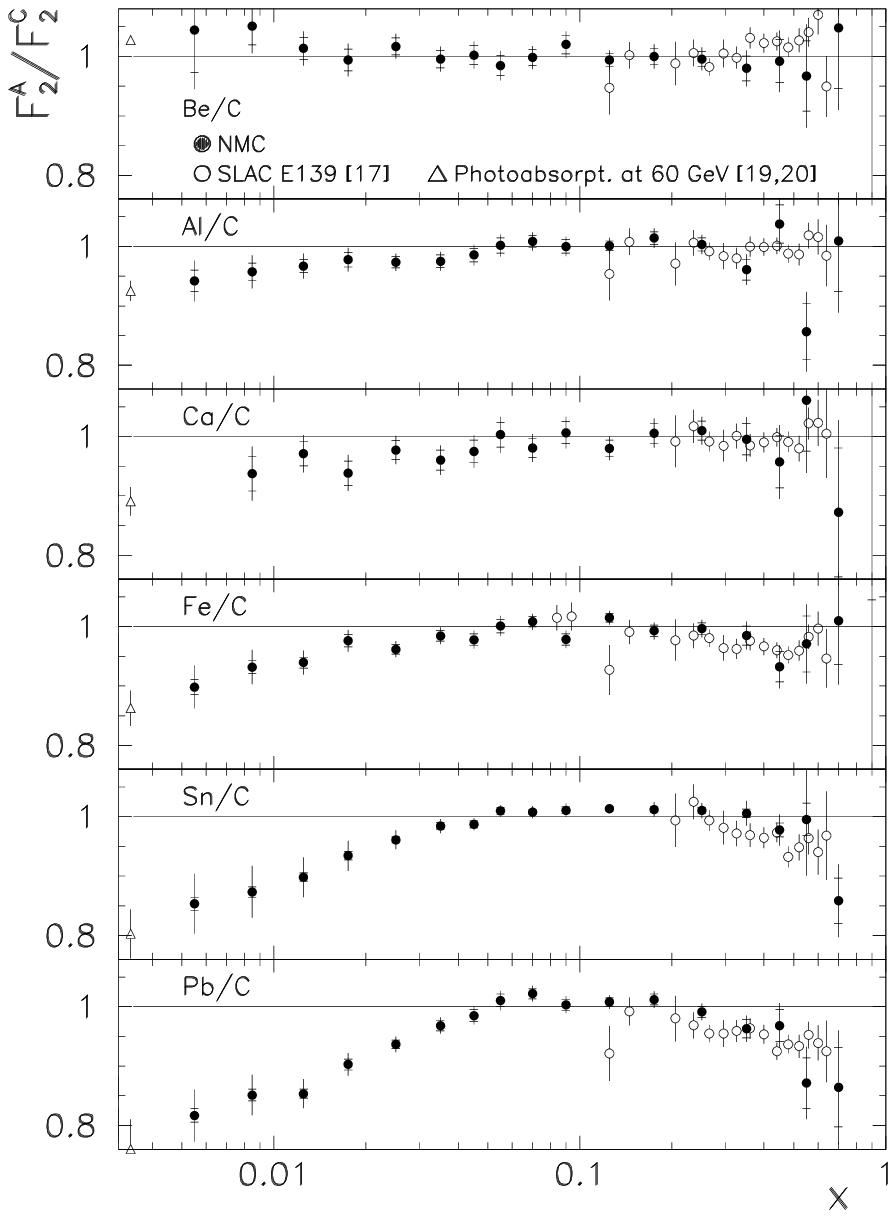,height=65mm,width=60mm}
\end{minipage}
\\
\caption{Left: E665 results. a) $F_2^d/F_2^p$ for three different
techniques of extracting the ratio. The curve shows a prediction of
Bade\l ek and Kwieci\'nski, \protect\cite{dshad}.
b) $2F_2^n/F_2^p=2F_2^d/F_2^p - 1$
as a function of $x$. The NMC data  at $Q^2= $4 GeV$^2$ are also shown.
Errors are statistical. The systematic uncertainty is represented
by the hatched area in Fig. b) (from \protect\cite{npe665}).\hspace*{3cm}
Right: NMC results on $F_2^A/F_2^C$, averaged over $Q^2$
(open symbols) together with earlier results of SLAC (closed symbols).
Inner/outer error bars represent the statistical/total errors. The
SLAC-E139 data for silver and gold were used for the comparison with the
tin and lead data of the NMC, respectively. The photoproduction cross
section data are given at a small value of $x$ for convenience (from
\protect\cite{ashad_nmc1}).}
\label{fig_np}
\end{figure}

\begin{figure}[htb]
\begin{minipage}[htb]{6.5cm}
\epsfxsize=5.5cm
\centering
 \epsfig{file=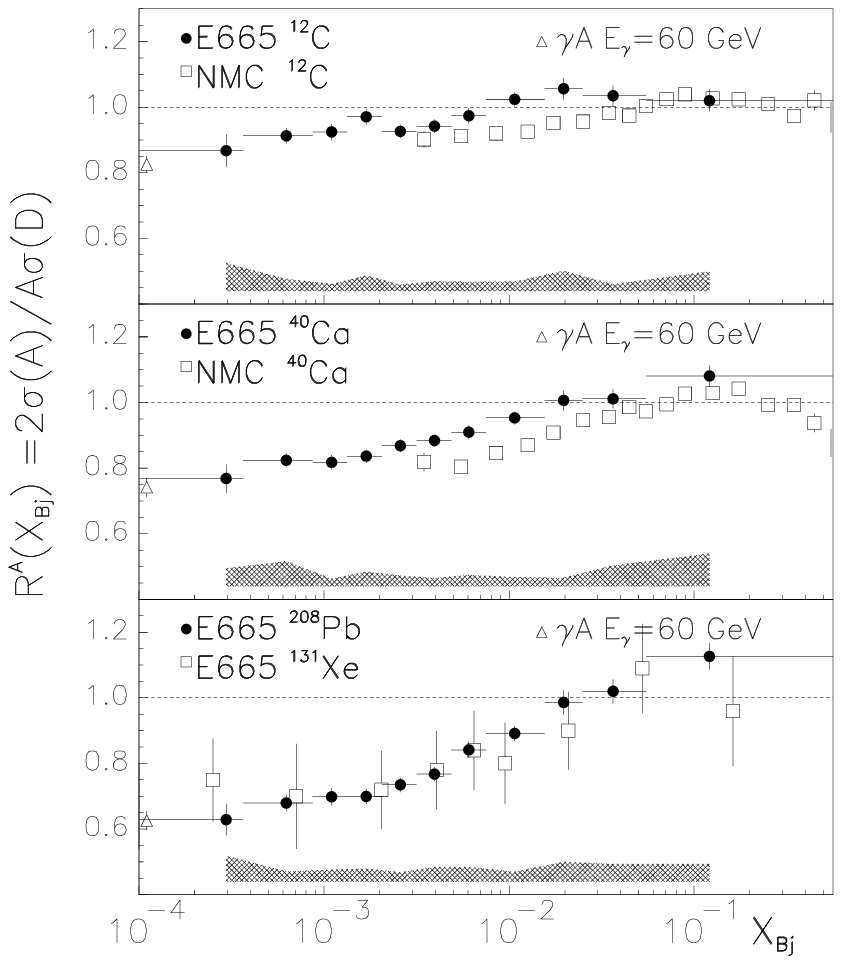,height=57mm,width=90mm}
\end{minipage}
\begin{minipage}[htb]{6.5cm}
\epsfxsize=5.5cm
\centering
 \epsfig{file=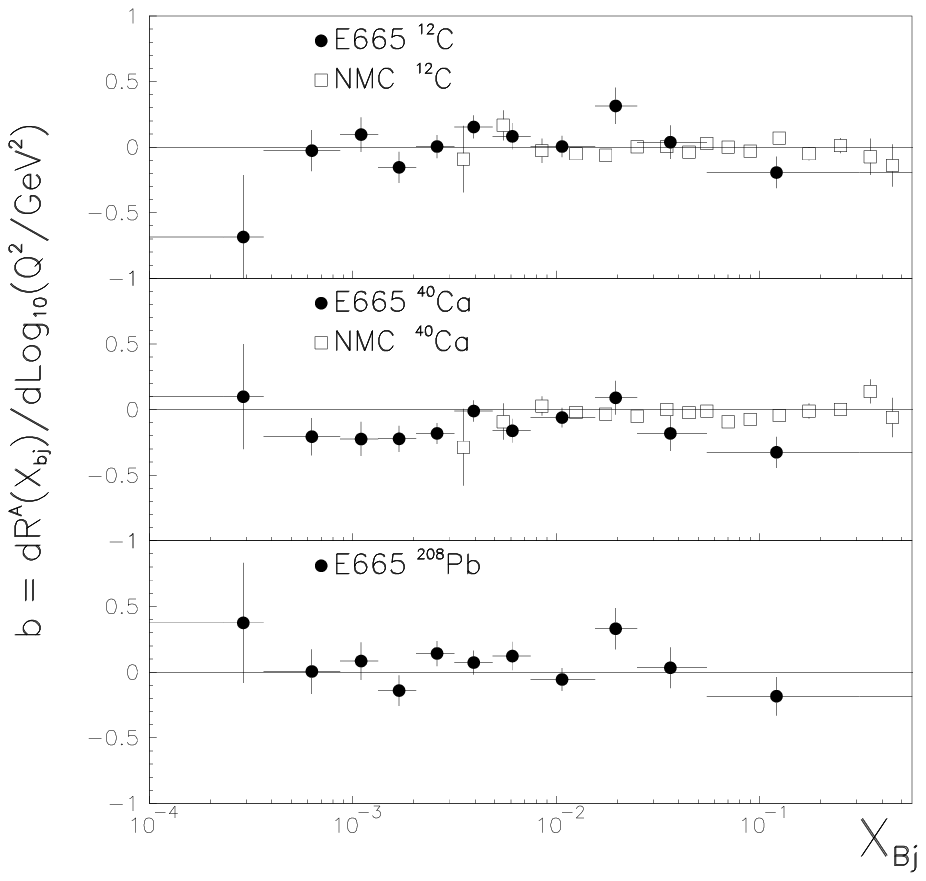,height=57mm,width=65mm}
\end{minipage}
\\
\caption{Left: E665 results on $\sigma ^A/\sigma ^D$. Errors are
statistical, the systematic ones are marked as shaded bands. Overall
normalisation uncertainties have not been included. The NMC results have
been shown for comparison where available (from
\protect\cite{ashad_e665}).\hspace*{3cm}
Right: $Q^2$ dependence of the E665 $F_2^A/F_2^D$ data (closed
circles). Errors result from the fit, in which only statistical errors were
considered. The NMC results are drawn as open squares (from
\protect\cite{ashad_e665}).}
\label{fig_ashad_e665}
\end{figure}
 
The ZEUS detector was improved for electron detection around the beampipe
by the addition of a scintillator strip detector
on the face of the rear calorimeter. This allowed the detection of the
scattered electron down to smaller angles (175.5$^0$ compared to 174$^0$ in
1993) with a large improvement of the angular resolution
(2 mrad compared to 7 mrad in 1993). The detector was also used to make
an event-by-event correction to the scattered electron energy
arising from the energy loss in the inactive material prior to the
calorimeter, thereby improving the energy resolution and reducing the 
energy scale uncertainty.
Due to the large statistics the electron energy scale 
and angular shift uncertainty for the 
the H1 results have been reduced to 1\% and 1 mrad respectively.
H1 could also -- due to the excellent accelerator
conditions at the end of the 1994 running period -- make use of the most inner
active elements of the calorimeter, and increase its low scattering
angle acceptance from 173$^0$ to 174$^0$. In 1993
H1 initiated a pilot project
 to shift the interaction vertex of the collisions towards
the forward (proton) region. They demonstrated that collisions produced
at a position shifted by about 70 cm downstream of 
the detector could still be
used for $F_2$ analysis, and allowed an increase in the acceptance from
8.5 GeV$^2$ to 4.5 GeV$^2$ for the 1993 data. In 1994 about 10 times more data
were accumulated using this method, and results have been shown by
both experiments.

Another way to access low $Q^2$ is by using a sample of deep inelastic events 
with an energetic photon (e.g. $E_{\gamma} > $4 GeV)
emitted collinear with the incident electron. These radiative events can be
interpreted as deep inelastic scattering events with a reduced incident energy
$E_r = E_e-E_{\gamma}$ 
which can be reconstructed through the detection of the
radiated photon in the small angle photon tagger of the luminosity
system of the experiments.
When using the electron method, the ``true" kinematic variables  $y_t$ and
$Q_t^2$ for such an $ep$ collision are
obtained by replacing in (\ref{kinematics1})
the nominal beam energy by the reduced energy $E_r$.
Both experiments have shown data using this process~\cite{H1f294,zeuslowq2}.
  
In summary, compared to the previous analyses,
the $F_2$ measurement has been extended to lower and
higher $Q^2$ (from $4.5-1600$~GeV$^2$ to  $1.5-5000$~GeV$^2$),
and to lower and higher $x$ (from $1.8 \times 10^{-4}-0.13$ to
$3 \times 10^{-5}-0.32$).
The final result is shown in Fig.~\ref{f2xall} as a function of $x$ and
in Fig.~\ref{sec2f5} as a function of $Q^2$.
The error bars of the data are now reduced to the 5\% to 10\% level (except
at high and very low $Q^2$). The normalization uncertainty
has been reduced to 1.5\% (2.5\%) for H1 (ZEUS).
The rise of $F_2$ with decreasing $x$ is confirmed with higher precision.
 This rise continues, albeit less strongly than at higher $Q^2$,
in the region of the
lowest $Q^2$ available.
Scaling violations are clearly seen in the plot of $F_2$ versus $Q^2$ and
will be interpreted in terms of QCD below.
 
To emphasize the rise of $F_2$ at low $Q^2$,
data from the eight lowest $Q^2$ bins are shown
and compared with recent $F_2$ parametrizations in Fig.~\ref{sec2f7}.
It  demonstrates that the rise of $F_2$ towards low $x$ is also present
in the low $Q^2$ region.
The measurement is in good agreement with recent data
from fixed-target experiments E665 and NMC  at higher $x$ values.
The curves represent GLAP QCD inspired predictions (GRV~\cite{GRV},
MRS~\cite{MRS} and CTEQ~\cite{CTEQ3}) and two Regge inspired predictions
(DOLA~\cite{DOLA} and CKMT~\cite{CKMT}).
The Regge inspired predictions, shown only for the lowest $Q^2$ bins, are
generally below the data.
 
The persistent rise of $F_2$ at low $x$ for small $Q^2$
indicates that the photoproduction  regime has not been
reached yet. This can also  be seen in Fig.~\ref{sec2f8}  which
shows  for the low $Q^2$ data the strong rise of
$F_2$   as a function of
$W$, the invariant mass of the $\gamma^* p$ system
(at low $x$, $W  \simeq \sqrt{Q^2/x}$). $F_2$ is
related to  the total cross section of the proton-virtual photon
interaction $\sigma_{tot}(\gamma^* p) $ via
\begin{equation}
\sigma_{tot}(\gamma^* p) \simeq \frac{4\pi^2 \alpha}{Q^2} F_2(W,Q^2).
\end{equation}
The $F_2$  growth  can be contrasted with the weak rise with
$W$   of the total real photoproduction cross 
section in the same range of $W$, as shown in Fig.~\ref{sec2f8}.
The different behaviour for $Q^2 = 0$ and data at a finite small $Q^2$
remains one of the interesting questions to be studied at HERA.  
  
\subsection{Fixed-target structure function data}
 
The NMC has presented their analysis of the proton and
deuteron structure functions~\cite{f2nmc},
in the range 0.006 $\leq x \leq$ 0.6 and 0.5 $\leq Q^2 \leq$ 75 GeV$^2$,
as shown for $F^d_2$ in Fig.~\ref{fig_f2nmc},
performed on the almost final sample of events. A clear scaling
violation pattern with slopes $d\ln F_2/d\ln Q^2$ positive at low $x$
and an ``approach to scaling" (i.e. a rise of $F_2$ from $Q^2 = 0$ to the 
scaling region) is visible. In this figure a
comparison of the NMC, SLAC~\cite{f2slac} and BCDMS~\cite{f2bcdms}
measurements is also shown. All three data sets are in good agreement
with each other.
They were thus used to obtain parametrizations of $F_2^p$ and $F_2^d$
and their uncertainties, using a 15--parameter function.
The low $x$ results of the EMC NA2 experiment have earlier been disproved
by the NMC measurements. The data confirm a characteristic
weak $x$ dependence of $F_2$ at low $Q^2$, observed for the first time
by the EMC NA28 experiment~\cite{f2na28} and also
interpreted in \cite{f2bbjk}
(see also Section 3).
 
New measurements of the proton and deuteron structure
functions for $x > $0.0001 have recently been presented by the E665
Collaboration and are shown in  Fig.~\ref{fig_f2e665}~\cite{f2e665}.
The lowest $Q^2$ and $x$ values in their data are 0.2 GeV$^2$, and 
8$\times$10$^{-4}$ respectively.  A clear pattern emerges from these data
at $Q^2$ values lower than a few GeV$^2$, namely a weak $x$,
and possibly a stronger than logarithmic $Q^2$ dependence, of $F_2$. 

The E665 region of $x$, i.e. $x>$10$^{-5}$, is now being investigated
by both the H1 and ZEUS Collaborations at HERA.
The most dramatic effect visible in the HERA large $Q^2$ data is a strong
increase of $F_2$ with decreasing $x$, while there is a  
rather weak $x$ dependence of the $F_2$ observed by  EMC NA28, NMC, and E665.
 
The fixed-target $F_2$ data have had great impact on the determination of
parton distributions (see e.g. \cite{mrs}). It is now seen that
these data join well to the results of HERA and thus make a joint QCD analysis 
possible in a large kinematic interval (see below).
In Fig.~\ref{sec2f9} a
detailed comparison of the structure function $F_2$ as function of $Q^2$
between H1, ZEUS,
E665 and NMC is shown for $x$ values around 0.2, 0.07, 0.05 and 0.012.
The data show a smooth continuation over the whole $Q^2$ region.
It also shows a (still) substantially different level of accuracy
between the HERA and
the fixed-target experiments. The former are still expected to improve
both in statistics and systematics in the next few years.
Apart from the above overall agreement, 
there exists however a discrepancy between the NMC and CCFR $F_2$ data at 
low $x$ (not shown). This was 
much discussed at the Workshop, but did not result in any new conclusion.

Both the NMC and E665 experiments have measured the deuteron to proton structure
function ratio, $F_2^d/F_2^p$, extending down to very low values of $x$.
In the case of  NMC the ratio has been measured directly, i.e. the measurement
of the absolute structure function is used only for calculation of 
the radiative corrections.
The data are usually presented as the ratio $F_2^n/F_2^p$ where $F_2^n$ is
defined as $2F_2^d-F_2^p$. This quantity would give the structure function of
the free nucleon in the absence of nuclear effects in the deuteron.
The results are presented on the left
in Fig.~\ref{fig_np}~\cite{npe665}. In both data sets the average $Q^2$
varies from bin to bin reaching down to $\langle Q^2\rangle =$ 0.2 GeV$^2$ 
at $x=$0.0008 for the NMC and $\langle Q^2\rangle =$ 0.004 GeV$^2$ at 
$x=$ 5$\times$10$^{-6}$ for E665.
The results of both experiments show that the ratio $F_2^n/F_2^p$ remains
below unity down to the smallest measured values of $x$. At low $x$ this
can be attributed to nuclear shadowing in the deuteron~\cite{dshad},
predicted to be only weakly $x$ dependent, as observed. It seems 
unlikely that the results can also indicate a difference in $F_2$ of protons
and neutrons at low $x$, since e.g.
in  Regge models the difference between the proton and neutron structure
functions vanishes with decreasing $x$~\cite{dshad,MELBAR,zoller}.
 
\begin{figure}[htb]
\begin{center}
 \epsfig{file=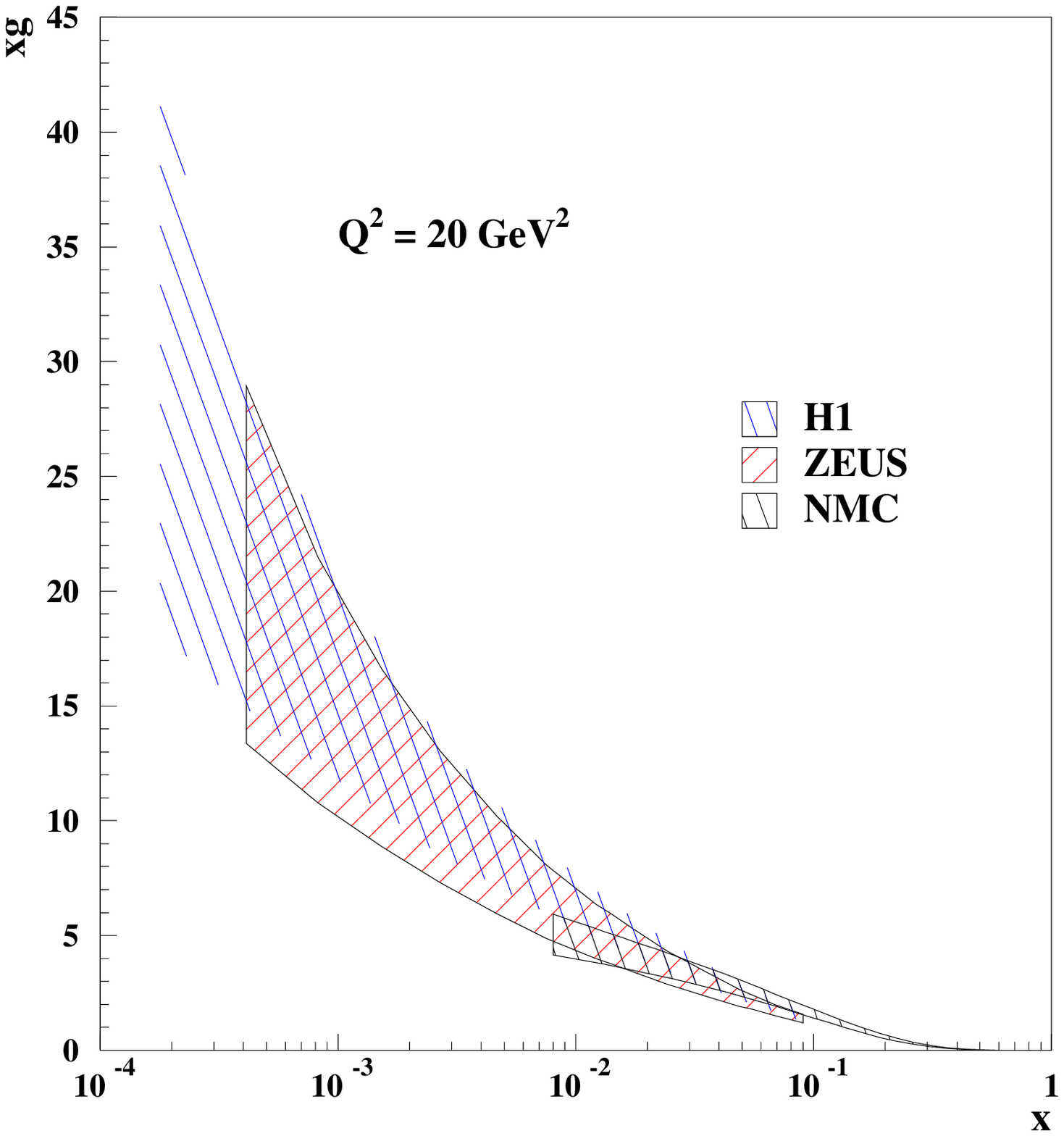,width=8cm,%
  bbllx=15pt,bblly=220pt,bburx=550pt,bbury=620pt}
\end{center}
\caption{The gluon density $xg(x)$ at 20 GeV$^2$ extracted from a NLO QCD
fits by the H1 \protect\cite{h1gluon}, ZEUS \protect\cite{zeusgluon} 
and NMC \protect\cite{nmcqcd} Collaborations.}
\label{sec2gluon}
\end{figure}

New data have appeared on  nuclear shadowing. NMC have performed a high
precision study of the $A$ dependence of nuclear shadowing 
in the range 0.004 $< x <$ 0.6 and 1.5 GeV$^2 < Q^2 <$ 60 GeV$^2$.
The results are shown on the right in Fig.~\ref{fig_np}~\cite{ashad_nmc1}.
These measurements of the ratios $F_2^A/F_2^C$ for
$A$ = Be, Al, Ca, Fe, Sn and Pb taken in conjunction with those on
D, He, Li, C and Ca~\cite{ashad_nmc2,ashad_nmc3} and
with earlier data of SLAC~\cite{ashad_slac}, show a detailed pattern
of the $x$ dependence of shadowing. The NMC data range from $A=2$ to $A=208$.
The functional dependence of $F_2^A/F_2^C$ on $A$ has been parametrized as
$F_2^A/F_2^C=cA^{(\alpha -1)}$ in each bin of $x$.
The amount of shadowing increases strongly with the mass number
$A$. Lower values of $x$ and $Q^2$ are covered by the nuclear data
of  E665 in the region $x >$0.0001 and $Q^2 >$0.1 GeV$^2$~\cite{ashad_e665},
as shown in Fig.~\ref{fig_ashad_e665} (left). A decrease in the amount
of shadowing observed by  E665 is presently under discussion.
Shadowing seems to saturate at $x$ about 0.004 as also indicated
by the NMC data on
the $F_2^{Li}/F_2^D$ and $F_2^{C}/F_2^D$ ratios measured down to $x=$0.0001
and $Q^2$=0.03 GeV$^2$~\cite{ashad_nmc3}.
No clear $Q^2$ dependence is visible in the E665 data in a wide interval of
$Q^2$, shown in  Fig.~\ref{fig_ashad_e665} (right), 
contrary to the preliminary NMC results in which  positive
$Q^2$ slopes for the $F_2^{Sn}/F_2^C$ ratio at $x <$ 0.1 are observed, as
shown in~\cite{ashad_nmc_q2_sn}. The shadowing region seems to have another
interesting feature: it contains a large fraction of large rapidity gap
(or diffractive) events, their fraction increasing with $A$~\cite{diffr_e665}.
 
\subsection{Parton distribution measurements}

Both the H1 and ZEUS experiments have performed 
next-to-leading order (NLO) QCD fits based on the Altarelli--Parisi (GLAP) 
evolution equations on the HERA and fixed-target $F_2$ data.
Fig.~\ref{f2xall} shows that the $F_2$ data can be well described by
such a QCD fit. Note that only data with $Q^2 \ge 5$ GeV$^2$ were used in the
fit. The fit result was evolved to lower $Q^2$ and used as a prediction in
the region $Q^2< 5$ GeV$^2$. The low $Q^2$ data are found to be well described.
This result suggests that, within the present accuracy, 
no (large) higher twist terms are required in this region,
contrary to the NMC fit, see below.
This may need to be reviewed when higher precision data at low $x$ become 
available.
 
The scaling violations from the HERA data  allow an estimate of the gluon
density $xg(x)$ at low values of $x$, while the fixed-target data
are needed to constrain the high $x$ region. The H1 QCD fit~\cite{h1gluon}
includes only proton data  from H1, NMC and BCDMS.
Additionally the momentum fraction carried by the gluon  is imposed to be
0.44. Apart from the ZEUS data, the ZEUS fit~\cite{zeusgluon}
includes data from NMC, both on proton and deuteron targets.
The results are shown in Fig.~\ref{sec2gluon}
for $Q^2 = 20$ GeV$^2$. The error bands shown
include a careful analysis of the systematics, taking into account
the correlation between different sources.
The results of the two experiments agree very well.
The resulting gluon distribution shows a clear rise with decreasing $x$.
Similar results have been found in~\cite{MRS}, which include also other data
than those from structure functions.
In the region $x>10^{-2}$ the extracted gluon densities agree with the
result obtained by the NMC.
 
\begin{figure}[htb]
\begin{minipage}[htb]{6cm}
\epsfxsize=5cm
\centering
 \epsfig{file=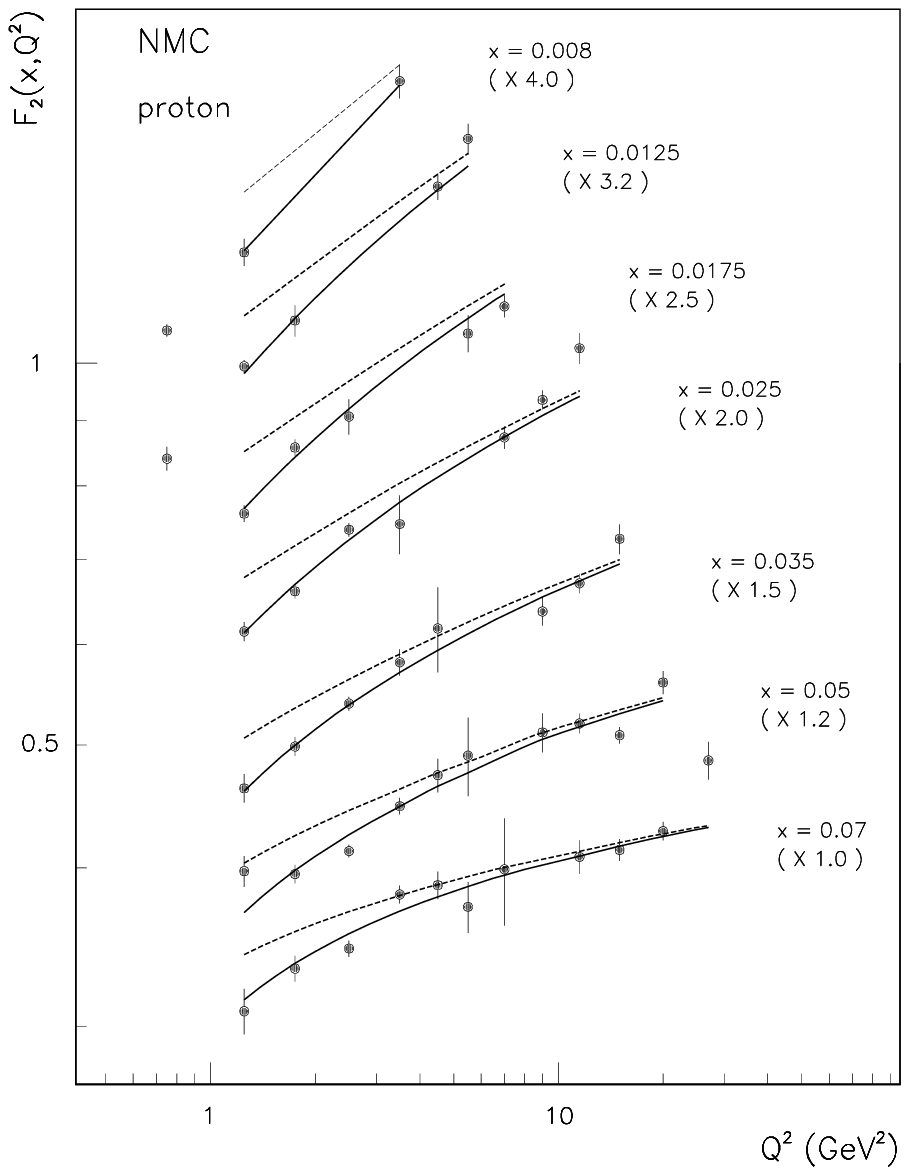,height=85mm,width=65mm}
\end{minipage}
\begin{minipage}[htb]{6cm}
\epsfxsize=5cm
\centering
 \epsfig{file=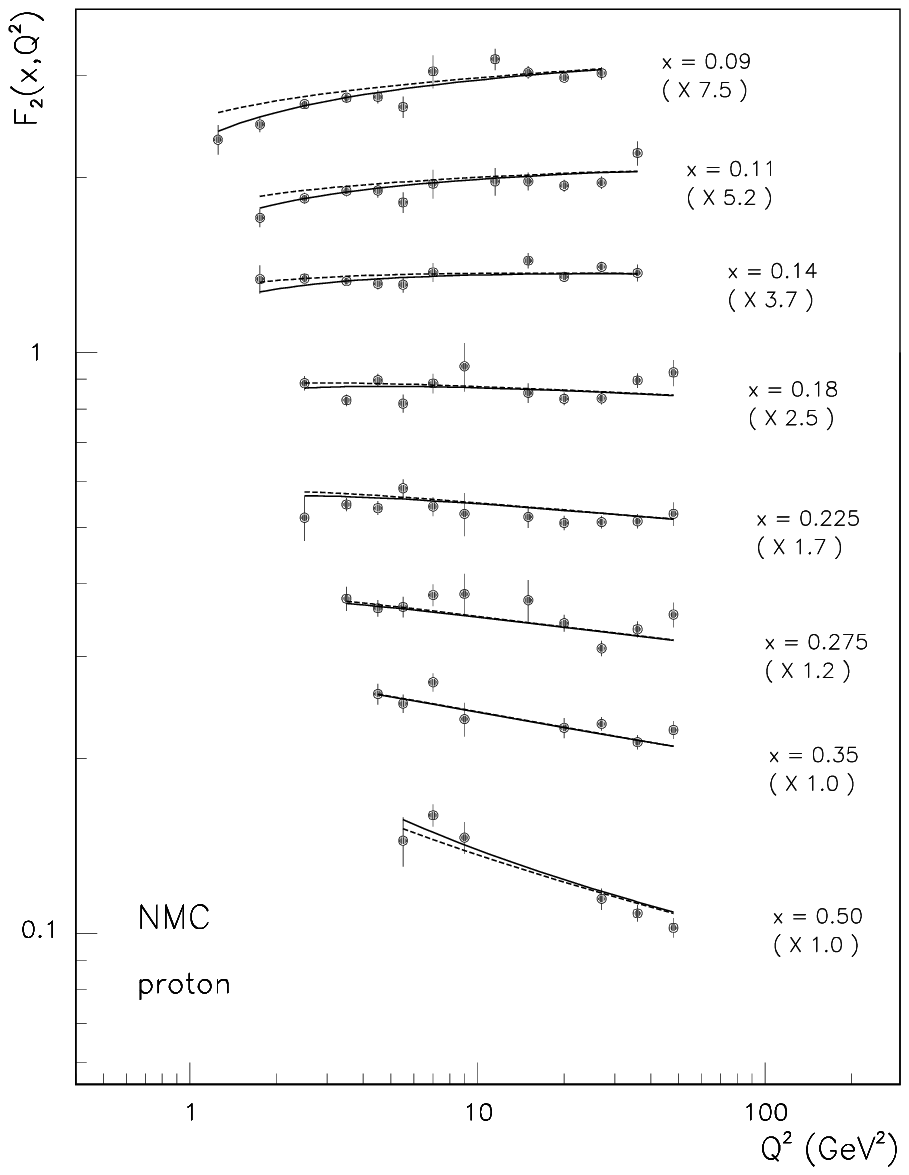,height=85mm,width=65mm}
\end{minipage}
\\
\caption{The results of the QCD fit to the
$F_2^p$ data. The solid line is the result of the QCD fit with higher
twist included. The dotted curve shows the contribution
of $F_2^{LT}$. The errors are statistical (from \protect \cite{nmcqcd}).}
\label{fig_qcd}
\end{figure}

NLO QCD fits have been performed by the NMC to their (earlier) accurate
measurements of the structure functions $F_2^p$ and $F_2^d$
down to low values of $x$~\cite{nmcqcd}.  The flavour singlet and
non--singlet quark distributions as well as the gluon distribution have been
parametrized at the reference scale equal to 7 GeV$^2$.  All the data with $Q^2
\geq $ 1 GeV$^2$ were included in the fit.  Besides the leading twist
contribution a higher twist term was also included using a factor 
$1 + H(x)/Q^2$ where $H(x)$ was determined from the SLAC and the
BCDMS measurements~\cite{ht}, averaged over the proton and
deuteron, and suitably extrapolated to lower values of $x$.  Results of the QCD
fit to the proton structure function data are shown in Fig~\ref{fig_qcd} and
clearly indicate the extension of the QCD analysis to the low $x$
and low $Q^2$ regions.  The contribution of higher
twists is however substantial at scales of about 1 GeV$^2$.
 
Additional information on the gluon density is extracted from hadronic 
final states. The H1 experiment has extracted the gluon density from 2-jet 
events. The method and results are given in~\cite{thompson,h1jetgluon}.
The results are limited to the region $x\ge 0.001$, and
agree  with the gluon extracted from scaling violations. 
The E665 experiment has used the energy--energy angular pattern of hadrons
produced in DIS to extract the gluon distribution function of the nucleon, 
down to $x$=0.005. 

An interesting point in the QCD analysis arises from the presence of
diffractive-like events in the deep inelastic scattering event sample.
A contribution of approximately 10\% from events with
a large rapidity gap towards the proton remnant, and with the characteristics
as measured in hadronic diffractive exchange, has been
established~\cite{diffdurham}.
These events possibly originate from a different production
mechanism than the one for the bulk of the deep inelastic data. 
So far no special account has been taken of this in the extraction of
partons from inclusive $F_2$ measurements; this topic was discussed at the
Workshop and should be considered further.

\subsection{$R(x,Q^2)$ measurements}
 
\begin{figure}[htb]
\begin{minipage}[htb]{6cm}
\epsfxsize=5cm
\centering
 \epsfig{file=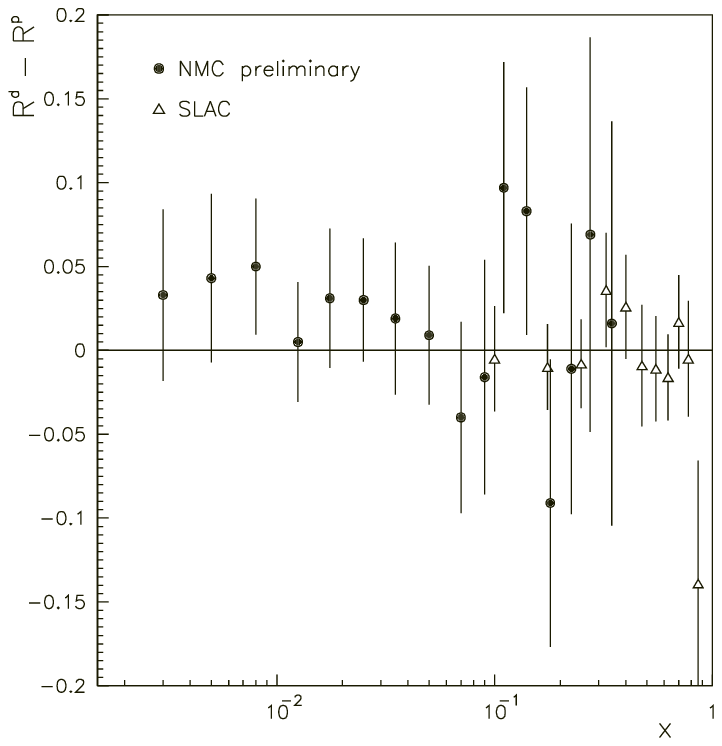,height=75mm,width=65mm}
\end{minipage}
\begin{minipage}[htb]{6cm}
\epsfxsize=5cm
\centering
 \epsfig{file=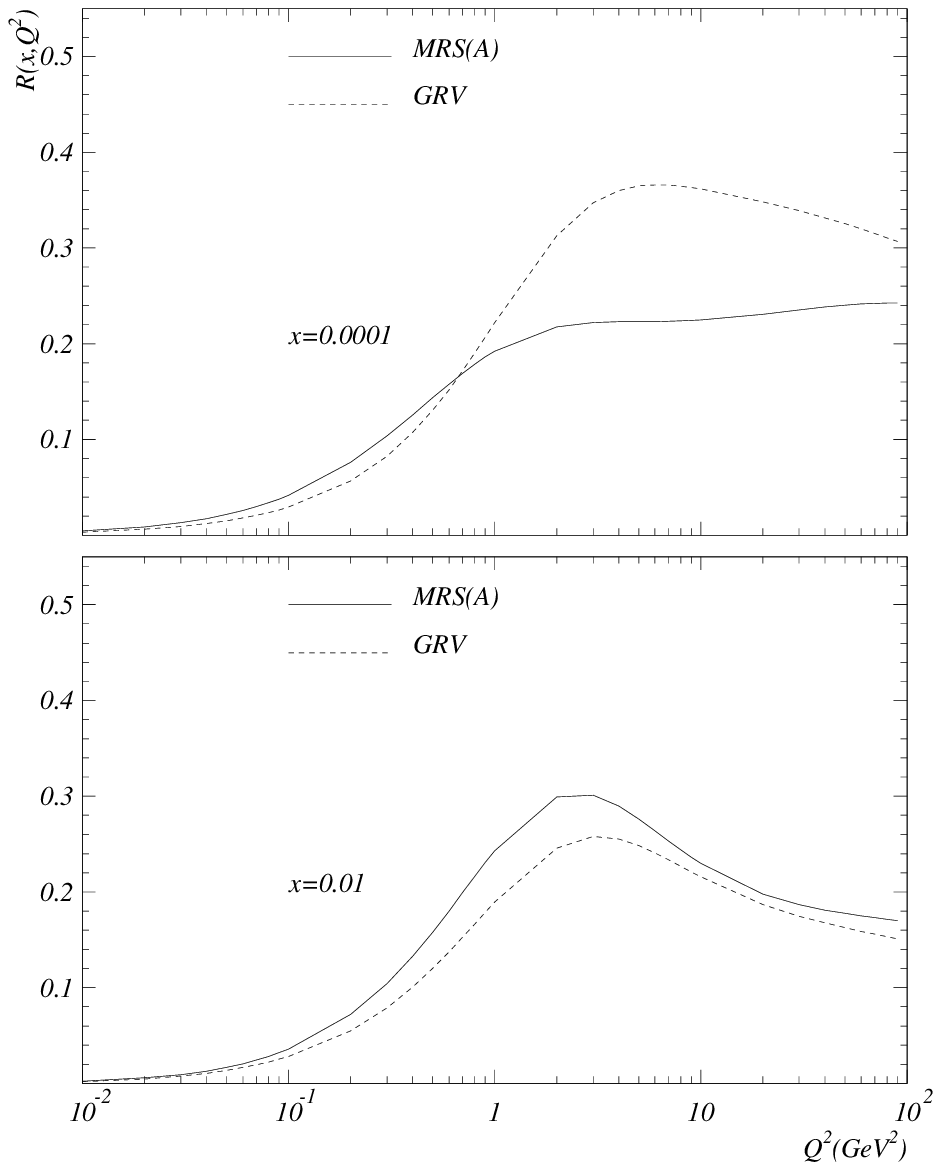,width=60mm}
\end{minipage}
\caption{Left: NMC (preliminary) results $R^d-R^p$ compared with the
QCD predictions (the curve) and with the results
of SLAC (open symbols), from
\protect \cite{rdp}.\hspace*{3cm}
Right: $R(x,Q^2)$ in the phenomenological model of
Bade\l ek, Kwieci\'nski and Sta\'sto, for two different parton
parametrisations (from \protect \cite{r_ksb}).}
\label{fig_rdp}
\end{figure}

Extraction of $F_2(x,Q^2)$ from the data needs information on $R(x,Q^2)$. In
particular the ratio of inelastic cross sections on different nuclei is 
only equal
to the corresponding structure function ratio, provided $R(x,Q^2)$ is the
same for these nuclei. Results of the NMC analysis on 
$R^{Ca}-R^C$~\cite{rcac} and
$R^d-R^p$~\cite{rdp}, shown in Fig~\ref{fig_rdp} (left),
demonstrate that neither of these quantities exhibit a significant
dependence on $x$ and that they are both compatible with zero.
The NMC reported preliminary measurements of $R(x,Q^2)$ for proton
and deuteron targets as a function of
$x$ in the range 0.006 $< x <$ 0.14~\cite{rdyring}.
The average $Q^2$ of these measurements ranges from 1.1 GeV$^2$ at the
smallest
$x$ to 15.5 GeV$^2$ at $x=$0.14. The results show a rise of $R$ with decreasing
(small) $x$.  Preliminary measurements of  $R(x,Q^2)$ on a heavy target
(Fe),
at $x > $0.01 and $Q^2 > $ 4 GeV$^2$ (at present) have also been reported
by the CCFR neutrino Collaboration~\cite{rccfr}.
In their data analyses the NMC and E665 experiments
assumed  $R$ was independent of the target atomic
mass $A$ and given by the SLAC parametrization~\cite{rworld} 
valid for $x > $0.1 and $Q^2 >$ 0.3 GeV$^2$. This
parametrization was then extrapolated (with 100$\%$ error)
to $Q^2 \rightarrow$ 0. Hence there is a need of a theoretical estimate
of $R$ (or $F_L$) in the region of low $x$ and low $Q^2$. Two ongoing
phenomenological studies are expected to deliver such estimates soon.
In these studies 
both the perturbative QCD contribution, which at low $x$ and low $Q^2$ is
dominated by the photon-gluon mechanism, and a non-perturbative term
are  taken into account. In \cite{r_ksb} the latter contribution is
determined phenomenologically (Fig.~\ref{fig_rdp} (right)) while in
\cite{rccfr} it is fitted to the low $Q^2$ data.
 
Also at HERA the study of $R$ is an issue of interest. 
For this purpose it has been
proposed to operate HERA at lower energies, to have the cross section
measurements at two $y$ values for a given $x,Q^2$ point.
Such a measurement could be done in 1996 or 1997.
A discussion on this topic is presented in ~\cite{mark}.
  
\subsection{Sum Rules}
Several sum rules have been formulated for different combinations of
structure functions. Strict QCD predictions, valid for $Q^2\rightarrow
\infty$, exist for those  involving only
flavour nonsinglet contributions: the Gross--Llewellyn-Smith and the Bjorken
sum rules. Experimental measurements of such sum rules provide a stringent 
test of 
fundamental QCD assumptions. They also in principle permit the extraction 
of 
the strong coupling constant, $\alpha_S$, from the data.
Due to the finite $Q^2$ of
the measurements, a predicted value of a sum rule is
usually presented in the form of a power series in $\alpha_S$, the
coefficients of which are directly calculated.

There is no strict QCD prediction for the sum rules containing the flavour
singlet contributions, i.e. the Gottfried and  Ellis--Jaffe sum rules. The
reason is that singlet contributions contain an 
``intrinsic" $Q^2$ dependence.
Testing them usually results in surprises which teach us a lot about the 
shortcomings of the  simple quark model.
 
All the sum rules involve integrations over the whole 0$\leq x \leq$1
interval. This means that due to the limited experimental acceptance,
interpolations from $x_{min}$ to 0 and from $x_{max}$ to 1 have to be
performed. Usually the former is more problematic due to a larger
contribution of the small $x$ region to integrals and/or to a poor theoretical
understanding of this kinematic region.
Thus the  extrapolation $x\rightarrow 0$ is
a major source of systematic errors in such sum rule tests. Another source is
produced by the  limited experimental acceptance in $Q^2$ at each $x$ value.
This usually means that a sum rule is measured at a certain $Q^2_0$,
common to all points but at values of $Q^2_0$ which are 
not sufficiently high to
exclude a contribution from nonperturbative effects (``higher twists").
  
\subsubsection{Tests of the flavour nonsinglet sum rules.}

The Bjorken sum rule involves the spin structure functions and will be
presented in Section 7. The Gross-Llewellyn-Smith sum rule, formulated 
within the Quark Parton Model, states that the integral over the 
valence quark densities is equal to 3
i.e. $\int_0^1 xF_3(x)dx/x$=3. The QCD corrections to this rule 
have been  calculated up
to $\alpha_S^3$~\cite{gls_corr}. The sum rule has been tested by the CCFR
Collaboration in conjunction with additional
 low energy data from bubble chamber experiments~\cite{gls_test}. 
The sum rule is fulfilled at the 10$\%$
accuracy level. The value of $\alpha_S$ has been obtained at $Q^2$=3
GeV$^2$; it corresponds to $\alpha_S(M_Z)$ = 
0.108$^{+0.003}_{-0.005}$(stat.)$\pm$0.004(syst.)$^{+0.004}_{-0.006}$(HT). 
The uncertainty due to the low scale of the measurement
(i.e. the presence of ``higher twists") dominates the statistical error.
 
\subsubsection{Tests of the flavour singlet sum rules.}

The NMC measurements of $F_2^d$ (fitted together with the SLAC and BCDMS data)
and of $F_2^n/F_2^p$ allow a determination of the Gottfried sum
i.e. $S_G = \int (F_2^p-F_2^n)dx/x$ where $F_2^p-F_2^n = 2F_2^d(1-F_2^n/F_2^p)/
(1+F_2^n/F_2^p)$. At $Q^2 = $4 GeV$^2$, neglecting 
any $Q^2$ dependence,
$S_G$ was found to be 0.235 $\pm$ 0.026~\cite{gott},
significantly below the simple quark--parton model value of 1/3.
This is evidence for a  flavour asymmetric sea in the nucleon 
($\bar d$ sea quarks carry
more momentum than $\bar u$), a fact confirmed by the NA51 
measurement of the Drell-Yan asymmetry in $pp$ and $pn$ collisions,
which gave: $\bar
u/\bar d$=0.51$\pm$0.04$\pm$0.05 at $x$=0.18 and $Q^2$=25 GeV$^2$~\cite{na51}.
Recently a non-negligible $Q^2$ dependence of $F_2^p-F_2^n$ as a function of
$x$ at low $Q^2$  has been reported by the NMC: both the 
position
of the maximum and the maximum value
of this function change with $Q^2$. This change becomes negligible
when higher twist contributions 
are separated out from the $F_2^d$ and $F_2^n/F_2^p$ 
measurements~\cite{antje_brus}.
The Ellis--Jaffe sum rule, involving the spin structure functions will be
discussed in Section 7.

\setcounter{section}{2}
\setcounter{equation}{0}
\renewcommand{\theequation}{3.\arabic{equation}}
\setcounter{figure}{0}
\renewcommand{\thefigure}{3.\arabic{figure}}
\setcounter{table}{0}
\renewcommand{\thetable}{3.\arabic{table}}

\section{Low $Q^2$, low $x$ insights from fixed-target data}

\subsection{Introduction and basic concepts}

Due to the experimental constraints the fixed-target 
studies of deep inelastic scattering at low $x$ necessarily were 
correlated with low $Q^2$ ($Q^2 \lapproxeq 1$ GeV$^2$).  There are 
two reasons why this kinematic region is of special interest.  First, 
as emphasized in section 2, the new HERA measurements at small $x$ 
highlight the importance of a theoretical understanding 
of the connection between the low $Q^2$ and high $Q^2$ behaviour.  The 
second is a practical reason; a unified treatment of low and high $Q^2$ 
is essential for the large $Q^2$ data analysis, since to implement 
radiative corrections we require a knowledge of structure functions 
for $Q_{meas}^2 \geq Q^2\geq 0$.

At low $Q^2$ there are constraints on the structure functions $F_i 
(x, Q^2)$ which follow from eliminating the kinematical singularities 
at $Q^2 = 0$ from the hadronic tensor $W^{\mu \nu}$.  It is easy to 
show that as $Q^2 \rightarrow 0$ we require
\begin{equation}
F_2 \; = \; O (Q^2) \quad \hbox{and} \quad F_L \; = \; O (Q^4).
\label{c1}
\end{equation}
Hence it is clear that Bjorken scaling, which holds approximately at high $Q^2$, 
cannot be a valid concept at low $Q^2$.

In dealing with low $Q^2$ data we need to introduce the concept of \lq\lq 
higher twists".  The operator product expansion leads to the representation
\begin{equation}
F_2 (x, Q^2) \; = \; \sum_{n = 0}^{\infty} \; \frac{C_n (x, Q^2)}{(Q^2)^n}
\label{c2}
\end{equation}
where the functions $C_n (x, Q^2)$ depend weakly (i.e.\ logarithmically) on 
$Q^2$.  The various terms in this expansion are referred to as leading 
$(n = 0)$ and higher $(n \geq 1)$ twists.  The QCD improved parton model
where
\begin{equation}
F_2 (x, Q^2) \; = \; x \sum_i \: e_i^2 \: [q_i (x, Q^2) + \overline{q}_i 
(x, Q^2)] \: + \: O (\alpha_S (Q^2)),
\label{c3}
\end{equation}
and which gives approximate Bjorken scaling, retains only the leading twist 
contribution.  Physically the higher twist effects arise from the struck 
parton's interaction with target remnants, thus reflecting confinement.  For 
$Q^2$ of the order of a few GeV$^2$, contributions of the \lq\lq higher twists" 
may become significant, see, for example, ref.\ \cite{nmcqcd}.  
Contrary to the common opinion higher twists, which are 
corrections to the leading (approximately scaling) term (\ref{c3}), can only 
be implemented for sufficiently large $Q^2$. Thus they cannot correctly 
describe the low $Q^2$ (i.e. nonperturbative) region since the expansion 
(\ref{c2}) gives a divergent series there. In particular the individual terms 
in this expansion violate the constraint (\ref{c1}).  In order to correctly 
describe this region the (formal) expansion 
(\ref{c2}) has to be summed beforehand, at 
large $Q^2$, and then analytically continued to the region of $Q^2\sim 0$.  
This is automatically embodied in certain models like the Vector Meson 
Dominance (VMD) model. To be precise the VMD model together with its
generalisation which gives (approximate) scaling at large $Q^2$ can be
represented in a form (\ref{c2}) for sufficiently large $Q^2$.

In practical applications to the analysis of experimental data
which extend to the moderate values of $Q^2$ one often includes the 
higher twists corrections in the following simplified way:
\begin{equation}
F_2(x,Q^2)=F_2^{LT}(x,Q^2)\left [1 + {H(x)\over Q^2}\right ]
\label{c4}
\end{equation}
where the $F_2^{LT}$ is the leading twist contribution to  $F_2$
and $H(x)$ is
determined from fit to the data. This simple minded expression
may not be
justified theoretically since in principle the higher twist
terms, i.e.
functions $C_n(x,Q^2)$ for $n\ge1$ in eq.(\ref{c2}) evolve
differently with
$Q^2$ than the leading twist term.

Here we give a brief overview of the parametrizations and the data in the 
low $x$, low $Q^2$ region.  We refer the reader to refs.\ \cite{rmp} and 
\cite{BCKK} for a more detailed review of the treatment of low $Q^2$ 
problems.

\subsection{Parametrisations of structure functions}
There exist several phenomenological parametrisations (fits) of
the structure
function $F_2$ which incorporate the $Q^2\rightarrow 0$
constraints 
as well as the Bjorken scaling behaviour at large $Q^2$
\cite{brasse,CHIO,DOLA,ALLM,NMCR,CKMT}. Certain
parametrisations \cite{ALLM,NMCR,CKMT} also contain the (QCD
motivated) 
scaling violations.  However, they usually are not linked with
the
conventional QCD evolution.  Nor is the low $Q^2$ behaviour
related
to the explicit vector meson dominance, known to dominate at low
$Q^2$.  
There exist parmetrisations which explicitly contain QCD
evolution:
\cite{GRV,f2bbjk,ss,MRSLQ}. 
Most of the parametrisations  essentially extend the parton model
formula for $F_2$ down to the low $Q^2$ region
modifying in a suitable way the parton distributions; 
the model \cite{f2bbjk} includes also the VMD
contribution besides the partonic one and is an absolute
prediction
(i.e. no fitting to the data). The low $Q^2$
modifications are typically the following ones:

\begin{enumerate}
\item
Instead of the variable $x$, modified variables $\bar x_i
=x(1+{Q^2_{0i}/Q^2})$
are used as arguments of the parton distributions where `$i$'
enumerates the type of the parton.
\item
Models which at large $Q^2$ include the QCD scaling violations,
 have the evolution in $Q^2$ either ``frozen" below
certain scale  $\bar Q^2_0$ which is of the order of 1 GeV$^2$ or
the
evolution in a ``shifted" variable $Q^2 + \bar Q^2_0$ is used.
\item
Parton distribution functions are multiplied by form factors
of the type ${Q^2/(Q^2+m_i^2)}$ which ensure vanishing of
the structure function $F_2(x,Q^2)$ at $Q^2=0$.
\end{enumerate}

Modifications are absent in the dynamical model 
 \cite{GRV} which, in principle, is
meant to describe the
structure functions only in the large $Q^2$ region even if the
QCD
evolution is extended down to very low scales.   Also
the recent parametrisation of the ``parton
distributions" \cite{MRSLQ} uses the variable
$x$ instead of $\bar x$.  This  model does not however extend
down to the
very low values of
$Q^2$ (i.e. for $Q^2 <$ 0.25 GeV$^2$) and, in particular, it does
not
accommodate the photoproduction.

Parametrisations of $F_2(x,Q^2)$ differ in their small
$x$ behaviour.  Most of them (except 
\cite{NMCR} and \cite{DOLA}) incorporate at
large $Q^2$ the steep rise of $F_2(x,Q^2)$ as the function of $x$ with 
decreasing $x$ which is much stronger than
implied (for instance) by the expectations based on the ``soft"
pomeron with intercept $\alpha_P$ = 1.08. This steep increase of
$F_2(x,Q^2)$
becomes very weak at low $Q^2$. Possible dynamical
origin of this effect is different in different models being
either attributed to the absorptive effects
{\cite{CKMT,ss,MRSLQ}, to the onset of the VMD mechanism
\cite{f2bbjk} or to the pure perturbative QCD effects related to
the change of the ``evolution length" \cite{GRV}.

There exists practically only one parametrisation of the
$R(x,Q^2)$
structure function for the nucleon, i.e. the SLAC
parametrisation, \cite{rworld},
based on measurements by SLAC, EMC, BCDMS and CDHSW and valid
at $x>$0.1
and $Q^2 > $0.3 GeV$^2$. Experimental analyses in DIS experiments
need to
know $R$ down to measured values of $x$ and for $0<Q^2<Q^2_{meas}$.
Two phenomenological studies deliver estimates
of $R$ in the unmeasured region.
Both the perturbative QCD contribution, which at low
$x$ and low $Q^2$ is
dominated by the photon-gluon mechanism, and a non-perturbative
term are there taken into account. In \cite{r_ksb} 
the latter contribution is determined phenomenologically while in
\cite{rccfr} it is fitted to the low $Q^2$ data.

\subsection{Experimental data}
The lowest values of $x$, correlated with lowest values of $Q^2$
($x \sim $10$^{-5}$ and $Q^2 \sim$0.001 GeV$^2$), were reached by the 
E665 Collaboration at Fermilab by
applying a special experimental technique which permits the
measurement of muon scattering angles as low as 1 mrad.
At HERA the lowest values of $Q^2$ (1.5 GeV$^2$)
were recently reached by two
methods: shifting the
interaction point in the proton beam direction in order to
increase
the acceptance of low $Q^2$ events and by using
radiative
events with hard photon emission collinear with the incident
electron, Fig.2.5, \cite{mark}. The
radiative
events
can be interpreted as non-radiative ones with reduced electron
beam energy.

During the last three years an abundance of new data reaching
$Q^2$
values smaller than 1 GeV$^2$ have appeared. These comprise:
results
on the proton and deuteron structure functions from NMC 
($x >$ 0.006, $ Q^2 >$~0.5 GeV$^2$) \cite{NMCF2} Fig.~\ref{fig_f2nmc}, 
and E665 ($x >$ 0.0001, $Q^2 >$~0.2 GeV$^2$) \cite{f2e665} 
Fig.~\ref{fig_f2e665}; results on 
the deuteron-to-proton structure function ratio,
$F_2^d/F_2^p$, measured by NMC and E665 for $x >$ 0.0008, $Q^2 >$~0.2
GeV$^2$ (NMC \cite{gott,NMCR4})
and $x >$ 0.000 005, $Q^2 > $ 0.004 GeV$^2$ (E665 \cite{E665R,npe665}),
Fig.~\ref{fig_np}(left), 
precise results from these two collaborations on $x$,
$A$ and $Q^2$ 
dependence of nuclear shadowing
\cite{ashad_nmc1,ashad_nmc3,ashad_nmc_q2_sn,ashad_nmc2,E665SHAD,ashad_e665}, 
Fig.~\ref{fig_np}(right)
and Fig.~\ref{fig_ashad_e665};
and measurements of $R^{Ca}-R^C$ \cite{rcac}, $R^d-R^p$ \cite{rdp}, 
Fig.~\ref{fig_rdp}(left), 
$R^p$ and $R^d$ at low $x$ by NMC \cite{rdyring} and $R^A$
at low $x$ by CCFR \cite{rccfr}. 

The above-mentioned data were presented and discussed in Section 2.
Here we
shall add only a few remarks connected with their low $Q^2$
behaviour.
The data on the nucleon $F_2$ display a weak $x$,
and possibly a stronger than logarithmic $Q^2$, dependence, 
at $Q^2$ lower than a few GeV$^2$. Observe that 
also the photoproduction cross section between the fixed-target and
HERA energies
increases rather weakly with energy, cf. Fig. 2.6. This should be
contrasted 
with measurements at HERA for $Q^2$ larger than a few GeV$^2$, 
which show
 a strong increase of $F_2$ with decreasing $x$, Fig. 2.3.
The QCD analysis of the NMC $F_2$ data, \cite{nmcqcd}, show
that the
contribution of higher twists of the form similar to
(\ref{c4}) is moderate even at scales about 1 GeV$^2$.
This is visible in Fig. 2.8
where the $F_2^p(x,Q^2)$ at low $Q^2$ is well
described by models \cite{f2bbjk,DOLA} directly containing higher
twist contributions. The higher twists seem also to account for
the $Q^2$
dependence of the $F_2^p-F_2^n$ function, cf.section 2.6. 
 
The $F_2^d/F_2^p$ ratio, Fig.~\ref{fig_np}(left), 
which stays always 
below unity down to the smallest measured values of $x$, reflects
nuclear shadowing in the deuteron, only weakly dependent on
$x$.
The data are well described by a model \cite{dshad}, which
contains the
VMD part, essential at low $Q^2$, and which relates shadowing
to the
diffractively produced final states. The agreement extends over
nearly five
orders of magnitude in $x$.
No clear $Q^2$ dependence is visible in the shadowing data in a
wide interval of $Q^2$, neither in $F_2^d/F_2^p$ nor in $F_2^A/F_2^D$
\cite{npe665,ashad_e665}, cf. Fig.~\ref{fig_ashad_e665}(right), 
except possibly for $F_2(Sn)/F_2(C)$ 
at $x < $0.1, \cite{ashad_nmc_q2_sn}. Shadowing thus appears as 
a leading twist phenomenon. 

\subsection{Outlook}
In this section we have listed the ideas and results concerning the
electroproduction structure functions in the region of low values
of $x$ and $Q^2$. $F_2(x,Q^2)$ should vanish linearly with $Q^2$ for $Q^2
\rightarrow 0$ (for fixed $\nu$), an important property which
follows from the conservation of the electromagnetic current. The purely
partonic description of inelastic lepton scattering has thus to break down
for low $Q^2$. At moderate $Q^2$ the higher twist
contributions to $F_2$ which vanish as negative powers of $Q^2$
are sometimes included in the QCD data analysis. One also
expects at low
$Q^2$ that the VMD mechanism should play an important role.

The small $x$ behaviour of $F_2(x,Q^2)$ is dominated by 
pomeron exchange. Analysis of the structure function in the
small $x$ region for both low and moderate values of $Q^2$
can clarify our understanding of the pomeron. 
At large $Q^2$ the problem is linked  with the QCD expectations
concerning deep inelastic scattering at small $x$ 
\cite{SMX}. Besides the structure functions (or total cross
sections) complementary information on the pomeron can also be
obtained
from the analysis of diffractive processes in the electro- and
photoproduction. This concerns both inclusive diffraction and
diffractive production of vector mesons \cite{levy}.

Descriptions of the low $Q^2$, low $x$ behaviour of $F_2$ range
from pure fits to experimental data to dynamically motivated models.

There now exists a wealth of measurements of $F_2$ in the low $Q^2$, low
$x$ region. These include the NMC and E665 results which extend
down to very low $x$ and $Q^2$ and display
characteristic \lq\lq approach
to scaling" behaviour, as well as the first results from HERA at $Q^2$
which extend down to 1.5 GeV$^2$. The data were QCD analysed,
showing its
applicability down to scales of the order of 1 GeV$^2$. Nuclear
shadowing 
was studied in great detail for targets ranging from $A=2$ to
$A=208$ 
by the NMC and E665 collaborations. Its $x$, $Q^2$ and $A$
dependence 
were precisely measured. Preliminary data on $R(x,Q^2)$ have also
been reported (NMC and CCFR).

The fixed-target (unpolarised) structure function measurement programme 
comes to an end in 1996. Many experiments 
contributed in a great and successful effort to learn about properties
of partons and strong interactions. Several aspects of this knowledge are
yet not understood. One of these and perhaps the most challenging
one is low $x$ dynamics and in particular its dependence on the probing
photon virtuality, $Q^2$.  The new possibilities concerning the study of 
this problem have
opened 
up with the advent of HERA.
The data collected there
show a very strong increase of $F_2$ with decreasing $x$ at high
$Q^2$. As the HERA data improve and extend to lower $Q^2$ it will
be informative to see in which region of $Q^2$ the strong increase
gives way to the slow rise evident in photoproduction, see 
Fig.~\ref{sec2f8}. Such data in the transition region open up the
possibility of a unified understanding of the underlying dynamics. 

\setcounter{section}{3}
\setcounter{equation}{0}
\renewcommand{\theequation}{4.\arabic{equation}}
\setcounter{figure}{0}
\renewcommand{\thefigure}{4.\arabic{figure}}
\setcounter{table}{0}
\renewcommand{\thetable}{4.\arabic{table}}

\section{QCD interpretation}
\subsection{Introduction}

The behaviour of the proton structure function $F_2 (x, Q^2)$ at
small $x$ reflects the behaviour of the gluon distribution, since
the gluon is by far the dominant parton in this regime.  Fig.\ 4.1
shows a sketch of the gluon content of the proton in the various
kinematic regions.  Perturbative QCD does not predict the
absolute value of the parton distributions, but rather determines
how they vary from a given input.  For instance from given
initial distributions at some scale $Q_0^2$, Altarelli-Parisi (GLAP)
\cite{GLAP} evolution enables us to determine the distributions at
higher $Q^2$.  GLAP evolution resums the leading $\alpha_S \ln
(Q^2/Q_0^2)$ terms where, in a physical gauge, the $\alpha_S^n
\ln^n (Q/Q_0^2)$ contribution is associated with a space-like
chain of $n$ gluon emissions in which the successive gluon
transverse momenta are strongly ordered along the chain, that is
$q_{T1}^2 \ll \ldots \ll q_{Tn}^2 \ll Q^2$.

\begin{figure}[htb]
\begin{center}
 \epsfig{file=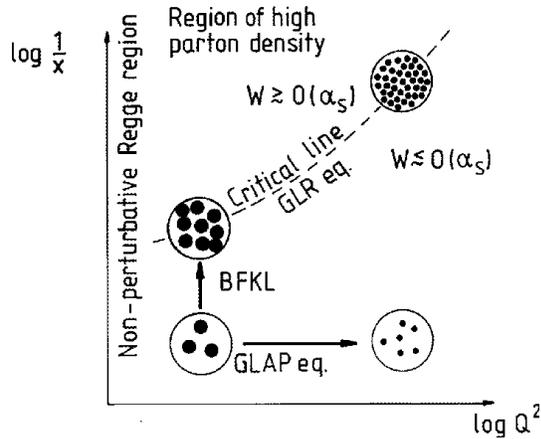,width=7cm}
\end{center}
\caption{
The gluon content of the proton as \lq\lq seen" in
various deep inelastic $(x, Q^2)$ regimes.  The critical line,
where gluon recombination becomes significant, occurs when $W
\approx O (\alpha_S)$.  $W$ is the ratio of the quadratic
recombination term to the term linear in the gluon density which
occur on the right hand side of the evolution equation.
}
\end{figure}

At sufficiently high electron-proton c.m.\ energy $\sqrt{s}$ we
encounter a second large variable, $1/x \sim s/Q^2$, and we must
resum the leading $\alpha_S \log (1/x)$ contributions.  The
resummation is accomplished by the BFKL equation \cite{BFKL}. 
In this case we have no ordering in $q_{Ti}^2$ along the chain,
but rather, as we evolve to smaller $x$, we have a diffusion or
random walk in $\ln q_{Ti}^2$.  The lack of strong ordering means
that we have to work in terms of the gluon distribution $f (x,
k_T^2)$ unintegrated over the gluon transverse momentum $k_T$. 
As we proceed to smaller $x$, via the BFKL equation, the gluon
density $f$ is predicted to increase as $x^{- \lambda}$ with $\lambda
\sim$ 0.5 (on account of the increased
$q_{Ti}$ phase space) and to possess a Gaussian-type distribution in $\ln
k_T^2$ which broadens as $\sqrt{\ln (1/x)}$.  This singular type of growth
in $x$, accompanied by the diffusion in $\ln k_T^2$, is the
characteristic property of the BFKL gluon density $f (x, k_T^2)$.

The increase of the gluon density with decreasing $x$ cannot
proceed indefinitely.  Eventually we reach the critical line
where gluon recombination effects become appreciable, see Fig.\
4.1.  At the onset these effects can be estimated by perturbative
QCD, but finally we enter a region of high density of weakly
interacting partons, where we have the unusual situation in which
we cannot use the normal methods of perturbation theory even
though $\alpha_S$ is small --- a region of much speculation and
interest.  At higher $Q^2$ we can evolve further in $x$ before we
reach the critical line, since the resolution goes as $1/Q$ and
the transverse area occupied by a parton $\sim 1/Q^2$.

\subsection{GLAP-based descriptions at small $x$}

Traditionally the parton distributions of the proton are
determined by global fits to a wide range of deep inelastic and
related data, see the review in these proceedings \cite{rgr}. 
The starting distributions are parametrized at some scale
$Q_0^2$, typically $Q_0^2 = 4 \:$ GeV$^2$, and evolved up in
$Q^2$ using next-to-leading order GLAP evolution.  The parameters
are then determined by fitting to all the available data.  A wide
range of precise data exist for $x \gapproxeq 0.05$ and so the
partons are well constrained in this region.

On the other hand for small $x$, $x \sim 10^{- 3}$, only the
structure function $F_2 (ep)$ is measured.  In principle these
HERA data should determine the small $x$ behaviour of the gluon
and the sea quark distributions.  Roughly speaking the data on
$F_2$ constrain the sea $S$ and the data on the slope $\partial
F_2/\partial \ln Q^2$ determine the gluon $g$.  For example if we
take 
\begin{equation}
F_2 \; \sim \; x S \; \sim \; A_S \: x^{- \lambda_S}
\label{eq:b1}
\end{equation}
\vskip-0.75cm
\begin{equation}
\frac{\partial F_2}{\partial \ln Q^2} \; \sim \; x g \; \sim \;
A_g \: x^{- \lambda_g}, 
\label{eq:b2}
\end{equation}
\noindent then we might expect to determine $\lambda_S$ and
$\lambda_g$.  The most recent HERA data are well described by
a parametrization with $\lambda_S = \lambda_g \sim 0.2 - 0.25$
with $Q_0^2 = 4 \:$ GeV$^2$, see MRS(A$^\prime$)
\cite{MRS}.  However, as the data improve we should be able to
determine $\lambda_S$ and $\lambda_g$ independently.  To be more
precise the slope (\ref{eq:b2}) gives information on $P_{qg}
\otimes g$.  The convolution means that the gluon is sampled at
higher $x$.  We will see the convolution also has implications
for the scheme dependence.

A satisfactory description of the data is also obtained from a
set of \lq\lq dynamical" partons evolved up from valence-like
distributions at a much lower scale $Q_0^2 = 0.34 \:$ GeV$^2$
\cite{GRV}.  The dynamical model was originally based on the
attractive hypothesis that at some low scale we just have the $u$
and $d$ valence quark distributions.  The gluon and sea quark
distributions are generated radiatively from the valence quark
distributions, via the processes $q \rightarrow qg$ and $g
\rightarrow q\overline{q}$.  Unfortunately the radiated
distributions are spiked towards small $x$ and must be
supplemented by inputting a contribution at larger $x$.  For
instance to describe prompt photon data valence-like gluons are
needed at the input scale $Q_0^2$, and to describe the NMC
measurements of $F_2$ valence-like sea quark distributions are
required.  It can be argued that $Q_0^2 = 0.34 \:$ GeV$^2$ is
much too low for perturbative QCD to be reliable.  The GRV
philosophy is that GLAP evolution preserves the leading twist and
so the parametrization becomes physical at some higher scale. 
The surprise is that the GRV partons appear to give a reasonable
description as low as $Q^2 \approx 1 \:$ GeV$^2$.

An idea of the potential steepness of the gluon distribution within 
GLAP evolution can be glimpsed from the double leading
logarithm (DLL) approximation which applies at small $x$ and
large $Q^2/Q_0^2$ ~\cite{DGPTWZ}. In this limit GLAP evolution of the gluon
gives
\begin{equation}
xg (x, Q^2) \; \sim \; xg (x, Q_0^2) \; \exp \: \left ( 2 \left [ \frac{36}{25} 
\; \ln  \left (\frac{t}{t_0} \right ) \; \ln  \left (\frac{1}{x} \right ) 
\right ]^{\frac{1}{2}} \right )
\label{eq:a1}
\end{equation}
\noindent where $t \equiv \ln (Q^2/\Lambda^2)$.  That is, $xg$
increases faster than any power of $\ln (1/x)$, but slower than a
power of $(1/x)$.  Moreover we see that $xg$ increases faster as
$x \rightarrow 0$ the longer is the $Q_0^2 \rightarrow Q^2$
evolution. The GRV analysis demonstrates that with a flat (or, to be precise, a
decreasing) input distribution at small $x$ such a rise is obtained with
the low starting point $Q_0^2 = 0.34$ GeV$^2$. The origin of the growth of
(\ref{eq:a1}) at small $x$ can be traced back to the singular behavior of the 
gluon
anomalous dimension at $n=1$ or $\omega=0$ (see below).
On the other hand, if the input gluon is singular,
$xg (x, Q_0^2) \sim x^{- \lambda}$ with $\lambda > 0$, as in
MRS/CTEQ, then this behaviour overrides the DLL form.  In fact
the present HERA data for $F_2$ cannot really distinguish between these
small $x$ gluon behaviours.  The data are shown in Fig. 2.4 as
a function of $\ln Q^2$ at given values of $x$.  From
(\ref{eq:b2}) we see that the slope of these points is a measure
of the gluon.  Also shown on the plot is the description
obtained from GRV partons \cite{GRV}. 
The GRV prediction is a little too steep at small $x$ but this could
be easily accommodated by a small increase in $Q_0^2$.  In
summary the existing data determine only one parameter; either
$\lambda (\approx 0.2-0.25)$ of the MRS/CTEQ global fits or
$Q_0^2 (\approx 0.5-1 \:$ GeV$^2)$ of the GRV-type
description. 

In principle, a possible check of whether the rise of $F_2$ at small $x$
is closer to the
behaviour (\ref{eq:a1}) or to a power-like growth may be obtained from the
double asymptotic scaling behaviour ~\cite{BFFR}. Namely, (\ref{eq:a1}) 
suggests that, instead 
of the variables $\ln(1/x)$ and $\ln(t/t_0)$, one might use the two variables
\begin{equation}
\sigma = \sqrt{ \ln(t/t_0) \ln(x_0/x) }, \hspace{1cm}
\rho =\sqrt{ \ln(t/t_0) / \ln(x_0/x)}.
\end{equation}
For example, by rescaling the observed $F_2$ by a factor $R'_F$ which 
eliminates power-like prefactors to the exponential in (\ref{eq:a1}) and higher 
order corrections in the exponent
the remainder should be proportional to the exponential, i.e.\ the logarithm
of $R'_F F_2$ should be linear in $\sigma$ and independent of $\rho$. 
Data \cite{DAS} show approximate agreement with this prediction, provided 
the scale $Q_0^2$ is chosen to be $1$ GeV$^2$ and $x_0 \sim$ 0.1.
The resummation of $\alpha_S 
\ln (1/x)$ contributions will, at sufficiently small $x$, give effects which 
violate the double scaling form.

There is, however, some warning against the use of next-to-leading order GLAP 
evolution in combination with a flat input for explaining the observed 
rise at small $x$. Consider, for simplicity, GLAP evolution for the gluon alone
\begin{equation}
\frac{\partial g (x, Q^2)}{\partial \ln Q^2} \; = \; \int_x^1 \;
\frac{dy}{y} \; P_{gg}  \left ( \frac{x}{y} \right ) \; g (y,
Q^2).
\label{eq:a7}
\end{equation}
\noindent In moment space this takes the factorized form
\begin{equation}
\frac{\partial \overline{g} (\omega, Q^2)}{\partial \ln Q^2} \; =
\; \gamma_{gg} (\omega, \alpha_S) \: \overline{g} (\omega, Q^2)
\label{eq:a8}
\end{equation}
\noindent 
where the anomalous dimension in leading order near $\omega=0$
\begin{equation}
\gamma_{gg} \; \equiv \; \int_0^1 \; dx \: x^\omega \: P_{gg} (x,
\alpha_S) \; \approx \; \frac{\overline{\alpha}_S}{\omega}
\label{eq:a9}
\end{equation}
\noindent 
In $x$-space this approximation is equivalent to retaining only 
the small $x$ approximation $P_{gg} \approx
\overline{\alpha}_S/x$ with $\overline{\alpha}_S \equiv 3
\alpha_S/\pi$.  This is the DLL approximation which generates a
gluon distribution of the form given in (\ref{eq:a1}). The general pattern 
of the terms that occur in the expansion of the anomalous dimensions and
the coefficient function functions as power series in $\alpha_s$ and the
moment index $\omega$ is summarised in Fig.\ 4.2. 
\begin{figure}[htb]
\begin{center}
 \epsfig{file=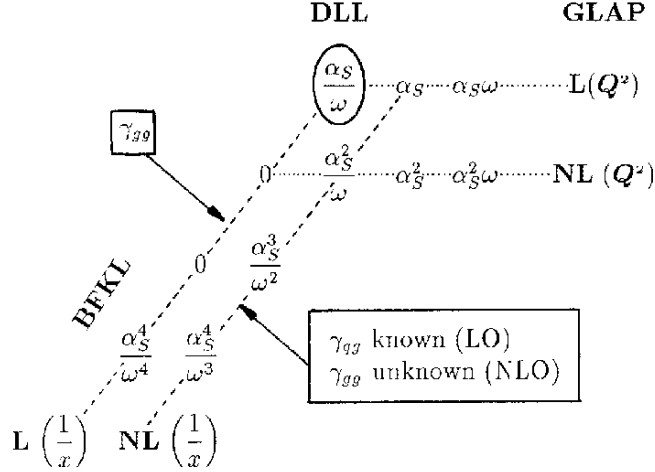,width=8.5cm}
\end{center}
\caption{
Possible terms in the perturbative expansion of
the anomalous dimensions (and coefficient functions).  Leading
order GLAP and BFKL evolution have only the DLL term in common.
}
\end{figure}
\noindent The terms
that are included in the full leading (and next-to-leading) order
GLAP evolution are shown connected by horizontal dotted lines. On the 
other hand in the vicinity of the point $\omega=0$ which governs the small $x$
limit it will be more accurate to resum, in the anomalous dimension,
all singular terms of the form
\begin{equation}
\gamma_{gg} \; \approx \; \sum_{n = 1}^\infty \; A_n \: \left (
\frac{\alpha_S}{\omega} \right )^\omega \; 
\label{eq:b9}
\end{equation}
which in $x$-space corresponds to 
\begin{equation}  
\sum_{n = 1} \; A_n \: \alpha_S \; \frac{(\alpha_S \log 1/x)^{n - 1}}{(n
- 1)!},
\label{eq:a10}
\end{equation}
\noindent that is an $\omega^{- n}$ behaviour translates into a
$(\log 1/x)^{n - 1}$ behaviour. These terms are included in the BFKL equation,
and the coefficients $A_n$ have been computed
\cite{J}. Interestingly the coefficients $A_2 = A_3 = A_5 = 0$. The anomalous 
dimension $\gamma_{gq}$ has a similar structure as $\gamma_{gg}$, i.e. the 
leading order consists of terms $(\alpha_s/\omega)^n$, whereas
for $\gamma_{qg}$ and $\gamma_{qq}$ the most singular terms are down by one
power of $\alpha_s$: $\alpha_S (\alpha_S/\omega)^n$.
For all four cases the coefficients of these leading order terms are known. 
In particular, for $\gamma_{qg}$ the coefficients are non-vanishing 
(unlike $\gamma_{gg}$), positive definite and large \cite{ch}.  As a 
consequence 
the BFKL increase\footnote{The reduction of the BFKL $k_T$-factorized 
(\ref{eq:a6}) 
to the 
GLAP collinear form is discussed in ref.\ \cite{kt}.} of $F_2 (\sim P_{qg} 
\otimes g)$ with decreasing $x$ appears to be much more due to the resummation 
in $P_{qg}$ (and in the coefficient functions) than that for $g (\sim P_{gg} 
\otimes g)$.  It is clear that 
for a consistent analysis we need higher order corrections to
$\gamma_{gg}$ and $\gamma_{gq}$: terms of the order 
$\alpha_s (\alpha_s /\omega)^n$. It is hoped that they can be derived from
the next-to-leading order corrections to the BFKL kernel which are being
computed by Fadin and Lipatov \cite{FL}.

Based upon our present (limited) knowledge of these singular terms of the
matrix of anomalous dimensions several numerical studies have been undertaken
\cite{pser}, incorporating resummation in various ways.
A general conclusion is that GLAP evolution with a flat 
input distribution is rather sensitive to whether resummation of the 
singular terms is included or not. The resummation in $\gamma_{qg}$, although
formally nonleading, seems to play a key role. On the other hand, if we start 
from a singular $x^{-\lambda}$ input with $\lambda \gapproxeq 0.3$ then the
result appears to be rather 
stable with respect to resummmation. A theoretical issue is the scheme
dependence of the resummation effects. Catani and Hautmann emphasized
at the Workshop
that accurate data on $F_L$ could remove the ambiguities and overconstrain
the problem. Indeed they showed that it is possible to obtain scheme
independent evolution equations of the form
\begin{eqnarray}
\frac{\partial F_2 (x, Q^2)}{\partial \ln Q^2} & \sim &
\Gamma_{22} \otimes F_2 + \Gamma_{2L} \otimes F_L \nonumber \\
\frac{\partial F_L (x, Q^2)}{\partial \ln Q^2} & \sim &
\Gamma_{L2} \otimes F_2 + \Gamma_{LL} \otimes F_L \nonumber
\end{eqnarray}
\noindent which inter-relate only physical observables, where the
anomalous dimensions $\Gamma_{ij}$ are computable in perturbative
QCD.   

\subsection{BFKL-based description of $F_2$ at low $x$}

This topic is well reviewed in \cite{SMX}, so we will be brief.  The
BFKL equation for the unintegrated gluon distribution $f (x,
k_T^2)$ may be written in the differential form
\begin{eqnarray}\fl
\label{eq:b10}
\frac{\partial f (x, k_T^2)}{\partial \ln (1/x)}  =  \frac{3
\alpha_S}{\pi} \; k_T^2 \; \int \; \frac{dk_T^\prime}{k_T^{\prime
2}} \; \left [ \frac{f (x, k_T^{\prime 2}) \: - \: f (x,
k_T^2)}{| k_T^{\prime 2} \: - \: k_T^2 |} \; + \; \frac{f (x,
k_T^2)}{(4k_T^{\prime 4} \: + \: k_T^4)^{\frac{1}{2}}} \right ] \\
\label{eq:a3}
\equiv  K \otimes f
\end{eqnarray}
\noindent where the $f (x, k_T^{\prime 2})$ term corresponds to
real gluon emissions, and where the cancellation of the
$k_T^\prime = k_T$ singularity between the real and virtual gluon
contributions is apparent.  The gluon distribution $f (x, k_T^2)$
may be calculated by integrating down in $x$ from a starting
distribution at, say, $x = x_0 = 0.01$.  The BFKL equation
effectively sums the leading $\alpha_S \log (1/x)$ contributions;
from (\ref{eq:a3}) we see that as $x \rightarrow 0$
\begin{equation}
f \; \sim \; \exp \: (\lambda \log (1/x)) \; \sim \; x^{-
\lambda}
\label{eq:a4}
\end{equation}
\noindent where $\lambda$ is the largest eigenvalue of the BFKL
kernel $K$.  For fixed $\alpha_S$ it can be shown \cite{BFKL} to
be $\lambda = (3 \alpha_S/\pi) \: 4 \ln 2$.  We need to know the NLO 
$\log (1/x)$ contributions to fully specify the running of $\alpha_S$.  
However, if a physically reasonable prescription for the running is 
assumed then it is found \cite{akms} that the solution has the form
\begin{equation}
f \; \sim \; C (k_T^2) \; x^{- \lambda}.
\label{eq:a5}
\end{equation}
For running $\alpha_S$ the integral in (\ref{eq:b10}) is weighted more to the 
infrared (non-perturbative) region, but it is found that $\lambda 
\approx 0.5$ is less sensitive to the treatment of the infrared 
region than the normalization $C$.

The BFKL predictions of the proton structure functions $F_i$ are
obtained using the $k_T$ factorization theorem \cite{fac,CE,LRS2}
\begin{equation}
F_i (x, Q^2) \; = \; \int_x^1 \; \frac{dx^\prime}{x^\prime} \;
\int \; \frac{dk_T^2}{k_T^2} \; f (x^\prime, k_T^2) \:
F_i^{\gamma g} \left ( \frac{x}{x^\prime}, k_T^2, Q^2 \right )
\label{eq:a6}
\end{equation}

\begin{figure}[htb]
\begin{center}
 \epsfig{file=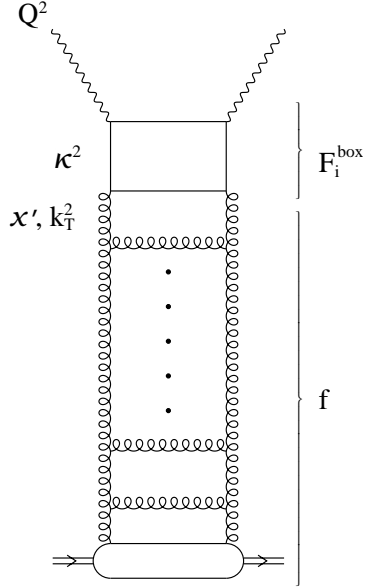,width=5cm}
\end{center}
\caption{
Pictorial representation of the
$k_T$-factorization formula of (\protect{\ref{eq:a6}}).  The variables
$x^\prime$ and $k_T$ that are integrated over are, respectively, 
the longitudinal fraction
of the proton's momentum and the transverse momentum carried by
the gluon which dissociates into the $q\overline{q}$ pair.
}
\end{figure}

\noindent where $F_i^{\gamma g}$ are the off-shell gluon
structure functions which at lowest order are given by the quark
box (and crossed-box) contributions to photon-gluon fusion, see
Fig.\ 4.3.  In the small $x$ region it is clearly important to
tackle the resummation of the $\alpha_S \log (1/x)$ contributions
and to go beyond (next-to-leading order) GLAP.  Indeed a
satisfactory BFKL prediction of the rise of $F_2$ at small $x$
was obtained prior to the HERA measurements \cite{akms}. 
However, there are ambiguities and limitations in the BFKL
description of $F_2$.  We list them below.
\begin{itemize}
\item[(i)] Due to the diffusion of the gluon distribution $f (x,
k_T^2)$ in $\ln k_T^2$, there is a significant contribution to
(\ref{eq:a6}) from the infrared region which is beyond the scope
of perturbative QCD and has to be included phenomenologically. 
The ambiguity is primarily in the overall normalization of $F_2$
rather than the $x$ dependence.  A physically reasonable choice
of the infrared parameters is found to give the experimental
normalization.

\item[(ii)] The BFKL equation only resums the leading order $\ln
(1/x)$ terms.  The next-to-leading order contributions are
essential for a stable prediction.  Sub-leading effects,
including energy-momentum conservation on the gluon emissions,
are expected to suppress the value of $\lambda$, see, for example,
\cite{KMSK}.  When known,
these will check whether or not the $\alpha_S (k_T^2)$
prescription for the running of $\alpha_S$ is correct.

\item[(iii)] An underlying soft Pomeron contribution has to be
included in the small $x$ region, determined by extrapolation of
the large $x$ values of $F_2$.

\item[(iv)] Shadowing corrections to the BFKL equation will eventually 
suppress the $x^{- \lambda}$ growth.  Although they have not yet been 
fully formulated, the evidence from the observed ratio of
diffractive to non-diffractive deep inelastic events, and from
the persistent rise of $F_2$ at very low $Q^2 \approx 2 \:$
GeV$^2$, indicates that shadowing effects are at most 10\% in the HERA
regime.

\item[(v)] We need a unified approach which incorporates both the
BFKL and GLAP resummations.  We discuss recent developments
below.
\end{itemize}
\indent The BFKL gluon distribution $f$ is characterised by a
steep rise with decreasing $x$ which is accompanied by a
diffusion in $\ln k_T^2$.  $F_2$ measures only the rise; it is
too inclusive an observable to probe the $k_T$ dependence.  For
this we need to explore the properties of the final state, such
as deep inelastic events containing an identified energetic
forward jet, see section 5.

\subsection{Unified evolution}

Because of the possible onset of the BFKL behaviour of $F_2$ in the HERA 
small $x$ domain it is important to study the validity of the GLAP evolution
in this region and to look for an interpolation between the two approaches.
First answers to these questions are contained in the studies of resummation
effects (see above). Including the $\log (1/x)$ resummation effects in GLAP, 
via the $(\alpha_S/\omega)^n$ and $\alpha_S (\alpha_S/\omega)^n$ terms in 
$\gamma_{gg}$ and $\gamma_{qg}$ respectively (and also in the coefficient 
functions) leads to a stronger increase of $F_2$ at 
small $x$, i.e. we are getting closer to BFKL. 

Ciafaloni, Catani, Fiorani, and Marchesini 
\cite{ccfm} have proposed a unified evolution equation which
embodies BFKL evolution at small $x$ and GLAP evolution at larger
$x$ (see also the review in these proceedings \cite{SMX}).  The
CCFM equation is based on the coherent radiation of
gluons, which leads to an angular ordering of gluon emissions. 
Outside the ordered region there is destructive interference
between the emissions.  For simplicity we concentrate on small
$x$.  Then the differential probability for emitting a gluon of
momentum $q$ is of the form
\begin{equation}
dP \; \sim \; \overline{\alpha}_S \: \Delta_R \; \frac{dz}{z} \;
\frac{d^2 q_T}{\pi q_T^2} \; \Theta (\theta - \theta^\prime)
\label{eq:a11}
\end{equation}
\noindent where successive emissions along the space-like gluon
chain occur at larger and larger angles.  $\Delta_R$ represents
the virtual gluon loop corrections which screen the $1/z$
singularity.  We can use (\ref{eq:a11}) to obtain a recursion
relation expressing the contribution of the $n$ gluon emission in
terms of that of $n - 1$.  On summing we find that the gluon
distribution satisfies an equation
\begin{eqnarray}
F (x, k_T^2, Q^2) & = & F^{(0)} (x, k_T^2, Q^2) \: + \nonumber \\
& & \\
& & \overline{\alpha}_S \; \int_x^1 \; \frac{dz}{z} \; \int \;
\frac{d^2 q}{\pi q^2} \; \Delta_R \: \Theta (Q - zq) \: F \:
\left ( \frac{x}{z}, | \vec{k}_T + \vec{q} |^2, q^2 \right ) \nonumber
\label{eq:a12}
\end{eqnarray}
\noindent where $F = f/k_T^2$.  The angular ordering, in the form
of the $\Theta$ function, introduces a dependence on an
additional scale (that turns out to be the hard scale $Q$ of the
probe), which is needed to specify the maximum angle of gluon
emission.

When we \lq\lq unfold" $\Delta_R$, so that the real and virtual
contributions appear on equal footing, and then take the leading
$\ln (1/x)$ approximation we find that (\ref{eq:a12}) reduces to
the BFKL equation with $F$ independent of $Q^2$ (since $\Theta (Q
- zq) \rightarrow 1$).  On the other hand, in the large $x$
region $\Delta_R \sim 1$ and $\Theta (Q - zq) \rightarrow \Theta
(Q - q)$, which leads to GLAP transverse momentum ordering.  If
we then replace $\overline{\alpha}_S/z$ by $P_{gg}$ we see that
(\ref{eq:a12}) becomes the integral form of the GLAP equation.

Explicit numerical solutions $F (x, k_T^2, Q^2)$ of the CCFM
equation have recently been obtained in the small $x$ region
\cite{kms}.  As anticipated, they show a singular $x^{- \lambda}$
behaviour, with $\lambda \approx 0.5$, a $k_T$ dependence which
broadens as $x$ decreases and a suppression of the gluon as
compared to BFKL at low $Q^2$.  Fig.\ 4.4 shows a comparison of the
CCFM $k_T$-factorization prediction for $F_2 = F \otimes
F_2^{\gamma g}$ with recent HERA data.  The DLL dot-dashed curve
is obtained by the same procedure except that we replace $\Theta
(Q - zq)$ by $\Theta (Q - q)$ and set $\Delta_R = 1$.  The
difference between the CCFM and DLL curves therefore show the
value of $x$ at which resummation effects become important.

\begin{figure}[htb]
\begin{center}
 \epsfig{file=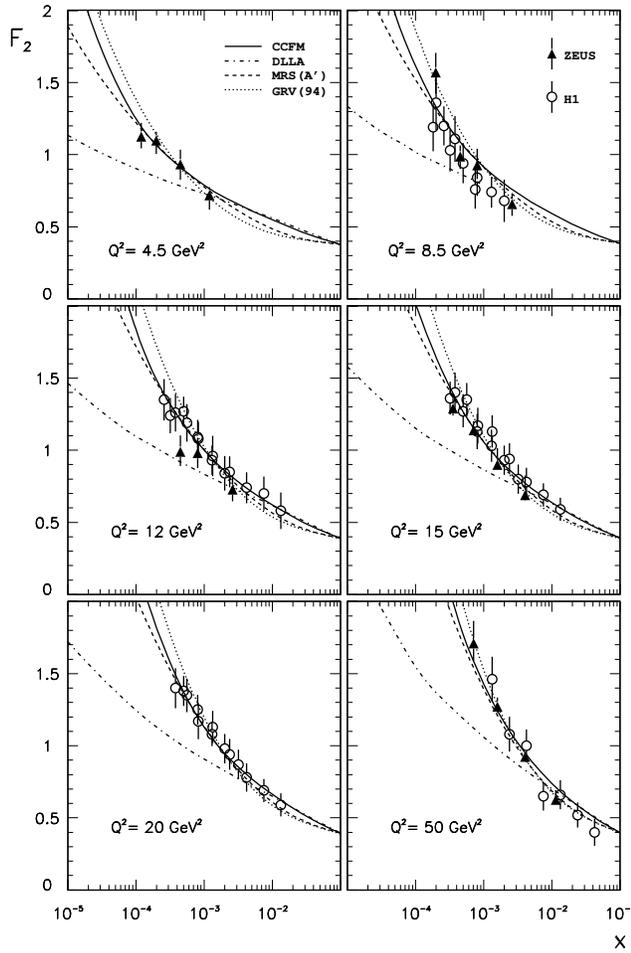,width=8.5cm}
\end{center}
\caption{
The continuous curve is the CCFM prediction of
$F_2$ obtained in ref.\ \protect{\cite{kms1}} compared with preliminary
1994 HERA measurements \protect{\cite{HERAPS}}.  
The DLL curve is obtained by the same procedure
except that we replace $\Theta (Q - zq)$ by $\Theta (Q - q)$ and
set $\Delta_R = 1$.  Also shown are the descriptions obtained by
GRV \protect{\cite{GRV}} and MRS(A$^\prime$) \protect{\cite{MRS}} 
partons.  The figure is taken from ref.\ \protect{\cite{kms1}}.
}
\end{figure}

\subsection{Open questions}

A completely satisfying explanation of the observed rise at small $x$ 
cannot avoid investigating the 
transition from the structure function $F_2$ at small $x$ to the 
photoproduction
cross section at high energies. Approaching this transition from the deep
inelastic kinematic region (large $Q^2$ and not so small $x$) where 
perturbative QCD can safely be used, we expect first to see the onset of 
more complicated partonic interactions (e.g.\ recombination effects) which
will screen the strong rise of the GLAP or BFKL approximation. So far, these 
contributions have not been observed. Numerical studies based upon the
nonlinear Gribov-Levin-Ryskin \cite{GLR} equation indicate that in the HERA 
region it
will be difficult to identify the screening effects: even if they are not
negligibly small in the HERA region, they could be masked by a change in the
input to the GLAP evolution equations.  Unfortunately, the GLR equations 
cannot
be used to make a reliable quantitative prediction of the magnitude of 
screening effects, due primarily to the lack of knowledge 
of the characteristic transverse area within which the gluons are concentrated 
\cite{GKR}, but also of the size of the enhancement due to QCD multigluon 
correlations \cite{BR,LRS}. So at present it is not quite clear yet
whether recombination effects are relevant at HERA and how large they are.

The conventional
GLAP analysis is constrained to leading twist only, and for large $Q^2$ this
approximation is well-justified. However, once we are approaching the low-$Q^2$
region, the neglect of higher-twist becomes more and more questionable. 
Interestingly the BFKL Pomeron does include higher-twist contributions, 
but only a small subset of the higher-twist operators that exist in QCD. In the
small-$x$ region, among the twist-four operators it is the four gluon operator 
which is expected to become important. Evolution equations of higher-twist
operators have been studied in ~\cite{BL}, and the anomalous dimension of the
twist-four gluon operator has been calculated in ~\cite{B,LRS}. But so far, the
results of these QCD calculations have not been used yet for numerical studies.
A phenomenological estimate of higher twist contributions at 
small $x$ and low $Q^2$ has been
done in ~\cite{MRSLQ}: it indicates that higher-twist effects are surely 
important when $Q^2$ less than 1 GeV$^2$, and when $Q^2$ increases that 
they disappear, for large $x$ more rapidly than for small $x$. Further 
studies of higher twist contributions at small $x$ are important 
and clearly needed.

It is quite clear that we are still quite far away from a satisfactory
understanding of the small-$x$ regime. The measurements of $F_2$ at small $x$ 
have considerably improved and indicate a fairly rapid transition from the 
\lq\lq soft" $x^{- 0.08}$ behaviour at $Q^2 = 0$ to a more singular $x^{- 0.2}$ 
behaviour by $Q^2 \sim 2 \:$ GeV$^2$.  Acceptable GLAP-based partonic 
descriptions (including next-to-leading order $\log Q^2$ terms) are 
possible down to at 
least $Q^2 \sim 1 \:$ GeV$^2$.  Equally well the small $x$ data can be 
described by BFKL leading order $\alpha_S \log (1/x)$ resummation, via $k_T$ 
factorization; though we await the next-to-leading $\log (1/x)$ contribution 
for a stable prediction.  We must develop a unified formalism which 
embraces both GLAP and BFKL.  The CCFM approach is a step in this direction.  
As noted above, more work is needed to quantify the effects of parton screening.  
Of course the partonic formalism must eventually fail as we go to lower and 
lower $Q^2$.  It remains a challenge to match this perturbative QCD approach 
to the successful non-perturbative Regge and Vector-Meson-Dominance 
description of proton structure for $Q^2 \sim 0$.

\setcounter{section}{4}
\setcounter{equation}{0}
\renewcommand{\theequation}{5.\arabic{equation}}
\setcounter{figure}{0}
\renewcommand{\thefigure}{5.\arabic{figure}}
\setcounter{table}{0}
\renewcommand{\thetable}{5.\arabic{table}}

\section{Hadronic Final States}

In order to make further progress to find the appropriate QCD approach to use 
in the small $x$ regime, hadronic final
states are analysed. 
The experiments at HERA measure the full hadronic final state, apart
from losses in the beampipe. Several results on hadronic final states
are reported in~\cite{thompson}.
In this section three topics will be
addressed. The first topic is  charged particle spectra as
measured in the current region of the Breit frame. Here one would expect
the timelike evolution of the parton shower to evolve as  in
$e^+ e^-$ annihilation. The second topic is the study of strangeness 
production and a possible discrepancy of the strangeness suppression
factor between DIS and $e^+ e^-$ experiments. Strange 
 particle data can also be interpreted in the light of
possible signals
for instanton production. Finally an overview of the hadronic final
state at low $x$ will be given with a discussion of possibilities of
detecting the onset of QCD $\alpha_S \log 1/x$ resummation effects.

\subsection{Breit frame: current region}

A natural frame to study the dynamics of the hadronic final state in DIS
is the Breit frame~\cite{FEYN}. In the  Quark Parton Model (QPM) the
$z$-component of the momentum of the incoming quark is $Q/2$ before and
$-Q/2$ after the interaction with the exchanged virtual boson.
In  $e^+ e^-$ annihilation the two quarks are produced with equal
and opposite momentum, $\pm \sqrt{s}/2$. 
In the direction of the struck quark (
called the current region) in the Breit frame
the particle
spectra are expected to have no dependence on $x$ and a dependence on
$Q^2$ similar to that observed in $e^+ e^-$ annihilation~\cite{Tbreit}
at energy $ s = Q^2$.
\begin{figure}[htb]
\begin{center}
 \epsfig{file=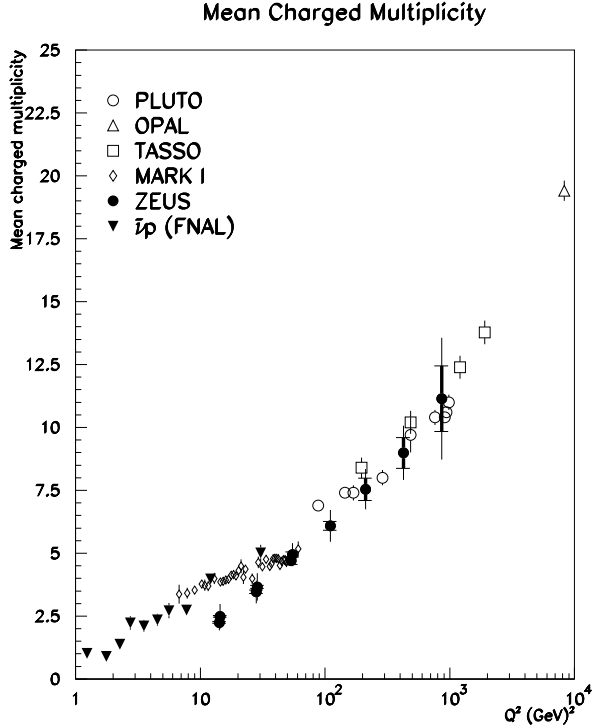,height=9cm} 
\end{center}
\caption{Twice the measured ZEUS multiplicity in
 the current region is compared to the results from HRS, 
 MARK I, OPAL, PLUTO and TASSO. Also show is twice the current 
 region multiplicity from FNAL DIS $\bar\nu P$ experiment. 
 All data have been corrected for $K^0_S$ and $\Lambda$ decays.}
\label{fig:HFS1}
\end{figure}

Both ZEUS~\cite{Zbreit} and H1~\cite{Hbreit} have published results of
the charged particle
scaled momentum and multiplicity spectra in the current region of the
Breit frame. The results indicate these fragmentation properties
of the struck quark in DIS are similar to those  from quarks created 
in $e^+ e^-$ annihilation for $Q^2 > 60$ GeV$^2$.
For $Q^2 < 60$ GeV$^2$ the mean multiplicity, $\langle n_{ch} \rangle$,
tends to be lower in DIS
than  in $e^+ e^-$ experiments~\cite{nche+e-}, see Fig.~\ref{fig:HFS1}.

In the low $Q^2$ regions measured at HERA 
it is possible to study the $x$ dependency by
comparisons with data from 
previous fixed-target DIS measurements~\cite{Musgrave},
which are 2 orders of magnitude higher in $x$ than 
the values probed at HERA. In Fig.~\ref{fig:HFS1} it is shown that 
the DIS $\langle n_{ch} \rangle $ data exhibit a marked dependence on $x$ 
contrary to the
naive expectation. Also, in this higher $x$ region the agreement between
DIS and $e^+ e^-$  experiments seems to hold. This $x$ dependence
can be explained when mass effects are taken into account in the DIS
kinematics. For (2+1) jet kinematics, the longitudinal momentum of the 
outgoing quark pair is given by $ p^{rad}_Z = \frac{1}{2} (\hat{s} - Q^2)/Q$ 
where $\hat{s}$ is the square of the
invariant mass of the emitted $q\bar{q}$ or a $qg$ pair.
When $Q^2 \gg \hat{s}$, the radiation is emitted in the QPM
direction, \( p^{rad}_Z \sim p^{QPM}_Z \). However, at low $Q^2$
and  $x$,
$\hat{s}$ is likely to be bigger than $Q^2$ and the radiation will be
emitted in the direction opposite to the QPM direction, hence
depopulating the current region. Typically the
smallest resolvable value of $\hat{s}$ is around 20 GeV$^2$ in the
kinematic region of HERA, therefore at $Q^2$ around 10 GeV$^2$ the emitted
radiation has a \( p^{rad}_Z = Q/2 \) or more, whereas $p^{QPM}_Z =
-Q/2$.

\subsection{Strangeness Production}

The production of strangeness has been studied by the experiments at HERA
through the production of $K^0$ mesons and $\Lambda$ 
baryons~\cite{zeusk0,h1k0}.
 These particles
are reconstructed from the measurements of the charged tracks of the
decay particles. The acceptance is presently limited to the central region
of the detectors, i.e. in pseudorapidity $\eta$
$|\eta| < 1.3$, hence the strangeness production
measurement  mainly covers from the current quark fragmentation region.
Fig.~\ref{k0} shows $K^0$ multiplicities, corrected for detector effects, for
 non-diffractive
deep inelastic events (i.e. events with no visible gap in the direction of
the proton remnant, see~\cite{diffdurham}) as function of the pseudorapidity
$\eta$ and
the transverse 
momentum $p_t$. 
The data are compared with model predictions. 
The MEPS model is based on the LEPTO~\cite{lepto} 
program and contains first order QCD matrix elements, matched with
parton showers which approximate the higher orders, 
for the partonic simulation of the hadronic final state. The
CDM model is based on the ARIADNE~\cite{ariadne} program and provides an 
implementation
of the colour dipole model of a chain of independently radiating dipoles
formed by  emitted gluons (photon gluon fusion events are not described
by this picture and are added at a rate given by the QCD matrix elements).
The calculations are shown for standard strange quark suppression  
of 0.3, i.e. the production of quark-antiquark pairs is $u\overline{u}:
d\overline{d}:s\overline{s}= 1:1:0.3$,  tuned  essentially on $e^+e^-$ 
hadronic data. It appears that this predicts to many $K^0$'s, as is seen most 
prominently in the $\eta$ distribution.  The calculation
with a strangeness suppression of 0.2 is in better agreement with the data.

This difference is intriguing and could be  a sign of the breakdown of 
jet universality, e.g. due to a \lq\lq medium dependence" of the fragmentation, 
or different kind of gluon strings in $e^+e^-$ and $ep$ 
scattering~\cite{sjoestrand}. However before  speculating on such 
possibilities, additional  data are needed. It should be checked whether
it is  the $K^0$ rate which is lower, or whether all
 particle rates are affected, 
e.g. by studying  $K/\pi$ ratios. Other strange
particle species, especially anti-baryons, should be studied as well.

Strange particle production is also of interest for the study of 
QCD instantons. Instantons are related to non-perturbative effects
of QCD, and effects on the hadronic final state 
in $ep$ collisions at HERA were recently 
calculated~\cite{schremp}. Due to the expected isotropic decay of a
dense partonic system, events with an unusual structure are expected.
The uncertainty on the expected 
instanton rate is however large. Due to the
flavour symmetry in the instanton
decay, less strangeness suppression is expected,
leading to an average density of 0.4 $K^0$s per unit of pseudorapidity
in the region of the instanton \cite{gibbs}.
 The shape of the rapidity distribution
is expected to be different from DIS events without an instanton.
Clearly enhanced strangeness production will be one of the features to 
search for instantons, others being large $E_T$ in a limited rapidity
interval, isotropically spread energy, etc. Searching for instantons is one of
the exciting topics of hadronic final state studies at HERA.  

\begin{figure}[htb]  \centering  \unitlength 1mm
 \epsfig{file=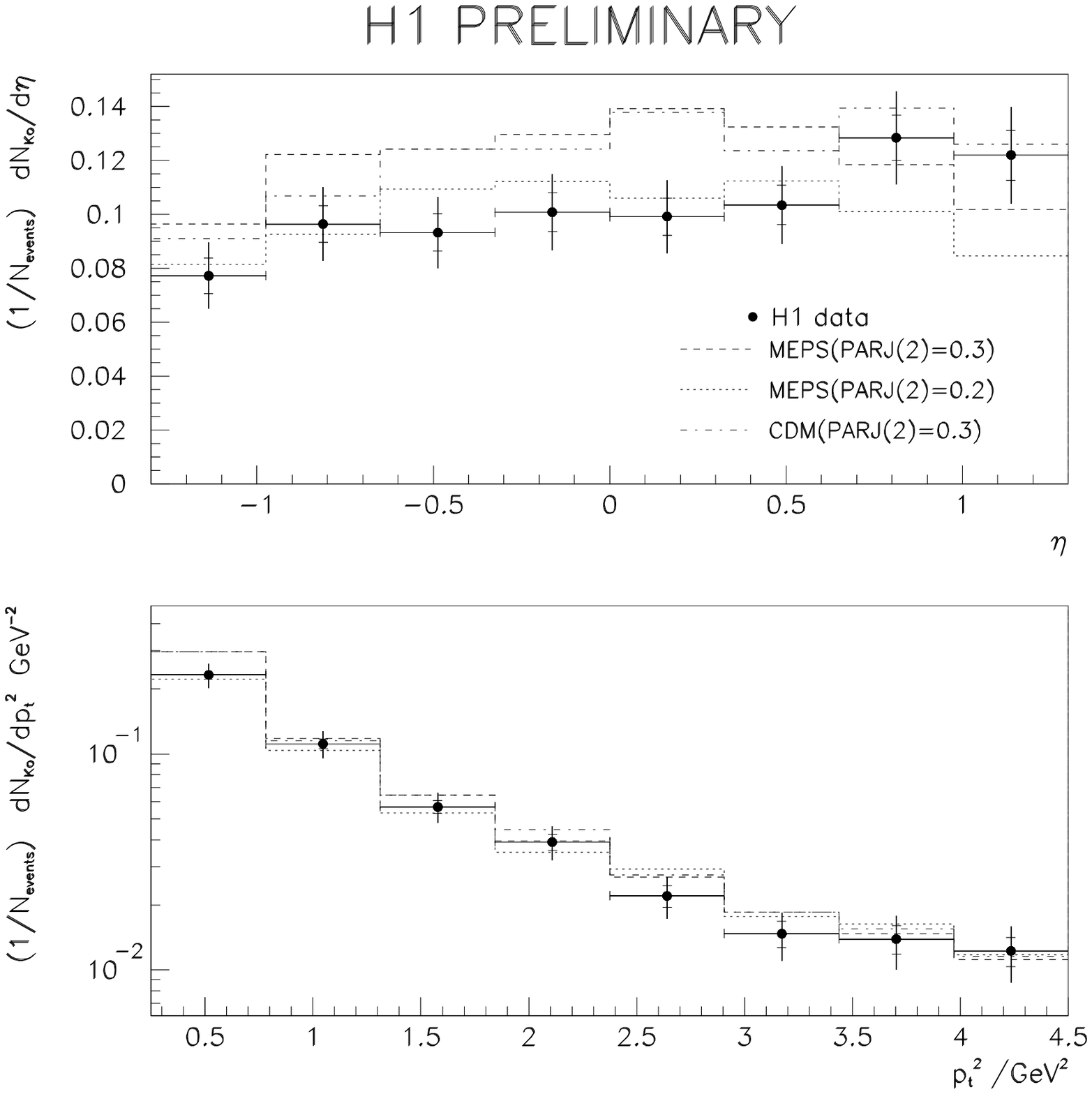,width=10.5cm,%
bbllx=0pt,bblly=10pt,bburx=560pt,bbury=510pt}
\caption{Corrected $K^0$ multiplicities in non-diffractive deep inelastic 
scattering
events as a function of $\eta$ (top) and $p^2_t$ (bottom) compared to the
predictions of the MEPS and CDM Monte Carlo program. The inner error bars 
are statistical and the outer error bars contain the full statistical and
systematical errors.}
\label{k0} 
\end{figure}

\subsection{Tests of QCD in the small $x$ regime}

In Section 2 it was shown that the 
the proton structure function $F_2$
exhibits a strong rise towards small Bjorken-$x$. 
Originally,
this rise caused much debate on whether the HERA data are still in 
a regime
where the QCD evolution of the parton densities can be described by the 
GLAP~\cite{GLAP} 
 evolution equations, or whether they extend into a new regime
where the QCD dynamics is 
described by the BFKL~\cite{BFKL}
evolution equation, but it was pointed out in Section 4 that 
present
 $F_2$ measurements 
do not yet 
allow us to distinguish between BFKL and conventional GLAP 
dynamics, and are perhaps
too inclusive a measure to be 
a sensitive discriminator.
Hadronic final states are expected to  give additional information and
could be more sensitive to the parton evolution.

For events at low $x$, hadron production in the 
region between the current jet and the 
proton remnant is expected to be 
sensitive to the effects of the BFKL or GLAP dynamics. 
At lowest order the BFKL and GLAP  evolution equations effectively resum the 
leading logarithmic $\alpha_S\ln1/x$ or
$\alpha_S\ln Q^2$ contributions respectively. 
In an axial gauge this
amounts to a resummation of ladder diagrams 
of the type shown in Fig.~\ref{HOTSPOT}.
This shows that before a  
quark is struck by the virtual photon,  
a cascade of partons may be emitted.
The fraction of the proton momentum carried by the emitted partons, 
$x_i$, and their transverse momenta,  
$k_{Ti}$,
are indicated in the figure.
In the leading $\log Q^2$ GLAP scheme 
this parton cascade follows a strong ordering in transverse
momentum 
$k_{Tn}^2 \gg k_{Tn-1}^2 \gg... \gg k_{T1}^2$, 
while there is only a soft 
(kinematical) ordering for the fractional momentum $x_n<x_{n-1}<...<x_1$.
In the BFKL scheme the cascade follows a strong ordering in fractional
momentum  
$x_n \ll x_{n-1} \ll... \ll x_1$, 
while there is no ordering in transverse
momentum.
 The transverse momentum follows a kind of random walk 
in $\log k_{T}$ space:
the value of $k_{Ti}$ is close to that of $k_{Ti-1}$, but it
can be both larger or smaller.

\begin{figure}[htb] \unitlength 1mm
\begin{center}
 \epsfig{file=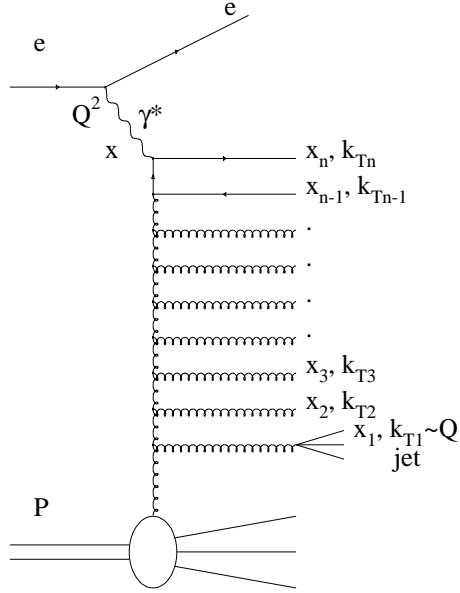,height=8cm}
\end{center}
\caption{Parton evolution in the ladder approximation. The
selection of DIS events containing a forward jet is illustrated.}
\label{HOTSPOT}
\end{figure}

Several measurements of the hadronic final state have been suggested to
exploit this difference at the parton level. The idea is to find observables 
which may reflect both of the BFKL characteristics of the unintegrated gluon 
distribution $f (x, k_T^2)$, that is the $x^{- \lambda}$ growth and the 
diffusion in $\log k_T^2$, as $x$ decreases.  Here we discuss those two 
measurements for which new information was provided during the Workshop: 
first, in section 5.4, we study the
transverse energy $(E_T)$ flow in the region away from the
current quark jet and second, in section 5.5, we investigate
the distribution of deep 
inelastic events containing an identified forward jet, that is, a measured 
jet as close as possible to, but distinct from, the proton remnants.

Apart from numerical calculations,
predictions for final state observables are also available 
as Monte Carlo models, based upon QCD phenomenology.
The CDM model description 
of gluon emission is similar to that of the BFKL evolution, 
because the gluons emitted by the dipoles
do not obey strong ordering in $k_T$ \cite{bfklcdm}. 
The CDM does not explicitly 
make use of the BFKL evolution equation, however.
The MEPS model is  based on GLAP dynamics; the emitted partons 
generated by the 
leading log parton showers are strongly ordered in $k_T$.

\subsection{Energy flow in the central region}

Due to the absence of $k_T$ ordering 
the BFKL approach is expected to give a larger
transverse energy, $E_T$, in the hadronic final state in the central
region of the hadronic centre-of-mass than the GLAP 
approach at low $x$.  This corresponds to the very forward region of
the detectors in the HERA laboratory frame. In this central region
of the hadronic center-of-mass
 Golec-Biernat \etal , \cite{Golec} show using the
GLAP approach that the partonic mean $E_T$,
$\langle E_T \rangle$, increases with increasing $x$, while for BFKL the
reverse is true, $\langle E_T \rangle$ decreases with increasing $x$. 
It should be noted  though that the $E_T$ spectra
are difficult to calculate in the GLAP framework 
because the $E_T$ weighting
emphasizes unsafe regions of phase space.

\begin{figure}[htbp]
\centerline{ \psfig{figure=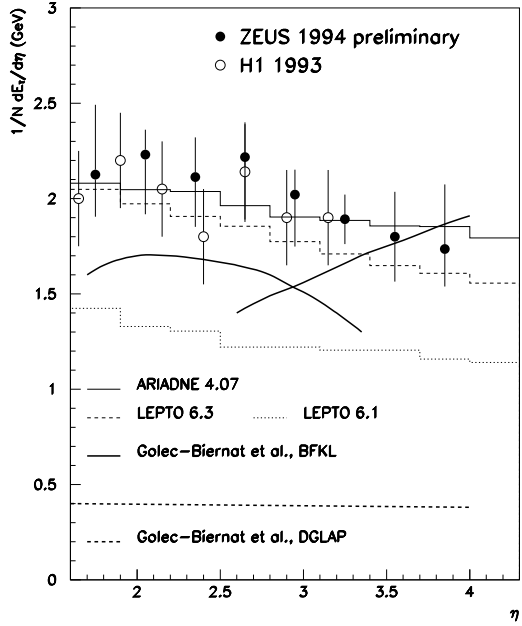,height=9cm} }
\caption{Mean transverse energy per unit rapidity
in the forward region of the HERA lab. system.}
\label{fig:Hforw}
\end{figure}

\begin{figure}[htbp]
\centerline{ \psfig{figure=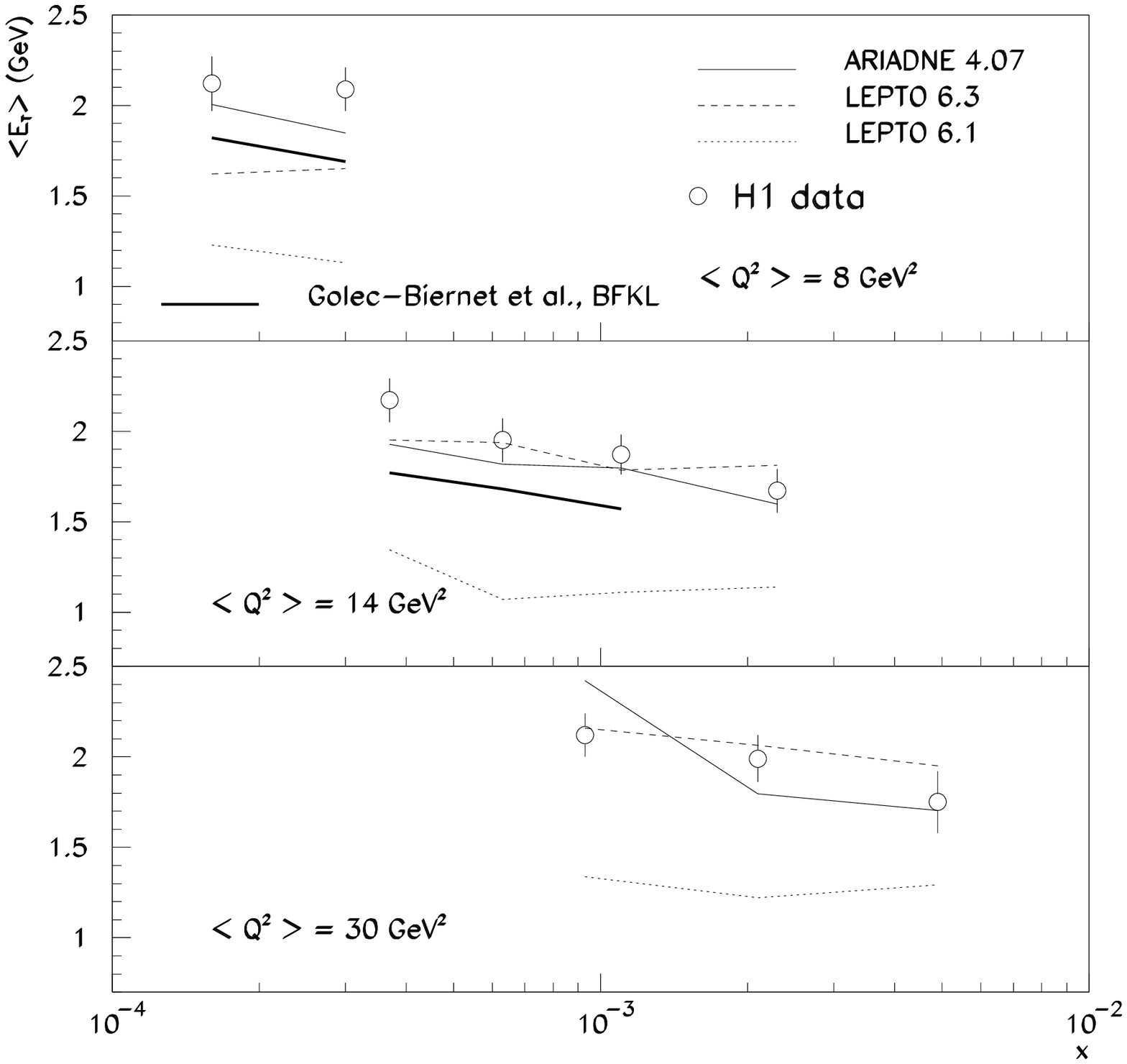,height=9cm}}
\caption{Transverse energy as a function
of $x$ for three different values of $Q^2$. The transverse energy is
measured in the CMS in the pseudorapidity range $-0.5$ to $+0.5$.}
\label{fig:Hetfncx}
\end{figure}

Fig.~\ref{fig:Hforw} shows the transverse energy flow
as a function of pseudorapidity, $\eta$,  in the laboratory
frame as measured by both H1~\cite{H1lowx} and ZEUS~\cite{Zlowx}. 
The level of $E_T$  is almost flat at $\approx 2$ GeV/(unit of
$\eta$). Also shown are the partonic calculations from \cite{Golec}
which  seem to indicate a preference towards the BFKL approach over
GLAP. Fig.~\ref{fig:Hetfncx} shows the mean transverse energy 
in the pseudorapidity range (as measured in the hadronic center-of-mass
system) 
$ -0.5 < \eta^* < 0.5$ by H1~\cite{H1lowx}. 
The data exhibit a rise in $\langle E_T \rangle$ with
decreasing $x$ as can also be seen from the BFKL partonic calculations 
shown in the same figure. To definitively comment
on whether these plots are signatures for BFKL it is necessary to
understand the underlying hadronisation.

In order to investigate hadronisation effects in this region 
two Monte Carlo generators have been studied: 
the CDM model (labeled ARIADNE 4.07 on the figure)
and LEPTO 6.1, both of which are based on the Lund string fragmentation
framework. 
The data shown in Figs.~\ref{fig:Hforw} and~\ref{fig:Hetfncx} are
reasonably described by CDM, whereas the overall $E_T$ predicted by
LEPTO 6.1 is far too low.
Unfortunately this
version of LEPTO is very sensitive to the cut-off applied to
avoid divergences in the matrix element --- the boson-gluon-fusion (BGF)
process, which is the dominant  ${\cal O}(\alpha_S)$ graph at low 
$Q^2$ and $x$,
creates two Lund strings as opposed to only one from the QCD Compton
process or a  ${\cal O}(\alpha_S^0)$ scattered quark. The effect of the
presence of strings  is an increased  amount of $E_T$ available
in the hadronisation phase. A newer version of the program, LEPTO 6.3, 
reduced this dependence on the matrix element cut-off by treating the scattering
off a sea quark in a manner similar to the 2 string scenario of the
BGF case. Using this version of the generator allows a better description
of the data, though there are
still some problems in describing the $x$ 
dependence in the lowest $Q^2$ region as shown in
 Figs.~\ref{fig:Hforw} and~\ref{fig:Hetfncx}.
However, the measurement seems to be rather sensitive to the non-perturbative
hadronisation phase and can, in its present form, no longer be considered 
as a direct sensitive test of BFKL dynamics.

\subsection{Deep inelastic events containing a forward jet}

Another possible signature of the BFKL dynamics is the behaviour of deep 
inelastic $(x, Q^2)$ events which contain a measured jet $(x_j, k_{Tj}^2)$ 
in the kinematic regime where $k_{Tj}^2 \simeq Q^2$ (so as to neutralise the 
ordinary gluon radiation which would have arisen from GLAP evolution) and 
where the jet has longitudinal momentum fraction $x_j$ as large as is 
experimental feasible $(x_j \sim 0.1)$. The aim is to observe events 
with $x/x_j$ 
as small as possible.  According to BFKL dynamics the differential structure 
function has a leading small $x/x_j$ behaviour of the form \cite{mueller}
$$
x_j \: \frac{\partial F_2}{\partial x_j \partial k_{Tj}^2} \; \sim \; 
\alpha_S (k_{Tj}^2) \: x_j \left [ g \: + \: \frac{4}{9} \: (q + \overline{q}) 
\right ] \; \left ( \frac{x}{x_j} \right )^{- \lambda}
$$
where the parton distributions are to be evaluated at $(x_j, k_{Tj}^2)$ --- 
where they are well known from global analyses.  The idea is to see if the 
DIS + forward jet data 
show evidence of the $(x/x_j)^{- \lambda}$ behaviour.
Jets are generally expected to be more robust against hadronisation effects
than is the $E_T$ flow.

We have studied DIS events at small $x$
which have a jet with large $x_j$. 
A cone algorithm is used to find jets, requiring an $E_T$ larger than 
5~GeV in a cone of radius 
$ R = \sqrt{\Delta\eta^2 + \Delta\phi^2} = 1.0$ in the space of
 pseudo-rapidity $\eta$ and azimuthal angle $\phi$
in the HERA frame of reference.
In order not to confuse the forward jet with the one at the top of the ladder   
the requirement $y > 0.1$ was imposed                                           
to ensure that the jet of the                                                   
struck quark is well within the central region of the detector                  
and is expected to have a jet angle larger than $ 60^0$.                       
Experimentally                                                                 
a cone algorithm is used to find jets, requiring an $E_T$ larger than          
5~GeV in a cone of radius                                                      
$ R = \sqrt{\Delta\eta^2 + \Delta\phi^2} = 1.0$ in the space of                
 pseudo-rapidity $\eta$ and azimuthal angle $\phi$                             
in the HERA frame of reference.                                                
Jets are accepted as forward jets if                                           
$x_{j}>0.025$, $0.5< k_{Tj}^2/Q^2<4$,                                       
$6^0 < \theta_{j}< 20^0$ and $k_{Tj}> 5 $ GeV,                              
where $\theta_{j}$ is the forward jet angle and                                   
$k_{Tj}$ is the transverse momentum of the jet.                                                                      
                                                                               
These selection criteria allow a study the cross section of forward           
jet production in the region                                                   
$Q^2 \approx$ 20 GeV$^2$ and                                               
$2\!\cdot\!10^{-4}\!<\!x\!<\!2\!\cdot\!10^{-3}$.                               
Hence the ratio $x_{j}/x$ is always larger than 10.                     

\begin{table}[t]
\begin{center}
\begin{tabular}{|c|c|c|c|c|} \hline
$x$ range & data & MEPS &CDM & $\sigma(ep\rightarrow$ jet$+X)$ \\
        &  events    & events & events & (pb) \\ \hline 
$2\cdot 10^{-4} - 1\cdot 10^{-3}$ &$271$ & 141 & 282 &
$709\pm42\pm166$ \\ \hline
$1\cdot 10^{-3} - 2\cdot 10^{-3}$ &$ 158 $ & 101 & 108 & 
$475 \pm 39 \pm 110$ \\ \hline
\end{tabular}
\end{center}
\caption[]
{\label{HOTTABLE} 
{\footnotesize
Numbers of observed DIS events with a selected forward jet,
corrected for radiative events faking this signature.
These may be directly compared with the
expectations from the Monte Carlo models.
The measured cross section $ep\rightarrow$ jet$+X$
for forward jets is also given.
The errors reflect the statistical and systematic uncertainties.}}
\end{table} 

The resulting number of 
events observed with at least one forward jet in the kinematical
region  $160^0<\theta_e<173^0$ and $E_e>12$ GeV
is given in Table~\ref{HOTTABLE} and compared to 
expectations of the MEPS and CDM models after detector simulation
and corrected for  background.
The measured cross section for forward jets
satisfying the cuts given above is also presented in
Table~\ref{HOTTABLE}. 
It has been corrected for detector effects using the CDM.
The ratio of the jet cross section for the low $x$ to the  high $x$ bin 
is $1.49\pm 0.25$.

\begin{figure}[htb] \unitlength 1mm                                            
\begin{center}                                                                
 \epsfig{file=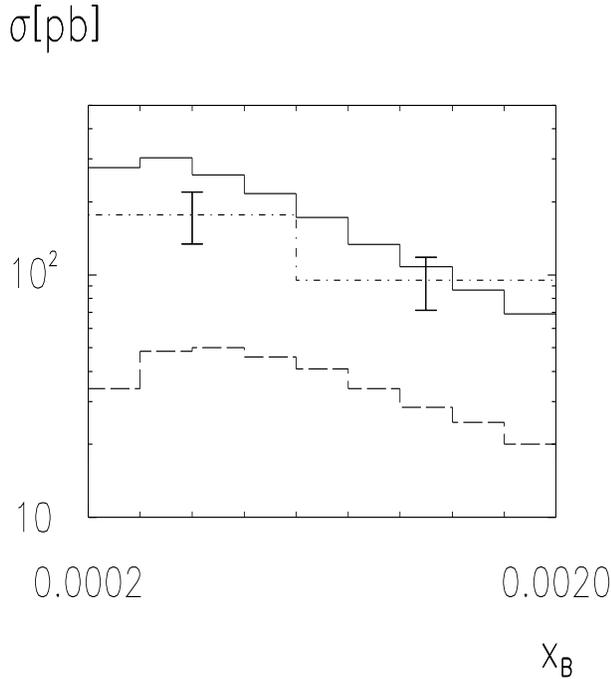,width=8cm}
\end{center}                                                                   
\caption{Comparison of the BFKL calculation and the approximate analytical
calculation of the three-parton matrix elements,
compared with the data of the H1 experiment.}
\label{fjets}                                                                  
\end{figure}                                                                   

In Fig.~\ref{fjets} the data are compared with a recent
calculation~\cite{bartels}. The calculation 
(at the parton level) has exactly the same cuts as the measurement.
The solid line is a BKFL calculation, while the dashed line is an
approximate analytical
three-parton matrix element calculation    
without the BFKL ladder. The results of H1 are compared with 
the calculations. The   BFKL curve agrees well with the data,    
while the lower curve is in clear disagreement.
Several corrections to this result should be considered, as discussed
in~\cite{bartels}( i.e. the
 calculation is at the parton level while the data are at the         
hadron level). Nevertheless the agreement with data is encouraging and
such measurements should be explored further with larger statistics 
event samples 
as they have a large potential to reveal BFKL effects in the data. 

A final general comment is in order.
The smaller the value of $x/x_j$ the larger is the BFKL effect and the more 
dominant is the leading $\log (1/x)$ formalism.  As with all BFKL predictions, 
the reliability can only be quantified when the sub-leading corrections are 
known.  Moreover by measuring properties of the final state we inevitably 
reduce the \lq\lq reach" of HERA.  For example in the present case we require 
$x/x_j$ to be as small as possible, yet experimentally jet recognition demands 
$x_j \lapproxeq 0.1$, and so we lose an order of magnitude in ``reaching small
$x$".

\setcounter{section}{5}
\setcounter{equation}{0}
\renewcommand{\theequation}{6.\arabic{equation}}
\setcounter{figure}{0}
\renewcommand{\thefigure}{6.\arabic{figure}}
\setcounter{table}{0}
\renewcommand{\thetable}{6.\arabic{table}}

\section{Charm production}

As we have seen above the gluon, the dominant parton at small $x$, has 
the least well determined distribution.  Due to the theoretical uncertainties 
it is difficult to make unambiguous determinations of the gluon from the 
scaling violations that are observed in the HERA measurements of $F_2$.  The 
longitudinal structure function $F_L$ gives, in principle, a much more direct 
determination of the gluon.  In practice a sufficiently accurate measurement of 
$F_L$, which requires 
different beam energies, is exceedingly difficult.  Here we note that $F_2$ 
itself contains an appreciable part which is directly sensitive to the gluon 
density, namely the contribution arising from charm production $F_2^c$.

\begin{figure}[htb]
\begin{center}
 \epsfig{file=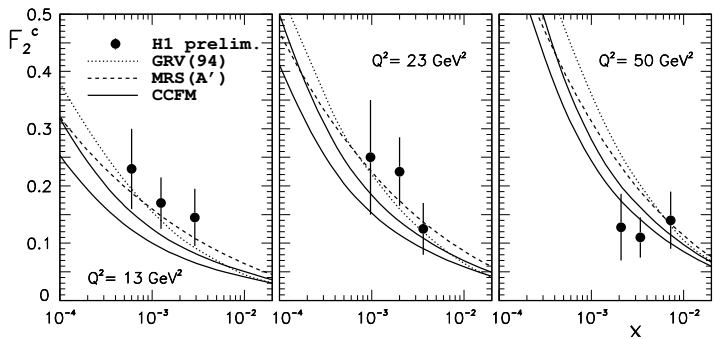,width=10cm}
\end{center}
\caption{Predictions for $F_2^c$ compared with preliminary H1 
\protect{\cite{H1CHARM}} data which became available after the Workshop.  
The curves show the predictions from 
MRS(A$^\prime$) partons \protect{\cite{MRS}} and from \protect{\cite{V}} 
using GRV partons (described in section 6.1), together with the CCFM 
prediction for $m_c = 1.4$ and 1.7 of ref.\ \protect{\cite{kms1}}.  The 
figure is adapted from ref.\ \protect{\cite{kms1}}.}
\end{figure}

At this Workshop both Vogt and Sutton presented predictions for $F_2^c$.  
The predictions are summarized in Fig.\ 6.1, which also contains preliminary 
measurements of $F_2^c$ that became available after the Workshop \cite{H1CHARM}.  
The figure shows two calculations based on GLAP-evolved partons, (GRV, MRS), 
together with a 
prediction obtained using the unintegrated gluon distribution obtained by 
solving the CCFM equation (see section 4.4).  All three calculations are 
quite different.  MRS treat charm as a parton and set $c (x, Q^2) = 0$ below 
some threshold $Q^2 < m^2$ and then use NLO GLAP evolution with $m_c = 0$.  The 
calculation \cite{V} using GRV partons 
treats charm as a heavy quark, not as a parton, and uses fixed-order (NLO) 
perturbation theory.  We discuss this approach in detail below.  Finally the 
CCFM predictions \cite{kms1} are obtained from the unintegrated gluon 
distribution using the $k_T$-factorization theorem.  They sum the LO $\log 
(1/x)$ contributions, but with the angular ordering constraints imposed.

\subsection{Fixed-order prediction for $F_2^c$}

At this Workshop, Riemersma presented the NLO QCD formalism for the 
$c\overline{c}$ contribution to $F_2$.  It is given by 
\cite{LRSN}

\begin{eqnarray}\fl
F_{2}^{\, c}(x,Q^2) = \frac{Q^2 \alpha_S(\mu ^2)}{4 \pi^2 m_{c}^{2}} 
\int_{ax}^{1} \! \frac{dy}{y} \, yg(y,\mu^2) \, e_{c}^{2} \left\{ c_{2,g}^{\,
(0)} + 4 \pi \alpha_S(\mu ^2) \left[ c_{2,g}^{\, (1)} + \bar{c}_{2,g}^{\, (1)} 
\ln \frac{\mu^2}{m_{c}^{2}} \right] \right\} \nonumber \\
\nonumber \\ 
+ \frac{Q^2 \alpha_{S}^{2}(\mu ^2)}{\pi m_c^{2}} \sum_{i=1}^{2n_f} 
\int_{ax}^{1} \! \frac{dy}{y} \, yq_{i}(y,\mu^2) \left[c_{2,i}^{\, (1)} + 
\bar{c}_{2,i}^{\, (1)} \ln \frac{\mu^2}{m_{c}^{2}} \right] 
\label{6.1}
\end{eqnarray}
\noindent where here the sum is over the light quark flavours $(n_f = 3)$.  The 
($\overline{\mbox{MS}}$) mass factorization scale $\mu $ has been put equal 
to the renormalization scale entering the strong coupling $\alpha_S$. The 
lower limit of integration over the fractional initial-parton momentum $y$ is
given by $y_{\rm min} = ax = (1 + 4m_{c}^{2}/Q^2) x $, corresponding to
the threshold $\hat{s} = 4m_{c}^{2} $ of the partonic center-of-mass
(c.m.) energy squared.  The coefficient functions $c_2$ can be 
expressed \cite{LRSN} in terms of the dimensionless variables 
\begin{equation}
  \xi  = \frac{Q^2}{m_{c}^{2}} \:\:\: , \:\:\:\:\: 
  \eta = \frac{\hat{s}}{4m_{c}^{2}} - 1 
       = \frac{\xi}{4} \left( \frac{y}{x} -1 \right) -1 \:\:\: .
\label{6.2}
\end{equation}

In leading order, ${\cal O}(\alpha_S)$, $F_{2}^{\, c}$
is directly sensitive only to the gluon density $g(y,\mu^2)$ via the well-known 
Bethe-Heitler process $\gamma^{\ast}g \rightarrow c\bar{c}$ \cite{RSN}.  
A comparison of the various contributions to $F_{2}^{\, c}$ in NLO shows that 
for the physically reasonable scales $\mu$, $\mu \simeq 2 m_c, \ldots
\sqrt{Q^2 + 4 m_{c}^{2}}$ (see below), the quark contribution in (\ref{6.1}) 
--- which is not necessarily positive due to mass factorization --- is very
small, typically about 5\% or less.  Therefore $F_{2}^{\, c}$ does
represent a clean gluonic observable also in NLO.

Before considering the phenomenological consequences of (\ref{6.1}), it
is instructive to recall the $\eta$-dependence of the gluonic
coefficient functions \cite{LRSN,BH}. They are displayed in Fig. 6.2 for two 
values of $Q^2$ typical for deep-inelastic small-$x$ studies at HERA, 
$Q^2 = 10, \: 100$ GeV$^2$. Here and in the following we take $m_c = 1.5$ 
GeV. The comparison of the NLO coefficients $c^{(1)}$ and $\bar{c}^{\, (1)}$ 
with the Bethe-Heitler result $c^{(0)}$ reveals that potentially large 
corrections arise from regions where $c^{(0)}$ is small, namely 
from very small and very large partonic c.m.\ energies.  These corrections 
are due to initial-state-gluon bremsstrahlung and the Coulomb singularity at 
small $\hat{s}$ (small $\eta $), and due to the flavour excitation 
process at $\hat{s} \gg 4m_{c}^{2}$ ($\eta \gg 1$). The large logarithms $\ln
[\hat{s}/(4m_{c}^{2})]$ originating from flavour excitation have been 
resummed --- at the expense of losing the full small-$\hat{s}$ information 
of (\ref{6.1}) --- by introducing a charm parton density, leading to the 
so-called variable-flavour scheme \cite{VFS}. The importance of these 
corrections in the HERA small-$x$ regime considered here will be investigated 
below.

The first question to be addressed in order to judge the
phenomenological usefulness of $F_{2}^{\, c}$ as a gluon constraint is
whether or not the available NLO expression (\ref{6.1}) is sufficient for 
obtaining results which are stable under variation of the (unphysical) mass
factorization scale. It has been argued in \cite{GRS} that one should use $\mu
\simeq 2m_c $. This is motivated by the conjecture that $\mu $ is
controlled by $\hat{s}$ and the fact that the integrand in (\ref{6.1}) is 
maximal close to the lower limit, $ \hat{s} \simeq 4 m_{c}^{2} $. The range of 
significant contributions in $\hat{s}$, however, broadens considerably 
with increasing $Q^2$, as we shall see below, hence $ \mu \simeq
\sqrt{Q^2 + 4m_{c}^{2}} $, chosen in \cite{LRSN,RSN}, appears at 
least equally reasonable. Therefore, we will estimate the
theoretical uncertainty of $F_{2}^{\, c}$ in NLO by varying the scale between 
$\mu = m_c$ and $ \mu = 2 \sqrt{Q^2 + 4m_{c}^{2}} $. 
The corresponding results for $F_{2}^{\, c}$ are shown in Fig.\ 6.3
for some fixed values of $x$ and $Q^2$, using the parton densities of
GRV \cite{GRV}, which are appropriate here as they treat charm as a massive 
quark and not as a parton. At small-$x$, $x < 10^{-2}$, the scale variation
amounts to at most $\pm 10\% $. 
\begin{figure}[ht]
\begin{center}
 \epsfig{file=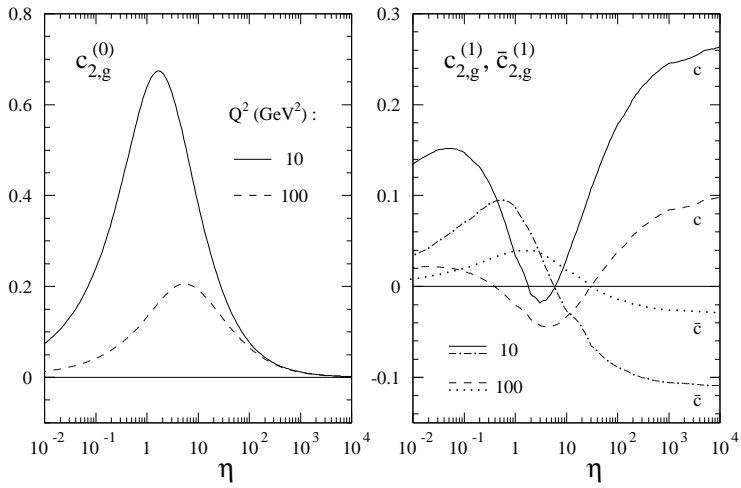,width=10cm}
\end{center}
\caption{The $\eta $-dependence of the coefficient functions
as parametrized in \protect{\cite{RSN}} for two values of $Q^2$ with 
$m_c = 1.5$ GeV.  The scheme-dependent quantity $c_{2,g}^{\, (1)}$ is 
given in the $\overline{\mbox{MS}}$ scheme.}
\label{fig6.2}
\end{figure}
\begin{figure}[hb]
\begin{center}
 \epsfig{file=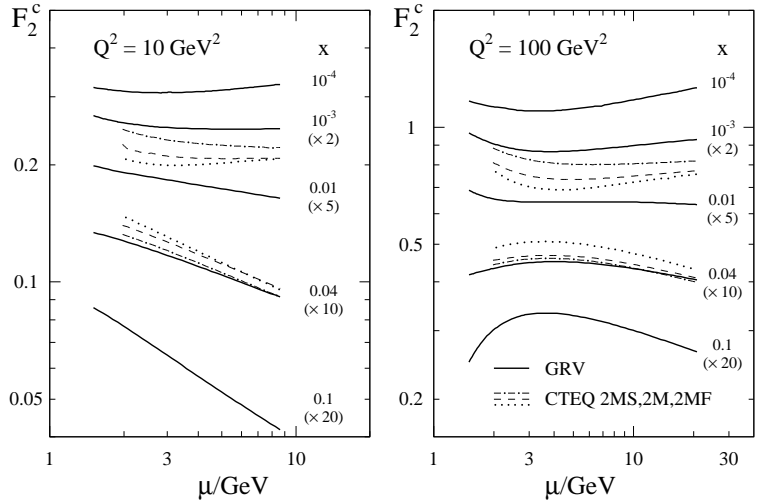,width=10cm}
\end{center}
\caption{The dependence of $F_2^c$ on the mass factorization scale
$\mu$ in the region \protect{$m_c \leq \mu \leq 2 (Q^2 + 4
m_c^2)^{0.5}$} at selected values  
of $x$ and $Q^2$.
The parton densities of GRV \protect{\cite{GRV}} (at all 
$x$) and CTEQ2 \protect{\cite{CTEQ2}} (at $x = 10^{-3}, 0.04$) have been 
employed.}
\label{fig6.3}
\end{figure}
Also displayed in Fig.\ 6.3 are the results
for the three CTEQ2 parton sets \cite{CTEQ2} at two selected values of
$x$.  Obviously the scale stability at small $x$ does not significantly
depend on the steepness of the gluon distribution. Consequently,
the NLO results of \cite{LRSN,RSN} seem to provide rather sound a
theoretical foundation for a small-$x$ gluon determination at
HERA, despite the large total c.m.\ energy $ s \gg 4m_{c}^{2} $ which
might suggest a destabilizing importance of $ \ln [\hat{s}/(4m_{c}^{2})] $ 
terms.  At large $x$, $ x \approx 0.1 $, on the other hand, the scale
dependence of $F_{2}^{\, c}$ is rather strong, especially at low $Q^2$.
Here, quantitatively reliable results require further theoretical input
by resumming the large small-$ \eta $ threshold contributions
mentioned above. However, $F_{2}^{\, c}$ is small in this region.

The next issue we investigate is the locality (in the momentum
fraction $y$) of the gluon determination via $F_{2}^{\, c}$. For this
purpose, we replace the upper limit in (\ref{6.1}) by a varying maximal gluon
momentum $y_{\rm max} < 1$. The contribution from initial-parton momenta 
smaller than $y_{\rm max}$ to $F_{2}^{\, c}(x,Q^2)$, denoted by $F_{2}
^{\, c}(x,Q^2, y_{\rm max}) $, 
\begin{figure}[ht]
\begin{center}
 \epsfig{file=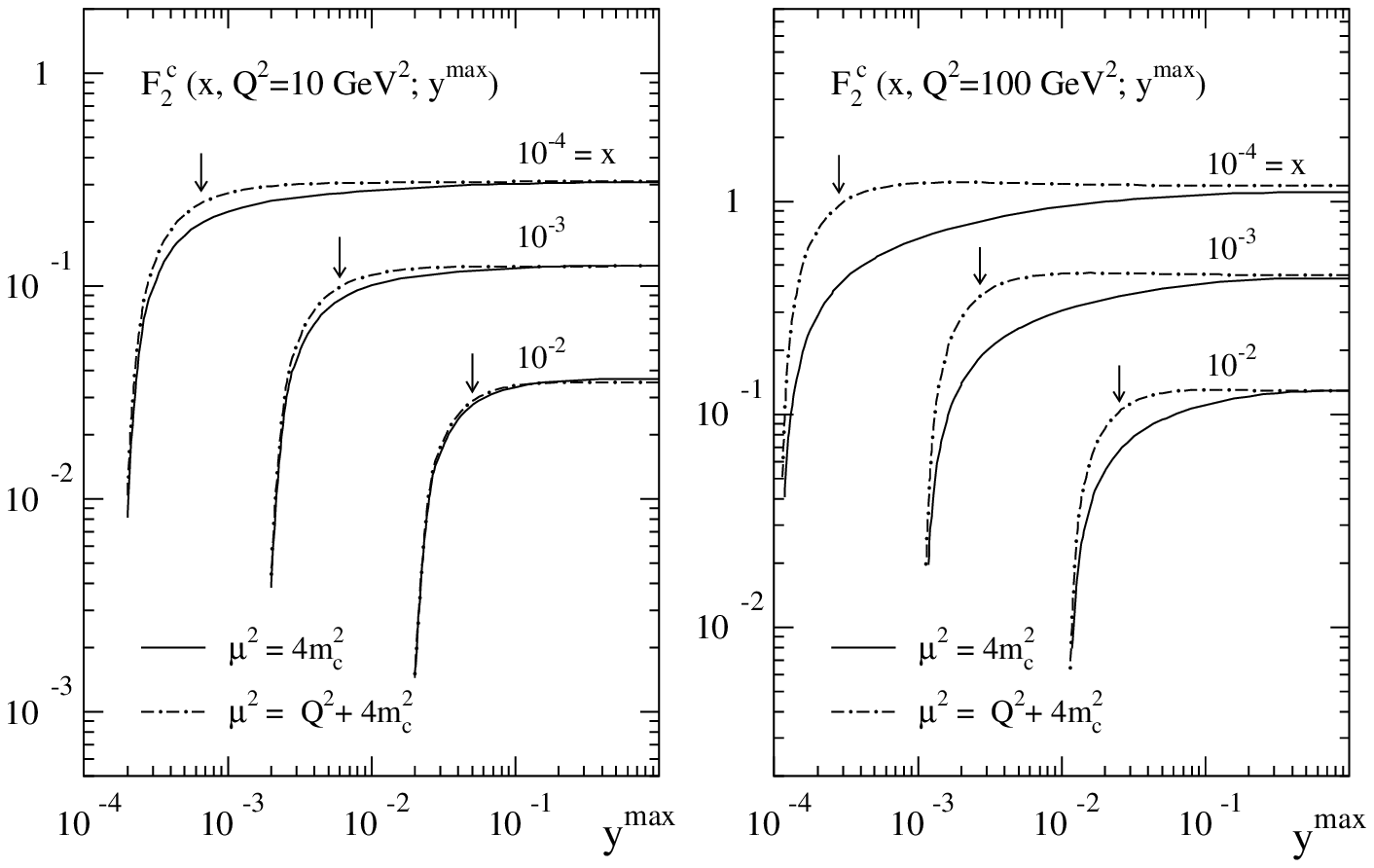,width=10cm}
\end{center}
\caption{The contribution of the initial-parton momentum
region $ax \leq y \leq y^{\rm max}$ to $F_{2}^{c}$ at small-$x$
for two choices of the scale $\mu$, using the parton densities of
\protect{\cite{GRV}}. The arrows indicate the values of $y^{\rm max}$
at which 80\% of the complete results are reached for $\mu = (Q^2 + 4 
m_{c}^{2})^{0.5}$.}
\end{figure}
\begin{figure}[hb]
\begin{center}
 \epsfig{file=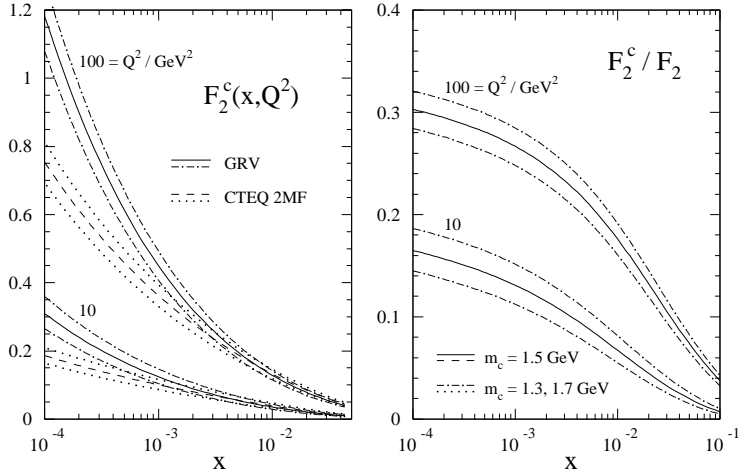,width=10cm}
\end{center}
\caption{The $x$-dependence of $F_{2}^{c}$ and $F_{2}^{\, c}/F_2$ at some 
fixed values of $Q^2$, as expected from the GRV gluon density 
\protect{\cite{GRV}}. Also shown are $F_{2}^{c}$ as obtained from the CTEQ 2MF 
parton set \protect{\cite{CTEQ2}} and the charm mass dependence of the 
predictions. $\mu = (Q^2 + 4 m_{c}^{2})^{0.5}$ has been employed.}
\end{figure}
is presented in Fig.\ 6.4 for the GRV parton
distributions \cite{GRV}. At scales $ \mu \approx \sqrt{Q^2 + 4m_{c}^{2}}$, 
about 80\% of $F_{2}^{\, c}$ originates in the region $y_{\rm min}\! =\! ax 
\leq y \: \lapproxeq \: 3 y_{\rm min} $.  Again the situation is 
very similar for the CTEQ2 parton densities \cite{CTEQ2}. Thus
$F_{2}^{\, c}$ allows for rather local a determination of $g(y,\mu^2
\approx \sqrt{Q^2 + 4m_{c} ^{2}})$.  The partonic c.m.\ energies in the 
region contributing 80\% to the structure function $F_{2}^{\, c}$ are given 
by $\eta \, \lapproxeq \, 5$ (20), corresponding to $\hat{s} \, \lapproxeq \, 
60\, (180) $ GeV$^2$, at $Q^2 = 10\, (100)$ GeV$^2$, respectively. This 
implies that for the $Q^2$ values under consideration here, the plateau 
region of $c_{2,g}^{\, (1)}$ and $\bar{c}_{2,g}^{\, (1)}$ at large $\eta$ 
(c.f.\ Fig.\ 6.2) does not play any important role. Resummation of 
large-$\eta $ flavour excitation logarithms is thus neither necessary nor 
appropriate, 
especially if it implies additional approximations in the more important
small-$\hat{s}$ region \cite{VFS}. A similar observation has already been 
made in \cite{GRS} for $\mu \simeq 2m_{c}$ at low $Q^2$. Note, however,
that the latter scale choice leads to a considerably wider important
range of $\hat{s}$ at high $Q^2$, somewhat in contrast to its
motivation described above.

The expected $x$-dependence of $F_{2}^{\, c}$ and its relative 
contribution to the total proton structure function $F_{2}$ are 
displayed in Fig.\ 6.5.  $F_{2}^{\, c}$ is large in the HERA
small-$x$ region (see Fig.\ 6.1).  The absolute magnitude amounts to about 
$0.2 \ldots 0.4$, i.e.\ $F_{2}^{\, c}$ is as large here as the total $F_2$ 
in the BCDMS/NMC fixed-target kinematical regime, making up up to a
quarter of $F_{2}$ measured at HERA. This is in contrast to
the bottom contribution $F_{2}^{\, b}$, which reaches at most $ 2
\ldots 3\% $.  The size of $F_{2}^{\, c}$ renders a reliable (fully
massive) treatment mandatory in any precise analysis of $F_2$ at small
$x$.  The sensitivity of $F_{2}^{\, c}$ to the gluon density is
illustrated by the difference of the CTEQ$\, $2MF (flat $xg(x,\mu ^2 = 2.6 $
GeV$^2$) \cite{CTEQ2} and the GRV (steep gluon) \cite{GRV} expectations.
There is quite some discriminative power of $F_{2}^{\, c}$, especially
close to the lower end of the $Q^2$ range considered here, $Q^2 \approx
10 $ GeV$^2$. Moreover, by measuring up to about 100 GeV$^2$ in $Q^2$,
the rapid growth of $yg(y \ll 1,\mu^2)$ predicted by the GLAP 
equations can be rather directly tested down to  $ y \simeq 10^{-3} $.
A theoretical obstacle to an easy accurate gluon determination
via the charm structure function is the dependence of $F_{2}^{\, c}$ on
the unknown precise value of the charm quark mass $m_c$. A 10\%
variation of $m_c$ to 1.35 GeV and 1.65 GeV as also considered in Fig.~4 leads
to a $ \pm 15 \ldots 25\% \, (5 \ldots 10\%)$ effect at $Q^2 = 10 \,
(100)$ GeV$^2$, respectively, with respect to the central curves.

\subsection{Outlook for $F_2^c$}

The measurement of $F_2^c$ at HERA should serve as a sensitive probe of the 
gluon at small $x$.  In section 6.1 we have seen that the fixed-order QCD 
predictions are stable, with scale variations of less than $\pm 10$\%, and offer 
a rather local measurement of the gluon.  Flavour 
excitation contributions from $\hat{s} \gg 4 m_c^2$ become important only at 
scales $Q^2$ higher than those relevant for the small $x$ observations at 
HERA.

Investigations of the effects of $\log (1/x)$ resummations have started (see 
Fig.\ 6.1 and also \cite{akms,kms1}) but as yet the theoretical framework is 
much more incomplete than the 
fixed-order approach.  Ultimately these effects may have to be incorporated 
in a detailed analysis of future precision data for $F_2^c$.

\subsection{$J/\psi$ production as a probe of the gluon}

It has long been advocated that inelastic $J/\psi$ photoproduction at HERA may 
serve as a measure of the gluon --- see, for example, \cite{MNS} or the review 
\cite{JST} which consider the process at leading-order accuracy.  Recently the 
higher-order corrections to this process have been calculated \cite{KZSZ,KRA}.  
A detailed analysis of the spectra in the high energy range at HERA shows that 
the perturbative calculation is not well-behaved in the limit $p_T \rightarrow 
0$, where $p_T$ is the transverse momentum of the $J/\psi$.  
\begin{figure}[ht]
\begin{center}
 \epsfig{file=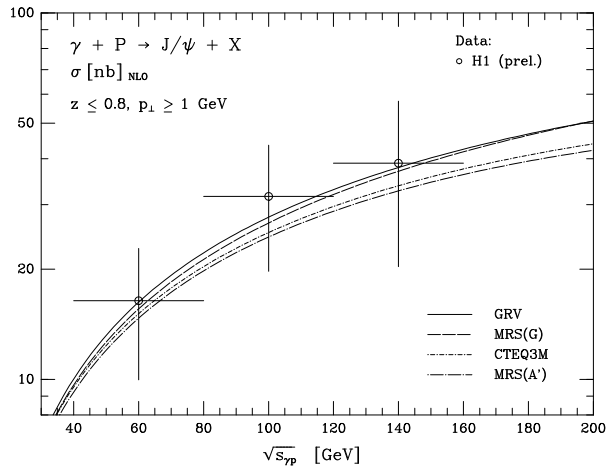,width=8cm,angle=-90}
\end{center}
\caption{Total cross section for inelastic $J/\psi$ photoproduction as a 
function 
of the photon-proton energy for different parametrizations of the parton 
distribution 
in the proton.  Experimental data from \protect{\cite{AID}}.  The figure is 
from \protect{\cite{KRA}}.}
\end{figure}
\begin{figure}[htb]
\begin{center}
 \epsfig{file=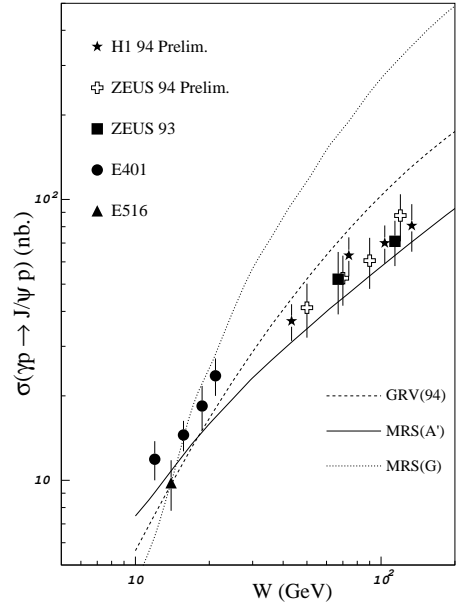,width=6cm}
\end{center}
\caption{The measurements of the cross section for diffractive $J/\psi$
photoproduction compared with the full perturbative QCD prediction obtained
from the three latest sets of partons. The figure is taken from
\protect{\cite{RRML}}.}
\label{fig6.7}
\end{figure}
No reliable prediction 
can be made in this singular boundary region without resummation of large 
logarithmic corrections caused by multiple gluon emission.  If the small $p_T$ 
region is excluded from the analysis, the next-to-leading order result accounts 
for the energy dependence of the cross section and for the overall 
normalization, 
see Fig.\ 6.6 \cite{KRA}.  However, since the average momentum fraction of the 
partons is shifted to larger values when excluding the small-$p_\perp$ region, 
the sensitivity of the prediction to the small-$x$ behaviour of the gluon 
distribution is not very distinctive.

Diffractive $J/\psi$ photoproduction appears to offer a more promising way to 
distinguish between the gluon distributions.  Since this is essentially an 
elastic process the cross section is a measure of the square of the gluon 
density.  To leading order the cross section is given by \cite{RYS,BFGMS}
\begin{equation}
\left . \frac{d \sigma}{d t} \: (\gamma^* p \rightarrow J/\psi p) \right |_0 \; 
= \; \frac{\Gamma_{ee} M_\psi^3 \pi^3}{48 \alpha} \: \frac{\alpha_S 
(\overline{Q}^2)^2}
{\overline{Q}^8} \; [xg (x, \overline{Q}^2) ]^{2}
\label{6.3}
\end{equation}
with $\overline{Q}^2 = \frac{1}{4} M_\psi^2$ and $x = M_\psi^2/W^2$, where $W$ 
is the $\gamma p$ c.m.\ energy.  In a study which originated at the Workshop 
\cite{RRML}, corrections to this formula have been calculated and comparisons 
with HERA data made, see Fig.\ 6.7.  It was emphasized that the $W$ dependence, 
rather than the normalisation, was the more reliable discriminator between the 
gluons.  The power of the method is evident from Fig.\ 6.7, which appears to 
favour the MRS(A$^\prime$) gluon.

\setcounter{section}{6}
\setcounter{equation}{0}
\renewcommand{\theequation}{7.\arabic{equation}}
\setcounter{figure}{0}
\renewcommand{\thefigure}{7.\arabic{figure}}
\setcounter{table}{0}
\renewcommand{\thetable}{7.\arabic{table}}

\section{Spin physics}
Interest in spin phenomena in deep inelastic scattering
revived in the eighties after the European
Muon Collaboration discovered that the spin dependent structure function
of the proton violates the Ellis-Jaffe sum rule and that the quarks probably
carry only a small part of the total proton spin. The problem of the
origin of the proton spin has led to an intense experimental and
theoretical activity but in spite of that it has not yet been answered
conclusively. 
Here we briefly review the status of
the spin effects in deep inelastic scattering; experimental results
and their theoretical 
interpretations. Finally its future prospects will be outlined. 

\subsection{Cross section asymmetries and sum rules}
The deep inelastic lepton-nucleon scattering
cross section is the sum
of a spin independent term $\overline{\sigma}$
and a term proportional to the lepton helicity $h_l$:
\begin{equation}
        \sigma = \overline{\sigma} + \frac{1}{2}h_l  \Delta\sigma.
\end{equation}

Only longitudinally polarised leptons will be considered and the spin
vector $s_l$ is thus related to the lepton four momentum vector $k$.
$\Delta \sigma$ gives
only a small contribution to the total deep inelastic cross section. It
depends on the two structure functions $g_1$ and $g_2$ and can be
expressed as
\begin{equation}
\label{delta_sigma}
        \Delta \sigma=\rm cos\psi\,\Delta \sigma_{\parallel}
        + \rm sin\psi\,\rm cos\phi\,\Delta \sigma_{\perp},
\end{equation}
with
\begin{eqnarray}
        \frac{{\rm d}^2\Delta\sigma_{\parallel}}{{\rm d}x{\rm d}Q^2}
      & = & \frac{16\pi\alpha^2 y}{Q^4} \Biggl[(1 -{y\over 2}
        -{\gamma^2 y^2\over 4})g_1 -{\gamma^2 y\over 2} g_2 \Biggr],
\nonumber\\
        \frac{{\rm d}^3\Delta\sigma_{T}}{{\rm d}x{\rm d}Q^2{\rm d}\phi}
      & = & -\cos \phi
           \,\frac{8\alpha^2 y}{Q^4}\,\gamma\,\sqrt{1-y-{\gamma^2y^2 
           \over 4}} \Biggl({y \over 2} g_1 + g_2 \Biggr).
\end{eqnarray}
In the above expressions, $\psi$ denotes the angle between the lepton
and the nucleon spin and $\phi$ the angle between the scattering plane and
the spin plane; furthermore $\Delta \sigma_{\perp}= \Delta
\sigma_{T}/\cos\phi$. $y=\nu/E$ and $\gamma=2Mx/\sqrt{Q^2}$ are the
usual kinematical factors. 

In experimental measurements, two asymmetries can be defined:
\begin{equation}
\label{A}
A_{\parallel} = 
{\Delta\sigma_{\parallel} \over 2 \overline{\sigma}}
\hspace{1cm}
{\rm and}
\hspace{1cm}
A_{\perp} = {\Delta\sigma_{\perp} \over 2 \overline{\sigma}}.
\end{equation}
These asymmetries are directly related to the virtual photon asymmetries, $A_1$
and $A_2$:
\begin{equation}
A_{\parallel} = D (A_1 + \eta A_2),  \hspace{1cm} A_{\perp}  =  D (A_2 -\xi
A_1),
\end{equation}
where
\begin{equation}
\label{a1a2}
A_1={g_1-\gamma^2g_2\over F_1}, ~~~~~~~~~~~A_2=\gamma{g_1+g_2\over F_1}.
\end{equation}
$D$, often called the depolarisation factor of the virtual photon,
depends on $y$ and the structure function $R=F_L/F_T$; factors $\eta$ and $\xi$
depend only on kinematic variables. $A_1$ and $A_2$ 
are often interpreted as virtual photon -- nucleon asymmetries. They
satisfy the bounds $|A_1|\leq 1$, $|A_2|\leq \sqrt R$. 

Within the QPM the spin dependent structure function $g_1$ is
given by
\begin{equation}
   g_{1}(x) = \frac{1}{2}{\displaystyle\sum_{i=1}^{n_f}} e_{i}^2
                        [ \Delta q_{i}(x) + \Delta \bar q_{i}(x) ],
\label{g1}
\end{equation}
with $\Delta q_{i}(x) = q_i^+(x) - q_i^-(x)$, where $q^{\pm}$ are the
distribution functions of quarks with spin parallel (antiparallel) to
the nucleon spin. 
Less obvious is the meaning of $g_2$
which contains a leading twist part, completely determined by $g_1$ and a
higher twist part, the meaning of which is subject to debate~\cite{jaf_g2}.
In QCD, $g_1$ evolves according to Altarelli--Parisi
equations, similar to the unpolarised ones. Corresponding coefficient-
and splitting functions have recently been calculated up to order 
$\alpha_S^2$ \cite{nlocorr}, permitting the next-to-leading order QCD
analysis of $g_1$ and thus a determination of the
polarised parton distributions, $\Delta q_i(x,Q^2)$.
Various groups have used the recent data to determine these
distributions, taking into account the
leading order QCD corrections\footnote{Several
analyses \protect{\cite{nlopart}} incorporating NLO corrections
\protect{\cite{nlocorr}} have appeared after this Workshop.}.
A comparison of different parametrizations 
\cite{compare} shows that
the polarised valence quark distributions $\Delta u_v(x,Q^2)$ and $\Delta
d_v(x,Q^2)$ can be determined with some accuracy from the data, while
the polarised sea quark and gluon distributions $\Delta
\bar{q}(x,Q^2)$ and $\Delta g(x,Q^2)$ are only loosely constrained by
the structure function measurements. 

Contrary to $g_1$ and $g_2$, definite theoretical predictions exist
for the first moment of $g_1$, $\Gamma_1 = \int _0^1 g_1(x)\; {\rm d}x$, which
measures
the expectation value of the axial vector current between two
nucleon states. Two sum rules exist for $\Gamma_1$.
The fundamental one was obtained by Bjorken \cite{bjorken}
 from the current algebra and isospin symmetry between the proton and the 
 neutron:
\begin{equation}
\label{bj_sr}
\Gamma_1^{\rm p} - \Gamma_1^{\rm n} =
\frac{1}{6} \left | \frac{g_A}{g_V} \right | = {1\over 6}(\Delta u - \Delta d)
\end{equation}
where $g_A$ and $g_V$ are the axial and vector weak coupling constants
in the neutron beta decay. The QCD corrections to this sum rule have been
computed up to the order $\alpha_S^3$ \cite{19} and the 
${\cal O}(\alpha_S^4)$ have been estimated \cite{20}. 

Separate sum rules, obtained by Ellis and Jaffe \cite{ellisjaffe},
hold for the proton and the neutron:
\begin{equation}
\label{ej_sr}
\Gamma_1^{p(n)} = \pm {1\over 12}\left | \frac{g_A}{g_V} \right | + 
{1\over 36}a_8 + {1\over 9} \Delta \Sigma
\end{equation}
Here $\Delta \Sigma = \Delta u+\Delta d+\Delta s$ is the flavour 
singlet axial coupling and $\Delta q$ denote first moments
of the spin dependent parton distributions in the proton,
$\Delta q=\int_0^1\Delta q_i(x)\; {\rm d}x$; $a_8$ (and 
$\left | {g_A}/{g_V} \right |$) are related to the symmetric and
antisymmetric weak flavour-SU(3) couplings in the baryon octet. If the
flavour-SU(3) is exact then $a_8$ can be predicted from measurements 
of hyperon
dacays. There is however no prediction for $\Delta \Sigma$, except for
$\Delta s$=0. In this case $\Delta \Sigma = a_8$, as was assumed in the
original formulation by Ellis and Jaffe~\cite{ellisjaffe}. 
QCD corrections to these sum rules have been calculated up to the order
$\alpha_S^2$ \cite{21} and the the ${\cal O}(\alpha_S 
^3)$ have been estimated \cite{22}. Due to the axial anomaly of the
singlet axial vector current, $\Delta \Sigma$ is intrinsically
$Q^2$--dependent. Depending on the factorization scheme applied
\cite{lamli} this results either in a scale--dependence of the sea
quark polarization or in an extra contribution to the Ellis--Jaffe sum
rule,  involving $\Delta g=\int_0^1[g^+(x)-g^-(x)]\; {\rm d}x$,
the gluonic equivalent of the quark distribution moments. Both
formulations are equivalent.

Higher twist effects in the $Q^2$ dependence of $\Gamma_1$ will not be
considered here. 

Evaluation of the $\Gamma_1$ requires knowledge of $g_1$ in the entire
interval from 0 to 1. Measurements cover a limited kinematic range and thus
extrapolations of $g_1$ to 0 and 1 are necessary. The latter is not
critical since $g_1\rightarrow$0 at $x\rightarrow$1 but the former is a
considerable problem since $g_1$ increases as $x$ decreases and its
behaviour at low $x$ is theoretically not understood, see sect.7.4. 
Therefore results on
$\Gamma_1$ depend on the assumptions made in the $x\rightarrow$0 
extrapolation.
Both SMC and SLAC experiments assume the Regge like behaviour of $g_1$,
i.e. that at $x\rightarrow$0, $g_1$
behaves like $x^\alpha$, 0$\leq \alpha \leq$ 0.5. A value $\alpha$=0 was
chosen and  $g_1$ was fitted to the two data points at lowest
$x$, allowing for variation of this behaviour within the Regge model.
This approach might however be inconsistent with QCD which predicts a
faster rise of $g_1$ at low $x$.

\subsection{Experiments}
New generation polarised electroproduction experiments are listed in
table \ref{table7.1}. 

The experiment of the Spin Muon Collaboration
(SMC) at CERN uses a naturally polarised muon beam  
and a cryogenic, solid state target. Experiments E142 -- E155 at
SLAC use an electron beam and liquid (solid) cryogenic targets. The
HERMES experiment at DESY uses an electron beam from the HERA collider
and internal gas targets. The scattered muon spectrometers in the SMC and
SLAC experiments have been used (with little change) in DIS experiments
preceeding the polarised programme, contrary to the HERMES apparatus. 
Electron and muon measurements are complementary: the former offers
very high beam intensities and thus statistics but its kinematic 
acceptance is limited to low values of $Q^2$ and moderate values
of $x$, the latter extends to higher $Q^2$ and 
down to low values of $x$ (an important aspect in the study of sum rules) but 
due to limited muon intensities the data
taking time has to be long to ensure a satisfactory statistics.

\begin{table}[htb]
\caption{New generation experiments on polarised deep inelastic charged
lepton -- nucleon scattering. The last column shows references to the principal
physics results obtained until now, (from \protect\cite{voss}).}
\begin{center}
\begin{tabular}{|l|c|c|c|c|c|}
\hline \hline
 Experiment & Beam & Year & Beam energy (GeV) & Target & References\\ \hline 
 \hline
SMC & $\mu^+$& 1992--5 & 100,190  & C$_4$D$_9$OD & \cite{45,48} \\
& & 1993 & 190 & C$_4$H$_9$OH & \cite{46,47} \\
& & 1996 & 190 & NH$_3$ & \\ \hline
E142 & $e^-$ & 1992 & 19.4 --25.5 & $^3$He & \cite{50} \\
E143 & $e^-$ & 1993 & 29.1 & NH$_3$, ND$_3$ & \cite{51,52} \\
E154 & $e^-$ & 1995 & 50 & $^3$He & \\
E155 & $e^-$ & 1996 & 50 & NH$_3$, ND$_3$ & \\ \hline
HERMES & $e^-$ & 1995-- & 30--35 & H, D, $^3$He & \\
\hline \hline
\end{tabular}
\end{center}
\label{table7.1}
\end{table}

The lowest $x$ in the results published by the SMC is about 10$^{-3}$ 
and corresponds to
$Q^2$ about 1 GeV$^2$. In the course of analysis are events having lower
$Q^2$ and reaching $x$ values of 10$^{-4}$. A special trigger has been
set up recently to extend the measurements down to $x$=10$^{-5}$, at
the expense of lowering $Q^2$ to 0.01 GeV$^2$. The upper limit of $Q^2$
in the SMC is about 100 GeV$^2$. The SLAC experiments' acceptance extends
from $x$ about 0.01 at $Q^2=1\;\mbox{GeV}^2$ up $Q^2 \approx
10\;\mbox{GeV}^2$ at
$x\sim$0.7. As in all fixed-target experiments, the low values of $x$
in the SMC and SLAC are correlated with low $Q^2$.

The cross section asymmetry measured in the polarised lepton -- polarised
nucleon experiments, $A_{exp}$, is related to the
asymmetries defined in eq.(\ref{A}) by $A_{exp}= fP_tP_bA$ where $P_t,P_b$
denote the target and beam polarisations and $f$, the target dilution
factor, accounts for the fact that only a fraction of
nucleons is polarised. The beam polarisation at the SMC has been measured
with a purpose-built polarimeter, using two independent methods:
polarised $\mu e$ scattering and an analysis of the energy spectrum of
electrons coming from the muon decay. The result is $P_\mu=-$0.790$\pm$0.025
at 190 GeV beam energy. The target, subdivided into two cells polarised
in opposite directions, was typically polarised up to 50$\%$ for the
deuteron and 85$\%$ for the proton target. The target spin directions
were reversed 5 times a day. Polarisation of the SLAC electron beam
reached 86$\%$ in the E143 and was randomly reversed. Polarisation of the
targets reached 80$\%$ for the proton and 25$\%$ for the deuteron one in
E143. Systematic uncertainties in the SMC and SLAC experiments are similar.

\subsection{Results of the measurements 
and spin structure of the nucleon}
Cross section asymmetries $A_1$ and spin dependent structure 
functions $g_1$ have
been measured for the proton and deuteron targets by the
SMC, \cite{45,48,46,47} and by the E143, \cite{51,52}. Information on the
neutron has been evaluated from the data on $^3$He (E142, \cite{50})
and from the data on the proton and deuteron (SMC, \cite{45,48}).
All data sets are in a very good mutual agreement even if $A_1$, 
extracted from data covering different $Q^2$ intervals, has
been assumed to be $Q^2$ independent. 

Results on $g_1$ for proton, deuteron and neutron are shown in Figs.
7.1, 7.2.
Here $g_1^n=2g_1^d/(1-1.5\omega _D) -
g_1^p$ where $\omega _D 
\sim$ 0.05 is the probability of the D-state of the deuteron. 
Conversion of $A_1$ to $g_1$, which was made under an assumption 
that $A_1$ scales, needs only information on
the structure function $F_1$ or, equivalently, $F_2$ and $R$ (cf.
eq.(\ref{a1a2})). The NMC parametrisation of $F_2(x,Q^2)$
\cite{NMCF2} and the SLAC 
parametrisation of $R(x,Q^2)$ \cite{rworld} have been used by both
SMC and SLAC. 

\begin{figure}
\begin{center}
 \epsfig{file=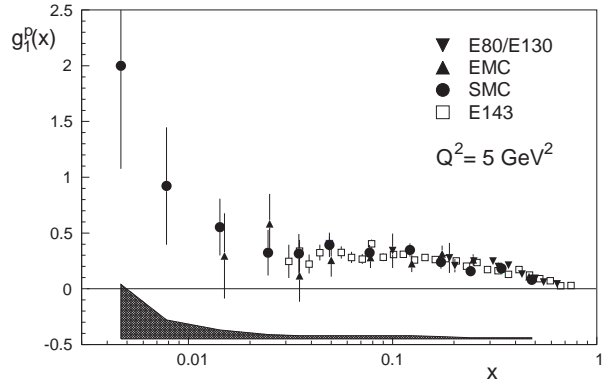,height=5cm}
\end{center}
\caption{The spin dependent structure function $g_1(x)$ of the proton at 
$Q^2$=5 GeV$^2$. the EMC data were reevaluated using the same $F_2$ and $R$
parametrisations as for the SMC and E143 data. Error bars are statistical; the
shaded area marks the SMC systematic errors. Figure taken from
\protect{\cite{voss}}.} 
\label{fig_g1p}
\end{figure}
\begin{figure}
\epsfxsize=13.cm
\centering
\leavevmode
\epsfbox[50 0 770 280]{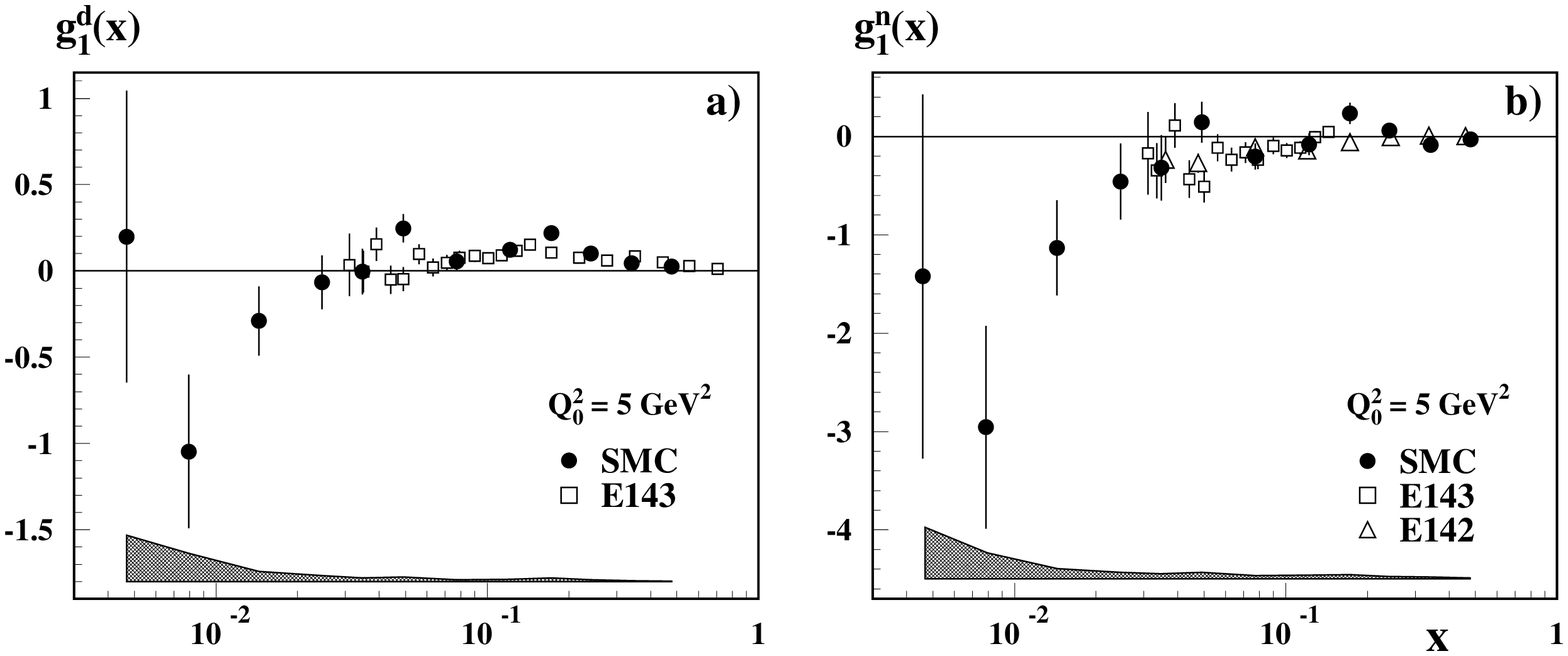}
\label{fig_g1d}
\caption{
The spin dependent structure functions (a) $g_1^d(x)$ and (b) $g_1^n(x)$, as a
function of $x$ at $Q_0^2$=5 GeV$^2$. error bars are statistical; the
shaded area marks the SMC systematic errors. Figure taken from
\protect{\cite{48}}.
} 
\end{figure}

The behaviour of the $g_1^p$ is different from that of $g_1^d$ and $g_1^n$,
especially at low $x$. This should be contrasted with the unpolarised 
case where a small difference between proton and neutron structure functions
can be explained by nuclear shadowing in the deuteron.
Measurements of the asymmetry, $A_2$, for the proton \cite{47} 
and the deuteron \cite{a2d} showed that this function
is significantly smaller than the bound $\sqrt{R}$ and consistent with zero.

The Bjorken sum rule seems to be fulfilled by the above data at the 10$\%$
level: its value measured by the SMC at $Q^2=$10 GeV$^2$ is $\Gamma_1^p - 
\Gamma_1^n = 
$0.199$\pm$0.038, to be compared with the QCD prediction (four flavours,
corrections up to $\alpha_S^3$): 0.187$\pm$0.003. The Ellis-Jaffe sum rule is
not confirmed by the data, the 
pattern of disagreement being similar in the proton
and deuteron results.  At $Q^2=5\;\mbox{GeV}^2$, the combined data of the 
EMC, SMC and SLAC give for $\Gamma_1^p, \Gamma_1^d$ and $\Gamma_1^n$
respectively: 
0.125$\pm$0.009, 0.041$\pm$0.005 and $-$0.037$\pm$0.008 as compared to the
predicted 0.167$\pm$0.005, 0.070$\pm$0.004 and $-$0.015$\pm$0.005. 
A most straightforward explanation of this violation may be a non--zero
polarisation of the strange sea.
Results of the
new SMC proton data analysis, with extended kinematic coverage and a NLO QCD
analysis to evolve the measured $g_1(x,Q^2)$ to a common value of $Q^2$,
confirm all the above conclusions.

The nucleon spin, $S_z={\textstyle\frac{1}{2}}$, can be decomposed as follows
\begin{equation}
S_z={\textstyle{1\over 2}}\Delta \Sigma + \Delta g + L_z
\label{spin}
\end{equation}
where $L_z$ is angular momentum due to the partons.
All the data sets (except perhaps E142), evaluated with a
consistent treatment of the QCD corrections at a common $Q^2$=5 GeV$^2$ 
and under the assumption that the flavour SU(3) is exact, 
show that the total quark contribution to the nucleon spin is small,
 $\Delta \Sigma \sim $0.2, and that the strange sea is indeed polarised: 
$\Delta s\sim-$0.1. SU(3)--breaking can decrease 
$\Delta s$ but leaves $\Delta \Sigma $ unchanged.
Choosing a factorization scheme in which the quarks polarisation is
scale independent, a $Q^2$ dependent gluonic contribution appears in the 
Ellis-Jaffe sum rule as a result of the anomalous dimension of the singlet
axial vector current \cite{lamli}. Then the Ellis-Jaffe assumption of
$\Delta s$=0 implies that at $Q^2$=5 GeV$^2$, $\Delta g\sim 3$ is needed
to restore the sum rule.

Finally we note the first measurements of the semi--inclusive spin
asymmetries for positively and negatively charged hadrons in the polarised
muon--proton and muon--deuteron scattering in the SMC experiment \cite{ww}.
The $x$ dependence of the spin distributions for the up and down valence quarks 
and for the non-strange sea quarks has been determined. 
The moments of the quark spin distributions
were obtained to be: $\Delta u_v$=1.01$\pm$0.24, $\Delta d_v=-$0.57$\pm$0.25;
moments for the non-strange sea quarks are consistent with zero over the whole
measured range of $x$.

\subsection{Theory predictions for $g_1$ at small $x$}
Observation of a difference  between the
proton and deuteron spin structure functions $g_1^{p,d}(x,Q^2)$ at
small $x$, Figs. 7.1, 7.2, indicates a sizeable {\it
non--singlet} contribution to $g_1$ in this region.
The dominant behaviour of $g_1^{n.s.}$ can be estimated from
a resummation of the leading QCD corrections. The
conventional Altarelli-Parisi
approach only resums terms of the form $(\alpha_S \ln
Q^2 \ln (1/x) )^k$ and yields
\begin{equation}
\label{eq:glap}
g_1^{n.s.}(x,Q^2) \sim \exp\left(\sqrt{\frac{2\alpha_S C_F}{\pi} \ln
\frac{Q^2}{\mu^2} \ln \frac{1}{x}} \right).
\end{equation}
Terms proportional to $(\alpha_S \ln^2(1/x))^k$, which arise due to
the presence of two fermions in the $t$-channel \cite{ggfl} are,
although more singular at small $x$, not included.

Ryskin outlined a calculation of the small $x$
behaviour of $g_1^{n.s.}$, which resums these contributions \cite{ber}.
This calculation not only includes the enlarged (compared to the
Altarelli-Parisi approach) phase space of ladder diagrams which leads to
the double logarithmic contributions, but also substantially differs
from similar resummations
performed for unpolarised structure functions in that both ladder
and non--ladder diagrams contribute to the most singular part of the
amplitude. Using the fact that the non--ladder gluon has to
be softer than all gluons within the ladder, these non--ladder
contributions can be resummed. This
procedure results in a nonlinear infrared evolution equation
\cite{kl} for the quark amplitudes of odd signature, which mixes
color--singlet and color--octet quark amplitudes.
By means of a Mellin transformation, this equation can be solved,
yielding
\begin{eqnarray}
g_1^{n.s.} & \sim & x^{-\omega^{(-)}_{n.s.}} \quad \mbox{with }
\omega_{n.s.}^{(-)
}
\simeq 1.04 \sqrt{2 \alpha_S C_F/\pi} \simeq 0.4 \quad (\mbox{for }
\alpha_S = 0.18). \nonumber \\
\end{eqnarray}
The non--ladder (bremsstrahlung) contributions give a relatively small
contribution (4\%) to $\omega_{n.s.}^{(-)}$; if only ladder diagrams 
are retained then $\omega_{n.s.}^{(-)} = \omega_{n.s.}^{(+)} = 
\sqrt{2 \alpha_S C_F / \pi}$ for fixed coupling. When running coupling
effects are included in the double logarithmic resummation generated
by the ladder diagrams then the effective slope is reduced to 0.2 - 0.3
\cite{JKS}. 

First preliminary results of a similar calculation for
the singlet contribution to
$g_1(x)$ at small $x$ were presented \cite{BER2}, indicating that
\begin{equation}
g_1^{s.} \sim x^{-\omega^{(-)}_{s.}} \qquad \mbox{with }
\omega_{s.}^{(-)}
\simeq 3 \; \omega_{n.s.}^{(-)} > 1 \quad (\mbox{for }
\alpha_S>0.12).
\end{equation}
As the above result contradicts conventional Regge--type extrapolations
of $g_1^{n.s.}$ in the Ellis--Jaffe sum rule, it shows the need for
extending the measurements of $g_1(x,Q^2)$ down to the lowest possible values
of $x$.

\subsection{Prospects for the future}
Understanding of the polarised structure functions has improved
dramatically in the recent years, thanks to the EMC, SMC and SLAC
measuremets. Several questions however remained unanswered. Among them is
the low $x$ behaviour of $g_1$, its $Q^2$ evolution, the gluon
polarisation and flavour decomposition of polarised parton distribution. 
The HERMES experiment, recently starting at HERA, using
a polarised electron beam and a polarised internal gas target will
especially address the last question from a presently unique
reconstruction of the hadronic final state.
To answer the remaining questions
a new generation of experiments, e.g. at the HERA {\it collider}, is needed. 
Prospects of spin physics at HERA were discussed at a workshop at
DESY--Zeuthen in August 1995. A polarised deep inelastic programme at HERA
could allow measurements over an extended kinematic range,
including low $x$ and high $Q^2$. Polarisation of the
proton beam is technically much more complicated than polarisation
of the electron beam, as the proton beam does not polarise naturally
\cite{herap}. 
Construction of the polarised proton beams of energy up to 250 GeV
in the RHIC collider rings has already been approved, a helpful step for HERA.
Various suggestions for dedicated
measurements of $\Delta g(x,Q^2)$, including the HMC/CHEOPS project at CERN,
were also discussed at the Zeuthen workshop \cite{dg}.  

\setcounter{section}{7}
\setcounter{equation}{0}
\renewcommand{\theequation}{8.\arabic{equation}}
\setcounter{figure}{0}
\renewcommand{\thefigure}{8.\arabic{figure}}
\setcounter{table}{0}
\renewcommand{\thetable}{8.\arabic{table}}

\section{Conclusions}
We believe that this report gives a fair summary of our present understanding
of the structure of the proton, as it is seen in deep-inelastic lepton-proton 
scattering. Much --- but not all --- of the recent experimental information
comes from HERA.  We have summarized the HERA and fixed-target data in 
the first part of this Report.

The most puzzling phenomenon continues to be the rise of $F_2$ at small $x$,
which seems to persist even at low values $Q^2$, and its connection with the
soft Pomeron seen in photoproduction at $Q^2=0$.
So far there is no real problem in describing the inelastic
data within the GLAP evolution scheme, i.e. it is possible 
to find starting distributions for quarks and gluons which evolve according to
the (next-to-leading) QCD evolution equations and successfully describe the
data. This alone, however, does not ${\it explain}$ the observed rise: either
one has to ascribe the origin of the rise to the input (gluon) distribution
(the BFKL scenario), or one has to allow a rather large $Q^2$ range of
the GLAP evolution. In the latter case, the GLAP evolution by itself leads
to a rise a small $x$. But in order to accommodate the observed rise at
low $Q^2$ the QCD evolution has to start at an even lower scale where -
on rather general grounds - one would have been hesitant to use a leading-twist 
perturbative QCD framework.
A further ambiguity is that, as yet, the \lq\lq BFKL scenario" is based only 
on the resummation of leading $\log (1/x)$ terms.  A quantitative description 
needs the computation of sub-leading effects.

The discussions of the Working Group, as described in this summary,
reflect the agreement among the participants that we 
need a thorough understanding of the validity of GLAP and BFKL descriptions 
(and 
a description which unifies both) in the low-$x$ and low-$Q^2$ region, as well 
as the 
interpolation between deep inelastic scattering and photoproduction, before
a deeper understanding of the interface between perturbative and 
non-perturbative 
physics can be reached. It seems clear that to gain further insight we must 
study more than just the fully inclusive 
total cross section as measured by $F_2$, i.e. we need to look more closely
into the final state of deep-inelastic scattering. 

The discussion of the spin structure of the proton also seems to be developing  
in the small-$x$ direction, somewhat analogous to the 
unpolarized case. Namely, the low-$x$ data points for both the proton and the
neutron spin structure functions $g_1$ indicate the possibility of a rise 
at small $x$. At the same time QCD calculations, based on leading and 
next-to-leading order Altarelli-Parisi splitting functions or, more recently, 
using the double-logarithmic approximation, predict at strong rise of $g_1$ 
at small $x$.  So there is little doubt that small-$x$ physics is becoming 
a field of particular interest also in the proton spin community.

\section*{Acknowledgements}

We thank all those who participated in the Proton Structure Working Group.  In 
particular we thank the people who gave talks to the Group, including Johannes 
Blumlein, Tim Carroll, Stefano Catani, John Dainton, Sandy Donnachie, Leonid 
Frankfurt, Mark Gibbs, Tim Greenshaw, Franseco Hautmann, Rob Henderson, 
Indumathi, Dave Kant, Valeri Khoze, Jan Kwieci\'nski, 
Genya Levin, Steve Riemersma, Misha Ryskin, Bill Seligman, Tara Shah, Torbjorn 
Sj\"{o}strand and Peter Sutton, 
besides the authors.  We especially thank Keith Ellis who, at 
very short notice, presented the summary talk for our group, and Mike Whalley, 
as well as Robin Devenish, for such excellent organisation before, during and 
after the Workshop.  We are grateful to Michael Kr\"{a}mer for useful 
information.  
We also thank Sharon Fairless and Rachel Lumpkin for their 
invaluable help in processing this article and PPARC for financial support.

\end{document}